\documentstyle[mprocl]{article}

\def\ra{\rightarrow}

\def\be{\begin{equation}}
\def\ee{\end{equation}}
\def\beq{\begin{equation}}
\def\eeq{\end{equation}}
\def\bea{\begin{eqnarray}}
\def\eea{\end{eqnarray}}

\newcommand{\matel}[3]{\langle #1|#2|#3\rangle}

\begin{document}

\begin{flushright}
UND-HEP-99-BIG01\\
January 1999
\end{flushright}

\vspace{1cm}

\title{CP VIOLATION AND BEAUTY DECAYS -- A CASE STUDY 
OF HIGH IMPACT, HIGH SENSITIVITY AND EVEN HIGH 
PRECISION PHYSICS 
\footnote{Invited Lectures given at the First APCTP Hadron 
Physics Winter School, Feb. 23 - 27, 1998, APCTP, Seoul, 
Korea}}

\author{I. I. BIGI}

\address{Physics Dept., University of 
Notre Dame du Lac, Notre Dame,\\ IN 46556, USA\\
E-mail: bigi@undhep.hep.nd.edu}   

\maketitle\abstracts{
The narrative of these lectures contains three main threads:  
(i) CP violation despite having so far been observed only in 
the decays of neutral kaons has been recognized as 
a phenomenon of truly fundamental importance. 
The KM ansatz constitutes the minimal implementation of CP 
violation: without requiring  
unknown degrees of freedom it can reproduce the known 
CP phenomenology in a nontrivial way.  
(ii) The physics of beauty hadrons -- in particular their 
weak decays -- opens a novel window onto 
fundamental 
dynamics: they usher in a new quark family (presumably 
the last one); they allow us to determine fundamental 
quantities of the Standard Model like the 
$b$ quark mass and the 
CKM parameters $V(cb)$, $V(ub)$, $V(ts)$ and $V(td)$; 
they exhibit speedy or even rapid $B^0 - \bar B^0$J
oscillations. 
(iii) Heavy Quark Expansions allow us to treat $B$ decays 
with an accuracy that would not have been thought 
possible a mere decade ago. 
These three threads are 
joined together in the following 
manner: (a) Huge CP asymmetries are {\em pre}dicted in 
$B$ decays, which represents a decisive test of the KM 
paradigm for CP violation. (b) Some of these 
predictions are made with high {\em parametric} 
reliability, which (c) can be translated into 
{\em numerical} precision through the judicious 
employment of novel theoretical technologies. 
(d) Beauty decays thus provide us with a rich and 
promising field to search for New Physics 
and even study some of its salient features. At the end of it 
there might quite possibly be a New Paradigm for 
High Energy Physics.  There will be some other threads 
woven into this tapistry: electric dipole moments, 
and CP 
violation in other strange and in charm decays.   
}

\tableofcontents

\begin{quotation} 
\noindent 
Schl\" aft ein Lied in allen Dingen, \\ 
die da tr\" aumen fort und fort, \\
und die Welt hebt an zu singen, \\
findst Du nur das Zauberwort.\\

\bigskip 

\noindent 
Sleeps a song in all things \\ 
that dream on and on \\ 
and the world will start to sing \\ 
if you find the magic word. \\ 
\begin{flushright}
{\em J. v. Eichendorff} 
\end{flushright}
\end{quotation}

\section{Prologue}
With very few symmetries in nature manifestly realized, 
why do I 
think that the breaking of CP invariance is very special -- 
more subtle, more fundamental and more profound than parity 
violation? 
\begin{itemize}
\item 
Parity violation tells us that nature makes a difference between 
"left" and "right" -- but not which is which! For the 
statement that neutrinos emerging from pion decays are 
left- rather than right-handed implies the use of 
positive instead of negative pions. "Left" and 
"right" is thus defined in terms of "positive"  
and "negative", respectively. This is like saying that 
your left thumb is on your right hand -- certainly 
correct, yet circular and thus not overly useful. 

On the other hand CP violation manifesting itself through 
\be 
\frac{{\rm BR}(K_L \ra l^+ \nu \pi ^-)}
{{\rm BR}(K_L \ra l^- \bar \nu \pi ^+)} \simeq 1.006J\neq 1 
\ee 
allows us to define "positive" and "negative" in 
terms of 
{\em observation} rather than {\em convention}, and 
subsequently likewise for "left" and "right". 
\item 
The limitation on CP invariance in the $K^0 - \bar K^0$ 
mass matrix 
\be 
{\rm Im} M_{12} \simeq 1.1 \cdot 10^{-8} \; \; {\rm eV} \; \; 
\hat = \; \; \frac{{\rm Im} M_{12}}{m_K} \simeq 2.2 \cdot  
10^{-17} 
\ee
represents the most subtle symmetry violation 
actually observed to date. 

\item 
CP violation constitutes one of the three essential 
ingredients in any 
attempt to understand the observed baryon number of the universe 
as a dynamically generated quantity rather than an initial 
condition \cite{DOLGOV2}. 

\item 
Due to CPT invariance -- which will be assumed throughout    
these lectures -- CP breaking implies a violation of 
time reversal invariance 
\footnote{{\em Operationally} one defines time reversal as the 
reversal 
of {\em motion}: $\vec p \ra - \vec p$, $\vec j \ra - \vec j$ for 
momenta $\vec p$ and angular momenta $\vec j$.}. That nature 
makes an 
intrinsic distinction between past and future on the 
{\em microscopic} level that cannot be explained by statistical 
considerations is an utterly amazing observation.   

\item 
The fact that time reversal represents a very peculiar 
operation can 
be expressed also in a less emotional way, namely through 
{\em Kramers' Degeneracy} 
\cite{KRAMERS}. Since the time reversal operator $\bf T$ 
has to be {\em anti}-unitary, ${\bf T}^2$ has eigenvalues 
$\pm 1$. Consider the sector of the Hilbert space with 
${\bf T}^2 = -1$ and assume the dynamics to conserve ${\bf T}$; 
i.e., the Hamilton operator $\bf H$ and ${\bf T}$ commute. It is 
easily shown that if $|E\rangle$ is an eigenvector of 
$\bf H$, so is 
${\bf T}|E\rangle$ -- with the {\em same} eigenvalue. Yet 
$|E\rangle$ and ${\bf T}|E\rangle$ are -- that is the main 
substance of this theorem -- orthogonal to each other! Each 
energy eigenstate in the Hilbert sector with ${\bf T}^2 = -1$ 
is therefore at least doubly degenerate. This degeneracy is 
realized 
in nature through {\em fermionic spin} degrees. Yet it is 
quite remarkable that the time reversal operator $\bf T$ already 
anticipates this option -- and the qualitative difference 
between fermions and bosons -- through ${\bf T}^2 = \pm 1$ -- 
{\em without} any explicit reference to spin! 

\end{itemize}
These lectures will be organized as follows: in Lectures 
I and II -- 
covered in Sect. \ref{CPPHENOM} - \ref{SUMMARYLIGHT} -- 
I will list the 
existing CP phenomenology, introduce the KM ansatz as the 
minimal implementation of CP violation and apply it to 
$\Delta S = 2,1$ transitions; in Lecture III -- 
Sect. \ref{BEAUTY} - \ref{CHARM} -- I describe in some detail 
CP violation in beauty and charm decays, 
both from the perspective of the KM ansatz as well as 
New Physics; in Lecture IV -- Sect. \ref{HQE} -- 
I describe Heavy Quark Expansions and their applications to beauty 
decays before 
giving a summary and presenting an outlook. 

While Lectures I - III represent 
a slight up-date of previous lectures \cite{VARENNA}, 
Lecture IV contains also a review of some very recent work 
on topical issues like quark-hadron duality.

\section{CP Phenomenology in $K_L$ Decays}
\label{CPPHENOM} 
\subsection{Symmetries and Particle-Antiparticle Oscillations}
A symmetry $\bf S$ can {\em manifestly} be realized in two 
different 
ways:    
\begin{itemize}
\item 
There exists a pair of degenerate states that transform 
into each 
other under $\bf S$. 
\item 
When there is an {\em un}paired state it has to be an eigenstate of 
$\bf S$. 
\end{itemize}
The observation of $K_L$ decaying into a $2\pi$ state -- which 
is CP even -- and a CP odd $3\pi$ combination therefore 
establishes CP violation only because $K_L$ and $K_S$ are 
{\em not} mass degenerate. 

In general, decay rates can exhibit CP violation in three different 
manners,  
namely through 
\begin{enumerate}
\item 
a {\em difference} in CP conjugate rates, like 
$K_L \ra l^- \bar \nu \pi ^+$ vs. 
$K_L \ra l^+  \nu \pi ^-$,
\item 
the {\em existence} of a reaction, like 
$K_L \ra \pi \pi$, 
 
\item 
a decay rate evolution that is {\em not a purely 
exponential} function of the proper time of decay; i.e., 
if one finds for a CP {\em eigenstate} $f$ 
\be 
\frac{d}{dt} e^{\Gamma t}{\rm rate} 
(K_{neutral}(t) \ra f) \neq 0 
\ee  
for all (real) values of $\Gamma$, then CP symmetry must be 
broken. This is easily proven: 
if CP invariance holds, the decaying state 
must be a CP eigenstate like the final state $f$; yet in that case 
the decay rate evolution must be purely exponential -- unless 
CP is violated. Q.E.D. 

\end{enumerate} 
That the first manner represents CP violation is obvious 
without further ado. The situation is a bit more 
subtle with respect to the other two:  
the second relies on the mass eigenstate not being a CP 
eigenstate and third one on flavour eigenstates not being 
mass eigenstates. That means the latter two categories 
involve particle-antiparticle oscillations in an essential way. 
 
The whole formalism of particle-antiparticle oscillations is 
actually a straightforward application of basic quantum mechanics. 
I will describe it in terms of strange mesons; the generalization to 
any other flavour or quantum number (like beauty or charm) is 
straightforward and will be given later. 

{\em As long as CP is conserved,} all relevant expressions can 
be given without having to tackle a differential equation 
explicitely. 
In the absence of weak forces one has two mass degenerate 
and stable mesons $K^0$ and $\bar K^0$ carrying definite 
strangeness $+1$ and $-1$, respectively, since the strong and 
electromagnetic  
forces conserve this quantum number. The addition of the weak 
forces changes the picture qualitatively: strangeness is no longer 
conserved, kaons become unstable and the new mass eigenstates 
-- being linear superpositions of $K^0$ and $\bar K^0$ -- no longer 
carry definite strangeness. The violation of the quantum number 
strangeness has lifted the degeneracy: we have two physical 
states $K_L$ and $K_S$ with different masses and lifetimes: 
$\Delta m_K = m_L - m_K \neq 0 \neq \Delta \tau = 
\tau _L - \tau _S$. 

The mass 
eigenstates $K_A$ and $K_B$ have to be CP 
eigenstates as pointed out 
above: $|K_A\rangle = |K_+\rangle$, $|K_B\rangle = |K_-\rangle$,  
where ${\bf CP}|K_{\pm}\rangle \equiv \pm |K_{\pm}\rangle$. 
Using the phase convention 
\be 
|\bar K^0 \rangle \equiv - {\bf CP} |K^0\rangle 
\ee 
the time evolution of a state that starts out as a $K^0$ is 
given by 
\be 
|K^0(t)\rangle = 
\frac{1}{\sqrt{2}} e^{-im_1t} e^{-\frac{\Gamma _1}{2}t} 
\left( |K_+\rangle + e^{-i\Delta m_Kt} e^{-\frac{\Delta \Gamma}{2}t}
|K_-\rangle \right) 
\ee 
The intensity of an initially pure $K^0$ beam traveling in 
vacuum  will then exhibit the 
following time profile: 
\be 
I_{K^0}(t) = |\langle K^0|K^0(t)\rangle |^2 = 
\frac{1}{4} e^{-\Gamma _1 t } 
\left( 1 + e^{\Delta \Gamma _Kt} + 2e^{\frac{\Delta \Gamma _K}{2} t} 
{\rm cos}\Delta m_K t\right)  
\ee 
The orthogonal state $|\bar K^0 (t)\rangle$ that was absent initially 
in this beam gets regenerated {\em spontaneously}: 
\be 
I_{\bar K^0}(t) = |\langle \bar K^0|K^0(t)\rangle |^2 = 
\frac{1}{4} e^{-\Gamma _1 t } 
\left( 1 + e^{\Delta \Gamma _K t} - 2e^{\frac{\Delta \Gamma _K}{2} t} 
{\rm cos}\Delta m_K t\right)  
\ee 
The oscillation rate expressed through $\Delta m_K$ and 
$\Delta \Gamma _K$ is naturally calibrated by the average 
decay rate $\bar \Gamma _K \equiv 
\frac{1}{2}( \Gamma _1 + \Gamma _2)$: 
\be 
x_K \equiv \frac{\Delta m_K}{\bar \Gamma _K} \simeq 0.95 
\; \; \; , \; \; \; 
y_K \equiv \frac{\Delta \Gamma _K}{2\bar \Gamma _K} \simeq 1  
\label{DMKDATA}
\ee 
Two comments are in order at this point: 
\begin{itemize}
\item 
In any such binary quantum system there will be two lifetimes. 
The fact that they differ so spectacularly for neutral kaons  
-- $\tau (K_L) \sim 600 \cdot \tau (K_S)$ -- is 
due to a kinematical accident: the only available nonleptonic 
channel for the CP odd kaon is the 3 pion channel, for which 
it has barely enough mass.  
\item 
$\Delta m_K \simeq 3.7 \cdot 10^{-6}$ eV is often related 
to the kaon mass: 
\be  
\frac{\Delta m_K}{m_K} \simeq 7 \cdot 10^{-15} 
\label{STRIKING} 
\ee 
which is obviously a very striking number. Yet 
Eq.(\ref{STRIKING}) somewhat overstates the point. 
The kaon mass has nothing really to do with the 
$K_L-K_S$ mass difference 
\footnote{It would not be much more absurd to relate 
$\Delta m_K$ to the mass of an elephant!} and actually is  
measured relative to $\Gamma _K$. There is however 
one exotic application where it makes sense to state 
the ratio $\Delta m_K/m_K$, and that is in the context 
of antigravity where one assumes matter and antimatter 
to couple to gravity with the opposite sign. The gravitational 
potential $\Phi$  
would then produce a {\em relative} phase between $K^0$ and 
$\bar K^0$ of  2 $m_K\Phi t$. In the earth's potential this would 
lead to a gravitational oscillation time of 
$10^{-15}$ sec, which is much shorter than the lifetimes or 
the weak oscillation time; $K^0 - \bar K^0$ oscillations could 
then not be observed \cite{GOOD}. 
There are some loopholes in this argument -- 
yet I consider it still intriguing or at least entertaining.

\end{itemize}

\subsection{General Formalism 
\label{GENFORM}}

Oscillations become more complex once CP symmetry is broken in 
$\Delta S=2$ transitions. The relevant formalism describes  
a general quantum mechanical situation. Consider a neutral 
meson $P$ with flavour quantum number $F$; it can denote 
a $K^0$ or $B^0$. The time evolution 
of a state being a mixture of $P$ and $\bar P$ is obtained 
from solving the (free) 
Schr\" odinger equation 
\be 
i\frac{d}{dt} \left( 
\begin{array}{ll}
P^0 \\
\bar P^0
\end{array}  
\right)  = \left( 
\begin{array}{ll}
M_{11} - \frac{i}{2} \Gamma _{11} & 
M_{12} - \frac{i}{2} \Gamma _{12} \\ 
M^*_{12} - \frac{i}{2} \Gamma ^*_{12} & 
M_{22} - \frac{i}{2} \Gamma _{22} 
\end{array}
\right) 
\left( 
\begin{array}{ll}
P^0 \\
\bar P^0
\end{array}  
\right) 
\label{SCHROED} 
\ee
CPT invariance imposes 
\be 
M_{11}= M_{22} \; \; , \; \; \Gamma _{11} = \Gamma _{22} \; . 
\label{CPTMASS}
\ee  
\begin{center} 
$\spadesuit \; \; \; \spadesuit \; \; \; \spadesuit $ \\ 
{\em Homework Problem \#1}: 
\end{center}
Which physical situation is 
described by an equation analogous to Eq.(\ref{SCHROED}) 
where however the two diagonal matrix elements differ 
{\em without} violating CPT? 
\begin{center} 
$\spadesuit \; \; \; \spadesuit \; \; \; \spadesuit $
\end{center} 
The subsequent discussion might strike the reader as overly 
technical, yet I hope she or he will bear with me since 
these remarks will lay important groundwork for a proper 
understanding of CP asymmetries in $B$ decays as well. 

The mass eigenstates obtained through diagonalising this matrix 
are given by (for details see \cite{LEE,BOOK}) 
\bea 
 |P_A\rangle &=& 
\frac{1}{\sqrt{|p|^2 + |q|^2}} \left( p |P^0 \rangle + 
q |\bar P^0\rangle \right) \\  
|P_B\rangle &=& 
\frac{1}{\sqrt{|p|^2 + |q|^2}} \left( p |P^0 \rangle -  
q |\bar P^0\rangle \right) 
\label{P1P2}
\eea 
with eigenvalues 
\bea 
M_A - \frac{i}{2}\Gamma _A &=& M_{11} - \frac{i}{2} \Gamma _{11} 
+\frac{q}{p}\left( M_{12} - \frac{i}{2} \Gamma _{12}\right) \\    
M_B - \frac{i}{2}\Gamma _B &=& M_{11} - \frac{i}{2} \Gamma _{11} 
- \frac{q}{p}\left( M_{12} - \frac{i}{2} \Gamma _{12}\right) 
\label{EV} 
\eea 
as long as 
\be 
\left( \frac{q}{p}\right) ^2 = 
\frac{M_{12}^* - 
\frac{i}{2} \Gamma ^*_{12}}
{M_{12} - 
\frac{i}{2} \Gamma _{12}} 
\label{Q/PSQ} 
\ee 
holds. I am using letter subscripts 
$A$ and $B$ for labeling the 
mass eigenstates rather than numbers $1$ and $2$ 
as it is usually done. For I want to 
avoid confusing them with the matrix indices 
$1,2$ in $M_{ij} - \frac{i}{2}\Gamma _{ij}$ for  
reasons that will become clearer later. 
 
Eqs.(\ref{EV}) yield for the differences in mass and width 
\bea 
\Delta M &\equiv& M_B - M_A = 
-2 {\rm Re} \left[ \frac{q}{p}(M_{12} - 
\frac{i}{2}\Gamma _{12})\right]  \\
\Delta \Gamma  &\equiv& \Gamma _A - 
\Gamma _B = 
-2 {\rm Im}\left[ \frac{q}{p}(M_{12} - 
\frac{i}{2}\Gamma _{12})\right] 
\label{DELTAEV} 
\eea
Note that the subscripts $A$, $B$ have been swapped in 
going from $\Delta M$ to $\Delta \Gamma$! This is 
done to have both quantities {\em positive} 
for kaons. 

In expressing the mass eigenstates $P_A$ and $P_B$ 
explicitely in terms of the flavour eigenstates -- 
Eqs.(\ref{P1P2}) -- one needs $\frac{q}{p}$. There 
are two solutions to Eq.(\ref{Q/PSQ}):  
\be 
\frac{q}{p} = \pm 
\sqrt{\frac{M_{12}^* - \frac{i}{2} \Gamma _{12}^*}
{M_{12} - \frac{i}{2} \Gamma _{12}}}
\label{Q/P} 
\ee
There is actually a more general ambiguity than this 
binary one. For antiparticles 
are defined up to a phase only: 
\be 
{\bf CP} |P^0 \rangle = 
\eta | \bar P^0 \rangle  \; \; \; {\rm with} 
\; \; |\eta| =1
\ee 
Adopting a different phase convention will change 
the phase for $M_{12} - \frac{i}{2} \Gamma _{12}$ 
as well as 
for $q/p$: 
\be 
|\bar P^0 \rangle \ra e^{i\xi}|\bar P^0 \rangle \; 
\Longrightarrow \; 
(M_{12}, \Gamma _{12}) \ra e^{i\xi} 
(M_{12}, \Gamma _{12}) \; 
\& \; 
\frac{q}{p} \ra e^{-i\xi} \frac{q}{p} \; , 
\ee
yet leave $(q/p) (M_{12} - \frac{i}{2} \Gamma _{12})$ 
invariant -- as it has to be since the eigenvalues, 
which are observables, depend on this combination, see 
Eq.(\ref{EV}). Also $\left| \frac{q}{p}\right|$ is an 
observable; its {\em deviation} from unity is one 
measure of CP violation in $\Delta F =2$ dynamics.

By {\em convention} most authors pick the 
{\em positive} sign in Eq.(\ref{Q/P}) 
\be 
\frac{q}{p} = +  
\sqrt{\frac{M_{12}^* - \frac{i}{2} \Gamma _{12}^*}
{M_{12} - \frac{i}{2} \Gamma _{12}}} \; . 
\label{QPPOS} 
\ee 
Up to this point the two states 
$|P_{A,B}\rangle$ are merely 
{\em labelled} by their subscripts. 
Indeed $|P_A\rangle$ and $|P_B\rangle$ switch places 
when selecting the minus rather than the plus sign in 
Eq.(\ref{Q/P}). 

One can define the labels $A$ and $B$ such that 
\be 
\Delta M \equiv M_B - M_A > 0
\label{DMPOS}
\ee 
is satisfied. Once this {\em convention} 
has been adopted, it becomes a sensible question 
whether 
\be 
\Gamma _B > \Gamma _A  \; \; \; {\rm or} \; \; \; 
\Gamma _B < \Gamma _A 
\ee 
holds, i.e. whether the heavier state is shorter or 
longer lived. 

In the limit of CP invariance there is more we can say: 
since the mass eigenstates are CP eigenstates 
as well, we can raise another meaningful question: 
is the heavier state CP even or odd? 
With CP invariance requiring 
arg$\frac{\Gamma _{12}}{M_{12}}=0$ 
we have $\left| \frac{q}{p}\right| =1$, i.e. 
$\frac{q}{p}$  
becomes a pure phase. 
It is then convenient to adopt a phase convention s.t. 
$M_{12}$ is real; it leads to 
\be 
\frac{q}{p} = \pm 1 
\ee 
Likewise we still have the freedom to choose between 
\be 
{\bf CP} |P^0 \rangle = + |\bar P^0 \rangle \; \; \; 
{\rm or} \; \; \; 
{\bf CP} |P^0 \rangle = - |\bar P^0 \rangle
\ee  
Let us consider various choices: 
\begin{itemize}
\item 
With $\frac{q}{p} = 1$ and 
${\bf CP}|P^0 \rangle = |\bar P^0 \rangle$ we have 
\bea 
|P_A \rangle &=& \frac{1}{\sqrt{2}} 
\left( |P^0\rangle + |\bar P^0 \rangle \right) = 
|P_+\rangle \\
|P_B \rangle &=& \frac{1}{\sqrt{2}} 
\left( |P^0\rangle - |\bar P^0 \rangle \right) = 
|P_-\rangle 
\label{P1P2++} 
\eea  
with $P_A$ and $P_B$ being CP even and odd, respectively: 
${\bf CP}|P_{\pm}\rangle = \pm |P_{\pm}\rangle$.  
\be 
M_{odd} - M_{even} = M_B - M_A = 
- 2 {\rm Re} \left[ \frac{q}{p}
\left( M_{12} - \frac{i}{2} \Gamma _{12} \right)
\right] = 
- 2 M_{12} 
\label{DM++} 
\ee 
\item 
Alternatively we can set $\frac{q}{p} = -1$ 
\bea 
|P_A \rangle &=& \frac{1}{\sqrt{2}} 
\left( |P^0\rangle - |\bar P^0 \rangle \right) =
|P_-\rangle \\
|P_B \rangle &=& \frac{1}{\sqrt{2}} 
\left( |P^0\rangle + |\bar P^0 \rangle \right) = 
|P_+\rangle
\eea 
while 
maintaining ${\bf CP}|P^0 \rangle = + |\bar P^0 \rangle$.  
$P_A$ and $P_B$ then switch roles; i.e., they are now 
CP odd and even, respectively. Accordingly: 
\be 
M_{odd} - M_{even} = M_A - M_B =  2 
{\rm Re} \left[ \frac{q}{p}
\left( M_{12} - \frac{i}{2} \Gamma _{12} \right)
\right] = 
- 2 M_{12} 
\label{DM-+}
\ee 
\item 
Finally let us consider choosing 
$\frac{q}{p} = 1$ together with 
${\bf CP}|P^0 \rangle = - |\bar P^0 \rangle$. $P_A$ and 
$P_B$ 
are again expressed by Eq.(\ref{P1P2++}), yet now are 
CP odd and even. Then 
\be 
M_{odd} - M_{even} = M_B - M_A = 2 M_{12} 
\label{DM+-} 
\ee
\item 
Eqs.(\ref{DM++},\ref{DM-+}) on one hand and Eq.(\ref{DM+-}) 
do not coincide on the surface. Yet we will see below that the 
theoretical expression for $M_{12}$ changes sign depending 
on the choice of ${\bf CP}|P^0 \rangle = \pm |\bar P^0\rangle$. 
Thus they all agree -- as they have to!  
\item 
It is attractive to write the general 
mass eigenstates in terms of the 
CP eigenstates as well: 
\bea 
|P_A \rangle &=& \frac{1}{\sqrt{1+|\bar \epsilon |^2}} 
\left( |P_+ \rangle + \bar \epsilon |P_-\rangle \rangle 
\right)  
\; \; \; , \; CP|P_{\pm}\rangle = \pm |P_{\pm}\rangle \\  
|P_B \rangle &=& \frac{1}{\sqrt{1+|\bar \epsilon |^2}} 
\left( |P_- \rangle + \bar \epsilon |P_+\rangle \rangle 
\right) 
\; ; 
\eea
$\bar \epsilon = 0$ means that the mass and 
CP eigenstates coincide, i.e. CP is conserved in 
$\Delta F=2$ dynamics driving $P - \bar P$ 
oscillations. With the phase between the 
orthogonal states $|P_+\rangle$ and 
$|P_-\rangle$ arbitrary, the phase of 
$\bar \epsilon$ can be changed at will and is not an 
observable; $\bar \epsilon$ can be expressed in terms of 
$\frac{q}{p}$, yet in a way that depends on the 
convention for the phase of antiparticles. For 
${\bf CP}|P\rangle = |\bar P\rangle$ one has 
\bea 
|P_+\rangle &=& \frac{1}{\sqrt{2}} 
\left( |P^0 \rangle + |\bar P^0 \rangle \right) \\
|P_-\rangle &=& \frac{1}{\sqrt{2}} 
\left( |P^0 \rangle - |\bar P^0 \rangle \right) \\
\bar \epsilon &=& 
\frac{1 - \frac{q}{p}}{1 + \frac{q}{p}} 
\eea  
whereas for ${\bf CP}|P\rangle = -|\bar P\rangle$ 
one finds 
\bea 
|P_+\rangle &=& \frac{1}{\sqrt{2}} 
\left( |P^0 \rangle - |\bar P^0 \rangle \right) \\
|P_-\rangle &=& \frac{1}{\sqrt{2}} 
\left( |P^0 \rangle + |\bar P^0 \rangle \right) \\
\bar \epsilon &=& 
\frac{1 + \frac{q}{p}}{1 - \frac{q}{p}} 
\eea 
\item 
The lack of orthogonality between $P_A$ and $P_B$ is a 
measure of CP violation in $\Delta F=2$ dynamics: 
\be 
\langle P_B | P_A\rangle = 
\frac{1 - \left| \frac{q}{p}\right| ^2} 
{1 + \left| \frac{q}{p}\right| ^2} = 
\frac{2{\rm Re} \bar \epsilon}{1+|\bar \epsilon |^2} 
\label{P1DOTP2} 
\ee 
\end{itemize}
Later we will discuss how to evaluate $M_{12}$ and thus 
also $\Delta M$ within a given 
theory for the $P-\bar P$ complex. The examples just listed 
illustrate that some care has to be applied in interpreting 
such results. For expressing mass eigenstates explicitely 
in terms of flavour eigenstates involves some conventions. 
Once adopted we have to stick with a convention; yet our 
original choice cannot influence observables.  

We had already referred to the fact that the 
{\em relative} phase between $\Gamma _{12}$ and 
$M_{12}$ represents an observable describing indirect 
CP violation. Therefore we adopt the notation 
\be 
M_{12} = \bar M_{12}e^{i\xi}\; , \; \; 
\Gamma _{12} = \bar \Gamma _{12}e^{i\xi}e^{i\zeta} \; \; 
{\rm and} \; \frac{\Gamma _{12}}{M_{12}} = 
\frac{\bar \Gamma _{12}}{\bar M_{12}}e^{i\zeta} \equiv 
r e^{i\zeta}
\label{MBARGBAR}
\ee
We restrict the angles $\xi$ and $\xi + \zeta$ to lie between 
$- \frac{\pi}{2}$ and $\frac{\pi}{2}$; i.r., the real 
quantities $\bar M_{12}$ and $\bar \Gamma _{12}$ are 
a priori allowed to be {\em negative} as well as 
{\em positive}! A relative minus sign between 
$M_{12}$ and $\Gamma _{12}$ is of course physically 
significant, while the absolute sign is not. Yet it turns out 
that the absolute sign provides us with a  
useful though dispensible bookkeeping device. 

Let me recapitulate the relevant points: 
\begin{itemize}
\item 
The labels of the two mass eigenstates $P_A$ and $P_B$ can 
be chosen such that 
\be 
M_{P_B} > M_{P_A} 
\ee 
holds. 
\item 
Then it becomes an {\em empirical} question whether 
$P_A$ or $P_B$ are longer lived: 
\be 
\Gamma _{P_A} > \Gamma _{P_B}  \; \; \; {\rm or} 
\; \; \; 
\Gamma _{P_A} < \Gamma _{P_B} \; \; ? 
\ee 
\item 
In the limit of CP invariance one can also raise the 
question whether it is the CP even or the odd state 
that is heavier. 
\item 
We will see later that within a {\em given theory} 
for $\Delta F =2$ dynamics  
one can calculate 
$M_{12}$, including its sign, if phase conventions 
are treated consistently. To be more 
specific: adopting a phase convention for 
$\frac{q}{p}$ and having ${\cal L}(\Delta F =2)$ one 
can calculate $\frac{q}{p} \left( M_{12} - 
\frac{i}{2}\Gamma _{12}\right) = 
\frac{q}{p}\matel{P^0}{{\cal L}(\Delta F =2)}{\bar P^0}$. 
Then one assigns the labels $B$ and $A$ such that 
$\Delta M = M_B - M_A = 
-2{\rm Re}\frac{q}{p} \left( M_{12} - 
\frac{i}{2}\Gamma _{12}\right)$ turns out to be 
{\em positive}! 
\end{itemize}

\subsection{The $K^0 - \bar K^0$ Complex}
For the kaon system I have already stated the 
observed values for $\Delta m_K$ and 
$\Delta \Gamma _K$ in Eq.(\ref{DMKDATA}); 
using the convention of Eq.(\ref{DMPOS}) --  
$\Delta m_K = M_B - M_A > 0$ -- the data tell 
us 
\be 
\Delta \Gamma _K = \Gamma _A - \Gamma _B > 0 \; ; 
\ee 
i.e., the (ever so slightly) heavier neutral kaon is 
(considerably) longer lived 
\footnote{The English language provides us with 
the convenient mnemonic that the subscript 
$L$ denotes both $l$onger in lifetime and 
$l$arger in mass.} 
and it is approximately the CP odd state. 
With CP violation small -- $\zeta _K = 
{\rm arg}\left( \Gamma _{12}^K/M_{12}^K\right) \ll 1$ -- 
one deduces from Eqs.(\ref{DELTAEV}) in the notation 
of Eq.(\ref{MBARGBAR}) with the convention of 
Eq.(\ref{QPPOS}): 
\be 
\Delta m_K \simeq - 2 \bar M_{12}^K \; \; , \; 
\Delta \Gamma _K \simeq 2 \bar \Gamma _{12}^K 
\label{DELTAMM12} 
\ee 

Let me add a few comments that apply specifically 
here: 
\begin{itemize} 
\item 
On very general grounds -- without recourse to any model -- 
one can infer that CP violation in the neutral kaon system 
has to be small. 
The 
{\em Bell-Steinberger relation} allows to place a bound on 
the scalar product of the two mass 
eigenstates, introduced in Eq.(\ref{P1DOTP2}) 
\cite{LEE,BOOK}: 
\be 
\langle K_L |K_S \rangle \leq \sqrt{2} 
\sum _f \sqrt{\frac{\Gamma _L^f\Gamma _S^f}{\Gamma _S^2} } 
\leq \sqrt{2} \sqrt{\frac{\Gamma _L}{\Gamma _S}} 
\simeq 0.06 
\label{BELL} 
\ee 
There is no input from any CP measurement. What is essential, 
though, is the huge lifetime ratio. 
\item 
There are actually two processes underlying the transition 
$K_L \ra 2\pi$: $\Delta S=2$ forces generate 
the mass eigenstates 
$K_L$ and $K_S$ whereas $\Delta S=1$ dynamics drive the decays 
$K \ra 2\pi$. Thus CP violation can enter in 
two a priori independant 
ways, namely through the $\Delta S=2$ and the $\Delta S=1$ 
sector. This distinction can be made explicit in terms of the 
transition amplitudes: 
\be 
\eta _{+-} \equiv 
\frac{A(K_L \ra \pi ^+ \pi ^-)}{A(K_S \ra \pi ^+ \pi ^-)} \equiv 
\epsilon _K+ \epsilon ^{\prime} \; , \; 
\eta _{00} \equiv 
\frac{A(K_L \ra \pi ^0 \pi ^0)}{A(K_S \ra \pi ^0 \pi ^0)} \equiv 
\epsilon _K- 2\epsilon ^{\prime}  
\ee
The quantity $\epsilon _K$ describes the CP violation 
common to the 
$K_L$ decays; it thus characterizes the decaying {\em state} and is 
referred to as {\em CP violation in the mass matrix} or 
{\em superweak CP violation}; $\epsilon ^{\prime}$ on the 
other 
hand differentiates between different channels and thus 
characterizes {\em decay} dynamics; it is called {\em direct 
CP violation}.   
\item  
{\em Maximal} parity and/or charge conjugation violation 
can be defined by saying there is no right-handed neutrino 
and/or left-handed antineutrino, respectively. Yet 
{\em maximal} CP violation {\em cannot} be defined in an 
analogous 
way: for the existence of the right-handed antineutrino 
which is the 
CP conjugate to the left-handed neutrino is already 
required by 
CPT invariance. 

\end{itemize} 

\subsection{Data}
The data on CP violation in neutral kaon decays are as follows: 
\begin{enumerate} 
\item 
{\em Existence} of $K_L \ra \pi \pi$:
\be 
\begin{array}{l}
{\rm BR}(K_L \ra \pi ^+ \pi ^-) = 
(2.067 \pm 0.035) \cdot 10^{-3} \\ 
{\rm BR}(K_L \ra \pi ^0 \pi ^0) = 
(0.936 \pm 0.020) \cdot 10^{-3} \\ 
\end{array}
\label{ETADATA}
\ee 
\item 
Search for {\em direct} CP violation: 
\be 
\frac{\epsilon ^{\prime}}{\epsilon _K} \simeq 
{\rm Re} \frac{\epsilon ^{\prime}}{\epsilon _K} = 
\left\{   
\begin{array}{ll}
(2.3 \pm 0.65) \cdot 10^{-3} & NA\; 31 \\
(1.5 \pm 0.8) \cdot 10^{-3} & PDG \; '96\;  average \\
(0.74 \pm 0.52 \pm 0.29) \cdot 10^{-3} & E\; 731 \\
\end{array} 
\right. 
\label{DIRECTCPDATA} 
\ee 
\item 
Rate {\em difference} in semileptonic decays: 
\be 
\delta _l\equiv 
\frac{\Gamma (K_L \ra l^+ \nu \pi ^-) - 
\Gamma (K_L \ra l^- \bar \nu \pi ^+)}
{\Gamma (K_L \ra l^+ \nu \pi ^-) + 
\Gamma (K_L \ra l^- \bar \nu \pi ^+)}  = 
(3.27 \pm 0.12)\cdot 10^{-3} \; , 
\label{SLDIFFDATA} 
\ee
where an average over electrons and muons has been taken. 
\item 
{\em T violation}: 
\be 
\frac{\Gamma (K^0 \Rightarrow \bar K^0) - 
\Gamma (\bar K^0 \Rightarrow K^0)} 
{\Gamma (K^0 \Rightarrow \bar K^0) + 
\Gamma (\bar K^0 \Rightarrow K^0)} = 
(6.3 \pm 2.1 \pm 1.8) 
\cdot 10^{-3} \; \; \; 
CPLEAR 
\label{CPLEAR}
\ee 
from a third of their data set \cite{CPLEARBLOCH}. 
It would be premature to claim this asymmetry has been 
established; yet it represents an intriguingly direct test 
of time 
reversal violation and is sometimes referred to as the Kabir 
test . It requires tracking the flavour identity of 
the {\em decaying} meson as a $K^0$ or $\bar K^0$ through its 
semileptonic decays -- $\bar K^0 \ra l^- \bar \nu \pi ^+$ vs. 
$K^0 \ra l^+  \nu \pi ^-$ -- and also of the {\em initially 
produced} kaon. The latter is achieved through correlations 
imposed by associated production. The CPLEAR collaboration 
studied low energy proton-antiproton annihilation 
\be 
p \bar p \ra K^+ \bar K^0 \pi ^- \; \; vs. \; \; 
p \bar p \ra K^-  K^0 \pi ^+ \; ; 
\ee 
the charged kaon reveals whether a $K^0$ or a $\bar K^0$ was 
produced in association with it. In the future the 
CLOE collaboration 
will study T violation in $K^0 \bar K^0$ production at 
DA$\Phi$NE: 
\be 
e^+ e^- \ra \phi (1020) \ra K^0 \bar K^0 
\ee
\item CP- and T-odd Correlations: 

\noindent 
The KTeV Collaboration at Fermilab has established the 
existence of a new rare $K_L$ decay mode 
\footnote{Obviously these data were not available at the 
actual lectures.}
: 
\be 
BR(K_L \ra \pi ^+ \pi ^- e^+ e^-) = 
(3.32 \pm 0.14 \pm 0.28 ) 
\cdot 10^{-7} 
\label{KTEVSEHGAL1} 
\ee  
With $\phi$ defined as the angle between the planes 
spanned by 
the two pions and the two leptons in the $K_L$ 
restframe:  
$$   
\phi \equiv \angle ( \vec n_l, \vec n_{\pi})
$$ 
\be  
\vec n_l = \vec p_{e ^+}\times \vec p_{e ^-}/
|\vec p_{e ^+}\times \vec p_{e ^-}| \; , \;  
\vec n_{\pi} = \vec p_{\pi ^+}\times \vec p_{\pi ^-}/ 
|\vec p_{\pi ^+}\times \vec p_{\pi ^-}|
\label{PHISEHGAL}
\ee    
one analyzes 
the decay rate as a function of $\phi$: 
\be 
\frac{d\Gamma}{d\phi} = \Gamma _1 {\rm cos}^2\phi + 
\Gamma _2 {\rm sin}^2\phi + 
\Gamma _3 {\rm cos}\phi \, {\rm sin} \phi 
\ee 
Since  
\be 
{\rm cos}\phi \, {\rm sin} \phi = 
(\vec n_l \times \vec n_{\pi}) \cdot 
(\vec p_{\pi ^+} + \vec p_{\pi ^-}) 
(\vec n_l \cdot \vec n_{\pi})/
|\vec p_{\pi ^+} + \vec p_{\pi ^-}| 
\ee
one notes that 
\be 
{\rm cos}\phi \, {\rm sin} \phi \; \; \; 
\stackrel{{\bf T},{\bf CP}}{\longrightarrow} \; \; \; 
- \; {\rm cos}\phi \, {\rm sin} \phi 
\ee    
under both T and CP transformations; i.e. the observable  
$\Gamma _3$ represents a T- and CP-odd correlation. 
It can be projected out by comparing the $\phi$ 
distribution integrated over two quadrants: 
\be 
A = 
\frac{\int _0^{\pi/2} d\phi \frac{d\Gamma}{d\phi} - 
\int _{\pi /2}^{\pi} d\phi \frac{d\Gamma}{d\phi}}
{\int _0^{\pi} d\phi \frac{d\Gamma}{d\phi}} = 
\frac{2\Gamma _3}{\pi (\Gamma _1 + \Gamma _2)} 
\ee
KTeV observes 
\be 
A = (13.5 \pm 2.5 \pm 3.0)\% \, , \; 
{\rm preliminary}
\label{KTEVSEHGAL2}
\ee 
This represents a new world record for the size of a 
CP asymmetry. 
\end{enumerate}

\subsection{Phenomenological Interpretation}
\subsubsection{Semileptonic Transitions}
CPT symmetry imposes constraints well beyond the equality of 
lifetimes for particles and antiparticles: certain 
{\em sub}classes of 
decay rates have to be equal as well. For example one finds 
\be 
\Gamma (\bar K^0 \ra l^- \bar \nu \pi ^+) = 
\Gamma (K^0 \ra l^+ \nu \pi ^-)  
\ee 
The rate asymmetry in semileptonic decays listed in 
Eq.(\ref{SLDIFFDATA}) thus reflects pure superweak CP violation: 
\be 
\delta _l = \frac{ 1 - |q/p|^2}{1+|q/p|^2} 
\ee
From the measured value of $\delta _l$ one then obtains 
\be 
\left| \frac{q}{p}\right| = 1 + (3.27 \pm 0.12) \cdot 10^{-3} 
\ee  
Since one has for the $K^0 - \bar K^0$ system specifically 
\be 
\left| \frac{q}{p}\right| \simeq 
1 + \frac{1}{2} {\rm arg}\frac{M_{12}}{\Gamma _{12}} 
\ee 
one can express this kind of CP violation through a phase: 
\be 
\Phi (\Delta S=2) \equiv {\rm arg}\frac{M_{12}}{\Gamma _{12}} = 
(6.54 \pm 0.24)\cdot 10^{-3} 
\label{PHI2}
\ee
The result of the Kabir test, Eq.(\ref{CPLEAR}), yields: 
\be 
\Phi (\Delta S=2) = (6.3 \pm 2.1 \pm 1.8)\cdot 10^{-3}\; ,  
\ee  
which is of course consistent with Eq.(\ref{PHI2}). 

Using the measured value of $\Delta m_K/\Delta \Gamma _K$ one 
infers 
\be 
\frac{M_{12}}{\Gamma _{12}} = - (0.4773 \pm 0.0023)
\left[ 1 - i (6.54 \pm 0.24)\cdot 10^{-3})\right] 
\ee  
\subsubsection{Nonleptonic Transitions}
From Eq.(\ref{ETADATA}) one deduces 
\be 
\begin{array}{l} 
|\eta _{+-}| = (2.275 \pm 0.019)\cdot 10^{-3} \\ 
|\eta _{00}| = (2.285 \pm 0.019)\cdot 10^{-3} \\ 
\end{array}
\ee 
As mentioned before the ratios $\eta _{+-,00}$ are 
sensitive also 
to direct CP violation generated by a phase between the 
decay amplitudes $A_{0,2}$ for $K_L\ra (\pi \pi )_I$, 
where the 
subscript $I$ denotes the isospin of the $2\pi$ system: 
\be 
\Phi (\Delta S=1) \equiv {\rm arg}\frac{A_2}{A_0} 
\ee
One finds 
\be 
\eta _{+-} \simeq  \frac{i \tilde x}{2\tilde x+i}
\left[ \Phi (\Delta S=2)  + 2 \omega \Phi (\Delta S=1) 
\right] \; , 
\ee  
with  
\be 
\tilde x \equiv  \frac{\Delta m_K}{\Delta \Gamma _K} = 
\frac{\Delta m_K}{\Gamma (K_S)} 
=\frac{1}{2} x_K \simeq 0.477 \; \; , \; \; 
\omega \equiv \left| \frac{A_2}{ A_0}\right| \simeq 0.05
\ee
where the second quantity represents the observed
enhancement of $A_0$ for which a name -- "$\Delta I=1/2$ rule" 
-- 
yet no quantitative dynamical explanation has been found. 
Equivalently one can write   
\be 
\frac{\epsilon ^{\prime}}{\epsilon _K} \simeq 2 \omega 
\frac{\Phi (\Delta S=1)}{\Phi (\Delta S=2)}
\ee 
The data on $K_L \ra \pi \pi$ can thus be expressed as 
follows \cite{WINSTEIN} 
\be 
\begin{array}{l} 
\Phi (\Delta S=2) = (6.58 \pm 0.26) \cdot 10^{-3} \\ 
\Phi (\Delta S=1) = (0.99 \pm 0.53) \cdot 10^{-3}  
\end{array}
\ee 
\subsubsection{Radiative Transitions}
The modes $K_{L,S} \ra \pi ^+ \pi ^- \gamma$ have been observed 
with 
\bea 
BR(K_L \ra \pi ^+ \pi ^- \gamma ) &=& (4.66 \pm 0.15) 
\cdot 10^{-5} \\ 
BR(K_S \ra \pi ^+ \pi ^- \gamma ) &=& (4.87 \pm 0.11) 
\cdot 10^{-3} 
\eea 
for $E_{\gamma} > 20$ MeV. Two mechanisms can drive these 
channels and an analysis of the photon spectra indeed reveals 
the intervention of both:
\begin{itemize}
\item 
Bremsstrahlung off the pions through an E1 transition: 
\be 
K_L \stackrel{\Delta S=1}{\longrightarrow} \pi ^+ \pi ^- 
\stackrel{E1}{\longrightarrow} \pi ^+ \pi ^- \gamma \; , \; 
K_S \stackrel{\Delta S=1}{\longrightarrow} \pi ^+ \pi ^- 
\stackrel{E1}{\longrightarrow} \pi ^+ \pi ^- \gamma 
\ee 
where only the first step in the $K_L$ decay is 
CP violating.  
\item 
Direct photon emission of the M1 type 
\be 
K_L \stackrel{M1\& \Delta S=1} {\longrightarrow} 
\pi ^+ \pi ^- \gamma \; , \; 
K_S \stackrel{M1\& \Delta S=1} {\longrightarrow} 
\pi ^+ \pi ^- \gamma \; , 
\ee 
which is CP conserving [violating] for the 
$K_L [K_S]$ process. 
\end{itemize} 
In analogy to $\eta _{+-}$ one defines a ratio 
of E1 amplitudes 
\be 
\eta _{+-\gamma} = \frac{T(K_L \ra \pi ^+ \pi ^- \gamma, E1)}
{T(K_S \ra \pi ^+ \pi ^- \gamma, E1)}
\ee
that measures CP violation. Without {\em direct} 
CP violation one has $\eta _{+-\gamma} = \eta _{+-}$. 

The interference of the CP violating E1 and conserving 
M1 amplitudes for $K_L \ra \pi ^+ \pi^- \gamma$ will yield 
a circularly polarized photon. To be more explicit: it 
yields a triple correlation between the pion momenta and the 
photon polarization 
\be 
P_{\perp}^{\gamma} = \langle \vec \epsilon _{\gamma} 
\cdot (\vec p _{\pi ^+} \times \vec p_{\pi ^-})\rangle 
\ee 
which is CP-odd; its leading contribution is proportional 
to $\eta _{+-}$ entering in the E1 amplitude. 

This polarization can be probed best for 
off-shell photons 
\be 
  K_L \ra \pi ^+ \pi ^- \gamma ^* 
  \ra \pi ^+ \pi ^- e^+ e^- 
\ee  
by measuring the correlation between the $e^+e^-$ 
and $\pi ^+ \pi ^-$ planes measured through the angle 
$\phi$, see Eq.(\ref{PHISEHGAL}). 

This transition has been analyzed several years ago in    
\cite{SEHGALWANN}. The transition amplitude reads as 
follows: 
$$ 
  T(K_L \to \pi ^+ \pi ^- e^+ e^-) = 
  e |T(K_S \to \pi ^+ \pi ^-)| \cdot 
$$
\be 
  \cdot \left[ 
  g_{E1} \left( 
  \frac{p_+^{\mu}}{p_+ \cdot k} - \frac{p_-^{\mu}}{p_- \cdot 
  k} 
  \right) 
  + g_{M1} \epsilon _{\mu \nu \alpha \beta} k^{\nu}
  p_+^{\alpha}p_-^{\beta} 
  \right] 
  \frac{e}{k^2} \bar u(k_-)\gamma _{\mu}v(k_+) 
\ee 
with $k= k_+ + k_-$; the two couplings $g_{E1,M1}$ are 
  given by 
\be 
  g_{E1} = \eta _{+-}e^{i\delta _0(m_K^2)} \; \; , \; \; 
  g_{M1} = 0.76 i e^{i\delta _1(s_{\pi})} 
\ee 
  with $\delta _{0,1}$ denoting the s- and p-wave 
  $\pi \pi$ phase shifts; the coefficient 0.76  is 
obtained from the observed branching ratio for the 
M1 transition. These expressions lead to the 
following predictions 
\footnote{These are predictions in the 
old-fashioned way: they were stated before there were data.}: 
\be  
  BR(K_L \to \pi ^+ \pi ^- e^+ e^-) \simeq 
  3 \cdot 10^{-7} 
\ee  
  depending on the cut one places on the $e^+e^-$ 
invariant mass and 
\be 
  A \simeq (14.3 \pm 1.3) \% 
\ee
The main theoretical uncertainty resides in what one 
assumes for the hadronic form factors. In 
\cite{SEHGALWANN} a phenomenological ansatz was 
employed; evaluating them in 
chiral perturbation theory yields similar numbers 
\cite{ELWOOD}. 

These predictions are in full agreement with the 
KTeV data, see 
Eqs.(\ref{KTEVSEHGAL1},\ref{KTEVSEHGAL2}). 

The discovery of such a large CP asymmetry is a significant 
result be it only to show that CP violation is not 
uniformly tiny in $K_L$ decays. One should note, though, 
that $A$ is driven by $\eta _{+-}$  entering through 
$K_L \ra \pi ^+ \pi ^- \ra \pi ^+ \pi ^- \gamma ^* 
\ra \pi ^+ \pi ^- e^+ e^-$; its size thus is not the 
prediction of a specific model. Two more comments are in 
order here: 
\begin{itemize}
\item 
{\em Direct} CP violation can affect $A$ as well; its 
size depends on the specifics of the dynamics underlying 
CP violation. Yet its contributions to $A$ averaged over 
all final states are tiny, namely $< 10^{-3}$ 
for the KM ansatz \cite{HEILLIGER}; it is hard to see 
how they could be significantly larger for other models. 
While such contributions could be considerably larger in 
certain parts of phase space, no promising avenue has been 
pointed out yet. 
\item 
The correlation $A$ is 
  clearly 
  T-odd. Yet with the time reversal operator 
  being {\em anti}unitary a T-odd correlation can 
  arise even with 
  T-invariant dynamics, if complex phases are present 
\cite{BOOK}; 
  final state interactions can generate such phases. 
One might be tempted to argue that an observation of 
  $A > 0.05$ {\em directly} 
establishes T violation: for {\em electromagnetic}  
final state interactions cannot 
generate an effect of that size while {\em strong} 
  final state interactions can affect neither the photon 
  polarization nor the orientation of the $\pi - \pi$ plane.
There is a third possibility, though, in this special 
situation: CP violation induces an E1 amplitude 
proportional to 
  $\eta _{+-}$; it contains an observable phase 
  $\phi _{+-}$ that depends on the 
  underlying physics. CPT invariance constrains 
  $\phi _{+-}$ to be close to $45^o$. In that scenario 
$\eta _{+-} \neq 0$ of course implies T as well as 
CP violation. As a matter of principle at least 
one can then ask what happens if both CP and 
CPT invariance are broken while T remains 
 conserved; the phase of $\eta _{+-}$ is then 
  no longer constrained. One can then fit this phase and 
-- as a point of principle -- will 
find a solution for reproducing $A$ 
with T invariant dynamics. In practise one can then 
check whether the value of $\phi _{+-}$ thus 
obtained is consistent with the findings 
  from $K_L \rightarrow \pi \pi$ and 
  $K_L \rightarrow l^{\pm} \nu \pi ^{\mp}$. One thus 
interpretes $A$ as a novel probe of 
CPT invariance  

\end{itemize}

\subsubsection{Resume}
The experimental results can be summarized as follows: 
\begin{itemize}
\item 
The decays of neutral kaons exhibit unequivocally CP violation 
of the superweak variety, which is expressed through the 
angle $\Phi (\Delta S=2)$. The findings from 
semileptonic, nonleptonic and now even radiative transitions 
concur to an impressive and reassuring 
degree;  
\item 
Direct CP violation still has not been established. 
\item 
A theorist might be forgiven for mentioning that the 
evolution of the 
measurements over the last twenty odd 
years has not followed the straight line this brief summary 
might suggest to the uninitiated reader. 
\end{itemize}

{Theoretical Implementation of CP Violation}
\label{KM} 

\subsection{Some Historical Remarks}
Theorists can be forgiven if they felt quite pleased with the state 
of their craft in 1964:
\begin{itemize}
\item 
The concept of (quark) families had emerged, at least in a 
rudimentary form. 
\item 
Maximal parity and charge conjugation violations had been found in 
weak charged current interactions, yet CP invariance apparently 
held. Theoretical pronouncements were made ex cathedra why this 
had to be so!
\item 
{\em Pre}dictions of the existence of two kinds of neutral 
kaons with different lifetimes and masses had been 
confirmed by experiment \cite{PAIS}. 
\end{itemize}
That same year the reaction $K_L \ra \pi ^+ \pi ^-$ was 
discovered \cite{FITCH}! Two things should be noted here. 
The Fitch-Cronin experiment had predecessors: rather than 
being an isolated effort it was the culmination of a whole 
research program. Secondly there was at least one theoretical 
voice, namely that of Okun \cite{OKUN}, who in 1962/63 had listed a 
dedicated search for $K_L \ra \pi \pi$ as one of the most important 
unfinished tasks. Nevertheless for the vast majority of the 
community the 
Fitch-Cronin observation came as a shock and caused considerable 
consternation among theorists. Yet -- to their credit -- these data 
and their consequence, namely that CP invariance was broken, 
were soon accepted as facts. This was phrased -- though 
{\em not explained} -- in terms of the Superweak Model 
\cite{WOLFSW} later that same year. 

In 1970 the renormalizability of the $SU(2)_L\times U(1)$ 
electroweak gauge theory was proven. I find it quite amazing 
that it was still not realized that the physics known at that time 
could not produce CP violation. As long as one had to struggle 
with infinities in the theoretical description one could be forgiven 
for not worrying unduly about a tiny quantity like 
BR$(K_L \ra \pi ^+ \pi ^-) \simeq 2.3 \cdot 10^{-3}$. Yet no such 
excuse existed any longer once a renormalizable theory had been 
developed! The existence of the Superweak Model somewhat 
muddled the situation in this respect: for it provides merely 
a classification of the dynamics underlying CP violation rather 
than a dynamical description itself. 

The paper by Kobayashi and Maskawa \cite{KM}, 
written in 1972 and published in 1973, was the first 
\begin{itemize}
\item 
to state clearly that the 
$SU(2)_L\times U(1)$ gauge theory even with two complete 
families \footnote{Remember this was still before the $J/\psi$ 
discovery!} is necessarily CP-invariant and 
\item 
to list the possible extensions that could generate CP 
violation; among them -- as one option -- was the three (or more) 
family scenario now commonly referred to as the KM ansatz. They 
also discussed the impact of right-handed currents and of a 
non-minimal Higgs sector. 
\end{itemize}

\subsection{The Minimal Model: The KM Ansatz}
Once a theory reaches a certain degree of complexity, many potential 
sources of CP violation emerge. Popular examples of such a scenario 
are provided by models implementing supersymmetry or its 
local version, supergravity; hereafter both are referred to as 
SUSY. In my lectures I will however focus on the minimal 
theory that can support CP violation, namely the Standard Model 
with three families. All of its dynamical elements have been 
observed -- except for the Higgs boson, of course. 

\subsubsection{Weak Phases like the Scarlet Pimpernel}

Weak interactions at low energies are described by four-fermion 
interactions. The most general expression for spin-one 
couplings are 
$$ 
{\cal L}_{V/A} = \left( \bar \psi _1 \gamma _{\mu}
(a + b \gamma _5)\psi _2\right) 
\left( \bar \psi _3 \gamma _{\mu}
(c + d \gamma _5)\psi _4\right) + 
$$
\be  
+ \left( \bar \psi _2 \gamma _{\mu}
(a^* + b^* \gamma _5)\psi _1\right) 
\left( \bar \psi _4 \gamma _{\mu}
(c^* + d^* \gamma _5)\psi _3\right)
\ee 
Under CP these terms transform as follows: 
$$  
{\cal L}_{V/A} \stackrel{CP}{\Longrightarrow} 
CP {\cal L}_{V/A} (CP)^{\dagger} = 
\left( \bar \psi _2 \gamma _{\mu}
(a + b \gamma _5)\psi _1\right) 
\left( \bar \psi _4 \gamma _{\mu}
(c + d \gamma _5)\psi _3\right) + 
$$ 
\be 
+ \left( \bar \psi _1 \gamma _{\mu}
(a^* + b^* \gamma _5)\psi _2\right) 
\left( \bar \psi _3 \gamma _{\mu}
(c^* + d^* \gamma _5)\psi _4\right)
\ee 
If $a,b,c,d$ are real numbers, one obviously has 
${\cal L}_{V/A}= CP {\cal L}_{V/A} (CP)^{\dagger} $ and CP 
is conserved. Yet CP is {\em not necessarily} broken if these 
parameters are complex, as we will explain specifically 
for the Standard Model. 

{\em Weak Universality} arises naturally whenever the weak 
charged current interactions are described through a 
{\em single} non-abelian gauge group -- $SU(2)_L$ in the case 
under study. For the single {\em self}-coupling of the gauge bosons 
determines also their couplings to the fermions; 
one finds for the quark couplings to the charged $W$ bosons: 
\be 
{\cal L}_{CC} = 
g \bar U_L^{(0)}\gamma _{\mu}D_L^{(0)} W^{\mu} + 
\bar U_R^{(0)} {\bf M}_U U_L^{(0)} + 
\bar D_R^{(0)} {\bf M}_D D_L^{(0)} + h.c. 
\ee   
where $U$ and $D$ denote the up- and down-type quarks, 
respectively: 
\be 
U= (u,c,t) \; \; \; , \; \; \; \; D=(d,s,b) 
\ee 
and ${\bf M_U}$ and ${\bf M_D}$ 
their 3$\times$3 mass matrices. In general 
those will not be diagonal; to find the physical states, one has to 
diagonalize these matrices: 
\be 
{\bf M}^{diag}_U = {\bf K}^U_R{\bf M}_U ({\bf K}^U_L)^{\dagger} 
\; \; , \; \; 
{\bf M}^{diag}_D = {\bf K}^D_R{\bf M}_D ({\bf K}^D_L)^{\dagger}
\ee 
\be 
U_{L,R} = {\bf K}^U_{L,R} U^{(0)}_{L,R} \; \; , \; \; 
D_{L,R} = {\bf K}^D_{L,R} D^{(0)}_{L,R}
\ee  
with ${\bf K}_{L,R}^{U,D}$ representing four unitary 
3$\times$3 matrices. 
The coupling of these physical fermions to $W$ bosons is then given 
by 
\be 
{\cal L}_{CC} = 
g \bar U_L({\bf K}_L^U)^{\dagger}{\bf K}_L^D\gamma _{\mu}D  W^{\mu} + 
\bar U_R {\bf M}^{diag}_U U_L + 
\bar D_R {\bf M}^{diag}_D D_L + h.c. 
\ee   
and the combination $({\bf K}_L^U)^{\dagger}{\bf K}_L^D 
\equiv {\bf V}_{CKM}$ 
represents the KM matrix, which obviously has to be unitary 
like $K^U$ and $K^D$. Unless the 
up- and down-type mass matrices are aligned in flavour 
space (in which case they would be diagonalized by the 
same operators ${\bf K}_{L,R}$) one has ${\bf V}_{CKM} \neq 1$.  

In the neutral current sector one has 
\be 
{\cal L}_{NC} = g^{\prime} 
\bar U_L^{(0)} \gamma _{\mu}U_L^{(0)}Z_{\mu} = 
g^{\prime} 
\bar U_L \gamma _{\mu}U_LZ_{\mu}
\ee 
and likewise for $U_R$ and $D_{L,R}$; i.e. {\em no} 
flavour changing neutral currents are generated, let alone 
new phases. CP violation thus has to be embedded into the 
charged current sector. 

If ${\bf V}_{CKM}$ is real (and thus orthogonal), CP symmetry is 
conserved in the weak interactions. Yet the occurrance of 
complex matrix elements does not {\em automatically} signal 
CP violation. This can be seen through a straightforward 
(in hindsight at least) algebraic argument. A unitary 
$N\times N$ matrix contains $N^2$ independant real 
parameters; $2N-1$ of those can be eliminated through 
re-phasing of the $N$ up-type and $N$ down-type fermion 
fields (changing all fermions by the {\em same} phase obviously 
does not affect  ${\bf V}_{CKM}$). Hence there are $(N-1)^2$ 
real physical parameters in such an $N \times N$ matrix. 
For $N=2$, i.e. two families, one recovers a familiar result, 
namely there is just one mixing angle, the Cabibbo 
angle. For $N=3$ there are four real physical parameters, 
namely three (Euler) angles -- and one phase. It is the latter 
that provides a gateway for CP violation. For $N=4$ Pandora's 
box opens up: there would be 6 angles and 3 phases. 

PDG suggests a "canonical" parametrization for the $3\times 3$ CKM 
matrix: 
$$ 
{\bf V}_{CKM} = 
\left(  
\begin{array}{ccc} 
V(ud) & V(us) & V(ub) \\
V(cd) & V(cs) & V(cb) \\
V(td) & V(ts) & V(tb) 
\end{array} 
\right) 
$$
\be 
= \left( 
\begin{array}{ccc} 
c_{12}c_{13} & s_{12}c_{13} & s_{13}e^{-i \delta _{13}}  \\
- s_{12}c_{23} - c_{12}s_{23}s_{13}e^{i \delta _{13}} &
c_{12}c_{23} - s_{12}s_{23}s_{13}e^{i \delta _{13}} & 
c_{13}s_{23} \\
s_{12}s_{23} - c_{12}c_{23}s_{13}e^{i \delta _{13}} &
- c_{12}s_{23} - s_{12}c_{23}s_{13}e^{i \delta _{13}} &
c_{13}c_{23} 
\end{array}
\right) 
\label{PDGKM} 
\ee 
where 
\be 
c_{ij} \equiv {\rm cos} \theta _{ij} \; \; , \; \;  
s_{ij} \equiv {\rm sin} \theta _{ij}
\ee  
with $i,j = 1,2,3$ being generation labels. 

This is a completely general, yet not unique parametrisation: a 
different set of 
Euler angles could be chosen; the phases can be shifted around 
among the matrix elements 
by using a different phase convention. 
In that sense one can refer to 
the KM phase as the Scarlet Pimpernel: 
"Sometimes here, sometimes 
there, sometimes everywhere!"  

Using just the observed hierarchy 
\be 
|V(ub)| \ll |V(cb)| \ll |V(us)| , |V(cd)| \ll 1
\label{HIER}
\ee  
one can, as first realized by Wolfenstein, expand 
${\bf V}_{CKM}$ in powers of the Cabibbo angle $\theta _C$: 
\be 
{\bf V}_{CKM} = 
\left( 
\begin{array}{ccc} 
1 - \frac{1}{2} \lambda ^2 & \lambda & 
A \lambda ^3 (\rho - i \eta + \frac{i}{2} \eta \lambda ^2) \\
- \lambda & 1 - \frac{1}{2} \lambda ^2 - i \eta A^2 \lambda ^4 & 
A\lambda ^2 (1 + i\eta \lambda ^2 ) \\ 
A \lambda ^3 (1 - \rho - i \eta ) \\
& - A\lambda ^2 & 1 
\end{array}
\right) 
+ {\cal O}(\lambda ^6) 
\label{WOLFKM}
\ee  
where 
\be 
\lambda \equiv {\rm sin} \theta _C
\ee 
For such an expansion in powers of $\lambda$ to be 
self-consistent, 
one has to require that $|A|$, $|\rho |$ and $|\eta |$ 
are of order 
unity. Numerically we obtain  
\be 
\lambda = 0.221 \pm 0.002  
\label{KMNUm1} 
\ee  
from $|V(us)|$,  
\be 
A = 0.81 \pm 0.06 
\label{KMNUM2}
\ee  
from $|V(cb)| \simeq 0.040 \pm 0.002|_{exp} \pm 0.002|_{theor}$ 
and 
\be 
\sqrt{\rho ^2 + \eta ^2} \sim 0.38 \pm 0.11 
\label{KMNUM3} 
\ee 
from $|V(ub)| \sim (3.2 \pm 0.8)\cdot 10^{-3}$. The numbers 
for $|V(cb)|$ and $|V(ub)|$ have changed quite considerably over the 
last few years; in particular there has been a 
substantial reduction in the uncertainties. 
This reflects -- in addition to more sensitive data of course -- 
the arrival of novel theoretical technologies for dealing with 
heavy flavour decays. These methods will be briefly 
described in Sect.\ref{HQE}.

We see that the CKM matrix is a very special unitary matrix: 
it is almost diagonal, it is almost symmetric and the matrix 
elements get smaller the more one moves away from the 
diagonal. 
Nature most certainly has encoded a profound message in 
this peculiar pattern. Alas -- we have not succeeded yet in 
deciphering it! I will return to this point at the end of my 
lectures.  

\subsubsection{Unitarity Triangles}

The qualitative difference between a two and a three family scenario 
can be seen also in a less abstract way.  
Consider $\bar K^0 \ra \pi ^+ \pi ^-$; it can proceed through 
a tree-level process 
$[s\bar d] \ra [d \bar u][u\bar d]$ , in which case its 
weak couplings are 
given by $V(us)V^*(ud)$. Or it can oscillate first to $K^0$ 
before decaying; i.e., on the quark level it is the 
transition $[s\bar d] \ra [d\bar s] \ra [d \bar u][u\bar d]$ 
controlled by 
$\left( V(cs)\right) ^2 \left( V^*(cd)\right) ^2 V^*(us)V(ud)$. 
At first sight it would seem that those two combinations of 
weak parameters are not only different, but should also exhibit a 
relative phase. Yet the latter is not so -- if there are two families 
only! In that case the four quantities $V(ud)$, $V(us)$, $V(cd)$ 
and $V(cs)$ have to form a unitary $2\times 2$ which leads to the 
constraint 
\be 
V(ud)V^*(us) + V(cd)V^*(cs) = 0 
\label{UNIT2FAM}
\ee  
Using Eq.(\ref{UNIT2FAM}) twice one gets  
$$  
\left( V(cs) V^*(cd)\right) ^2 V^*(us)V(ud) = 
- |V(cd)V(cs)|^2 V(cs)V^*(cd) = 
$$ 
\be 
= |V(cd)V(cs)|^2 V^*(ud) V(us) \; ; 
\ee   
 i.e., the two combinations $V^*(ud)V(us)$ and 
$\left( V(cs)\right) ^2 \left( V^*(cd)\right) ^2 V^*(us)V(ud)$ are 
actually parallel to each other with {\em no} 
relative phase. A penguin 
operator with a charm quark as the internal fermion 
line generates another contribution to 
$K_L \ra \pi ^-\pi ^+$, this one controlled by $V(cs)V^*(cd)$. Yet 
the unitarity condition Eq.(\ref{UNIT2FAM}) forces this contribution 
to be antiparallel to $V^*(ud)V(us)$; i.e., again no relative phase. 

The situation changes fundamentally for three families: the weak 
parameters $V(ij)$ now form a $3\times 3$ matrix and the condition 
stated in  Eq.(\ref{UNIT2FAM}) gets extended: 
\be 
V(ud)V^*(us) + V(cd)V^*(cs) + V(td)V^*(ts) = 0 
\label{UNIT3FAM}
\ee  
{\em This is a triangle relation in the complex plane.} 
There emerge now 
relative phases between the weak parameters and the 
loop diagrams with internal charm and top quarks can generate 
CP asymmetries. 

Unitarity imposes altogether nine algebraic conditions on the 
matrix elements of ${\bf V}_{CKM}$, of which six are triangle relations 
analogous to Eq.(\ref{UNIT3FAM}). 
There are several nice features about this representation in terms of 
triangles; I list four now and others later: 
\begin{enumerate}
\item 
The {\em shape} of each triangle is independant of the phase 
convention adopted for the quark fields. Consider for example 
Eq.(\ref{UNIT3FAM}): changing the phase of any of the 
up-type quarks will not affect the triangle at all. Under 
$|s\rangle \ra |s\rangle e^{i \phi _s}$ the whole triangle will 
rotate around the left end of its base line by an angle 
$\phi _s$ -- yet the shape of the triangle -- in contrast 
to its orientation in the complex plane -- remains the same! 
The angles inside the triangles are thus observables;  
choosing an orientation for the triangles is then a matter 
of convenience. 
\item 
It is easily shown that all six KM triangles possess 
the same area. 
Multiplying Eq.(\ref{UNIT3FAM}) by the phase 
factor $V^*(ud)V(us)/|V(ud)V(us)|$, which does not change the 
area, yields 
\be 
|V(ud)V(us)| + \frac{V^*(ud)V(us)V(cd)V^*(cs)}{|V(ud)V(us)|} + 
\frac{V^*(ud)V(us)V(td)V^*(ts)}{|V(ud)V(us)|} = 0 
\ee  
$$  
{\rm area (triangle \; of \; Eq.(\ref{UNIT3FAM})})  = 
\frac{1}{2} |{\rm Im}V(ud)V(cs)V^*(us)V^*(cd)| = 
$$ 
\be 
= \frac{1}{2} |{\rm Im}V(ud)V(ts)V^*(us)V^*(td)|
\ee 
Multiplying Eq.(\ref{UNIT3FAM}) instead by the phase 
factors $V^*(cd)V(cs)/|V(cd)V(cs)|$ or $V^*(td)V(ts)/|V(td)V(ts)|$ 
one sees that the area of this triangle can be expressed in other ways 
still. Among them is 
\be 
{\rm area (triangle \; of \; Eq.(\ref{UNIT3FAM})})  = 
\frac{1}{2} |{\rm Im}V(cd)V(ts)V^*(cs)V^*(td)|  
\ee  
Due to the unitarity relation 
\be 
V^*(cd)V(td) + V^*(cb)V(tb) = - V^*(cs)V(ts) 
\label{UNIT3FAM2}
\ee 
one has  
\be 
{\rm area ( triangle \; of \; Eq.(\ref{UNIT3FAM})})  = 
\frac{1}{2} |{\rm Im}V(cd)V(tb)V^*(cb)V^*(td)|  
\ee  
-- yet this is exactly the area of the triangle defined by 
Eq.(\ref{UNIT3FAM2})! 
This is the re-incarnation of the original 
observation that there is a {\em single irreducible}  
weak phase for three families. 
\item 
In general one has for the area of these triangles 
$$  
A_{CPV}(\rm every \; triangle) = \frac{1}{2} J 
$$ 
\be  
J = {\rm Im}V^*(km) V(lm)V(kn)V^*(ln) = 
{\rm Im}V^*(mk) V(ml)V(nk)V^*(nl) 
\label{KMAREA}
\ee 
irrespective of the indices $k,l,m,n$; $J$ is obviously re-phasing 
invariant. 
\item  
If there is a representation of $V_{CKM}$ where 
all phases were confined to a $2\times 2$ 
sub-matrix exactly rather than approximately, then one can  
rotate all these phases away; i.e., CP is conserved in such a 
scenario! Consider again the triangle described by 
Eq.(\ref{UNIT3FAM}): it can always be rotated such that its 
baseline -- $V(ud)V^*(us)$  -- 
is real. Then Im$V(td)V^*(ts)$ = - Im$V(cd)V^*(cs)$ holds. 
If, for example, there were no phases in the third row and column, 
one would have Im$(V(td)V^*(ts)) =0$ and therefore 
Im$V(cd)V^*(cs) =0$ as well; i.e., $V(ud)V^*(us)$ and 
$V(cd)V^*(cs)$ were real relative to each other; 
therefore $J=0$, i.e. all six triangles had zero area meaning 
there are no relative weak phases! 
\end{enumerate}  

\subsection{Evaluating $\epsilon _K$ and $\epsilon ^{\prime}$ 
\label{THEPS}} 
In calculating observables in a given theory -- in the case under 
study $\epsilon _K$ and $\epsilon ^{\prime}$ 
within the KM Ansatz -- one is faced with the `Dichotomy of the Two 
Worlds', namely 
\begin{itemize}
\item 
one world of {\em short}-distance physics where even the strong 
interactions can be treated {\em perturbatively} in terms of 
quarks and gluons and in which theorists like to work, and 
\item 
the other world of {\em long}-distance physics where one has 
to deal with hadrons the behaviour of which is controlled 
by {\em non}-perturbative dynamics and where, by the way, 
everyone, including theorists, lives. 
\end{itemize}
Accordingly the calculational task is divided into two 
parts, namely first determing the relevant 
transition operators in the short-distance world and then 
evaluating their matrix elements in the hadronic world. 

\subsubsection{$\Delta S=2$ Transitions }
Since the {\em elementary} interactions in the 
Standard Model can change strangeness at most by one unit, 
the $\Delta S=2$ amplitude driving $K^0 - \bar K^0$ 
oscillations is obtained by iterating the 
basic $\Delta S=1$ coupling: 
\be 
{\cal L}_{eff} (\Delta S=2) = {\cal L}(\Delta S=1) \otimes 
{\cal L}(\Delta S=1) 
\ee  
There are actually two ways in which the $\Delta S=1$ transition 
can be iterated: 

{\bf (A)} 
The resulting $\Delta S=2$ transition is described by 
a {\em local} operator. The celebrated box diagram makes 
this connection quite transparent. The contributions that do 
{\em not} depend on the mass of the internal quarks cancel against 
each other due to the GIM mechanism. Integrating over the internal 
fields, namely the $W$ bosons and the top and charm quarks 
\footnote{The up quarks act merely as a subtraction term here.} 
then yields a convergent result: 
$$  
{\cal L}_{eff}^{box}(\Delta S=2, \mu ) = 
\left( \frac{G_F}{4\pi }\right) ^2 \cdot 
$$ 
\be  
\left[  \xi _c^2 E(x_c) \eta _{cc} + 
\xi _t^2 E(x_t) \eta _{tt} + 
2\xi _c \xi _t E(x_c, x_t) \eta _{ct}
 \right]  [\alpha _S(\mu ^2)]^{- \frac{6}{27}} 
\left( \bar d \gamma _{\mu}(1- \gamma _5) s\right) ^2 
+ h.c. 
\label{LAGDELTAS2}
\ee  
with $\xi _i$ denoting combinations of KM parameters 
\be 
\xi _i = V(is)V^*(id) \; , \; \; i=c,t \; ; 
\ee  
$E(x_i)$ and $E(x_c,x_t)$ reflect the box loops with equal and 
different internal quarks, respectively \cite{INAMI}:  
\be 
E(x_i) = x_i 
\left(   
\frac{1}{4} + \frac{9}{4(1- x_i)} - \frac{3}{2(1- x_i)^2} 
\right) 
- \frac{3}{2} \left( \frac{x_i}{1-x_i}\right) ^3 
{\rm log} x_i 
\ee  
$$  
E(x_c,x_t) = x_c x_t 
\left[ \left( 
\frac{1}{4} + \frac{3}{2(1- x_t)} - \frac{3}{4(1- x_t)^2} \right) 
\frac{{\rm log} x_t}{x_t - x_c} + (x_c \leftrightarrow x_t) - 
\right. 
$$ 
\be 
\left. - \frac{3}{4} \frac{1}{(1-x_c)(1- x_t)} \right]  
\ee 
\be 
x_i = \frac{m_i^2}{M_W^2} 
\ee  
and $\eta _{ij}$ containing the QCD radiative corrections from 
evolving the effective Lagrangian from $M_W$ down to 
the internal quark mass. The factor $[\alpha _S(\mu ^2)]^{-6/27}$ 
reflects the fact that a scale 
$\mu$ must be introduced at which the four-quark operator 
$\left( \bar s \gamma _{\mu}(1- \gamma _5) d\right) ^2 $ is 
defined. This dependance on the auxiliary variable 
$\mu$ drops out when one takes the matrix element of this 
operator (at least when one does it correctly).  
Including next-to-leading log 
corrections one finds (for $m_t \simeq 180$ GeV) \cite{BURAS}: 
\be 
\eta _{cc} \simeq 1.38 \pm 0.20 \; , \; \; 
\eta _{tt} \simeq 0.57 \pm 0.01 \; , \; \; 
\eta _{cc} \simeq 0.47 \pm 0.04 
\ee                       

{\bf (B)} 
However there is also a {\em non}-local $\Delta S=2$ 
operator generated from the iteration of ${\cal L}(\Delta S=1)$. 
While it  
presumably provides a major contribution to $\Delta m_K$, 
it is not sizeable for $\epsilon _K$ within the KM ansatz 
\footnote{This can be inferred from the observation that 
$|\epsilon ^{\prime}/\epsilon _K|\ll 0.05$} and will be 
ignored here. 

Even for a local four-fermion operator it is non-trivial to 
evaluate an on-shell matrix element 
between hadron states since that is 
clearly controlled by non-perturbative dynamics. Usually one 
parametrizes this matrix element as follows 
(in the phase convention ${\bf CP}|K^0\rangle = 
|\bar K^0\rangle$): 
$$  
\matel{ K^0}{(\bar d \gamma _{\mu}(1-\gamma _5)s) 
(\bar d \gamma _{\mu}(1-\gamma _5)s)}{\bar K^0} = 
$$ 
\be 
= \frac{4}{3} B_K 
\matel{ K^0}{(\bar d \gamma _{\mu}(1-\gamma _5)s)}{0} 
\matel{0}{(\bar d \gamma _{\mu}(1-\gamma _5) s)}{\bar K^0} = 
-\frac{4}{3} B_K f_K^2m_K
\label{BAGFACT} 
\ee 
The factor $B_K$ is -- for historical reasons of no 
consequence now -- 
often called the bag factor; $B_K = 1$ is referred to as 
{\em vacuum saturation} or {\em factorization ansatz} since it 
corresponds to a situation where inserting the vacuum intermediate 
state into Eq.(\ref{BAGFACT}) reproduces the full result 
after all colour contractions of the quark lines have been included. 
Several theoretical techniques have been employed to estimate the 
size of $B_K$; their findings are listed in 
Table \ref{TABLEBAG}.  
\begin{table}
\begin{tabular} {|l|l|}
\hline   
Method & $B_K$ \\ 
\hline 
\hline 
Large $N_C$ Expansion & $\frac{3}{4}$\\
\hline 
Large $N_C$ Chiral Pert. with loop correction & $0.66 \pm 0.1$\\
\hline 
Lattice QCD & $0.84 \pm 0.2$ \\
\hline 
\end{tabular}
\centering
\caption{Values of $B_K$ from various theoretical techniques} 
\label{TABLEBAG} 
\end{table} 
These results, which are all consistent with each other and with 
several phenomenological studies as well, can be summarized as 
follows:  
\be 
B_K \simeq 0.8 \pm 0.2 > 0 
\label{BAG} 
\ee  
Since the size of this matrix element is determined 
by the strong interactions, one indeed expects $B_K \sim 1$. 

We have assembled all the ingredients now for calculating 
$\epsilon _K$. The starting point is given by 
\footnote{The exact expression is 
$|\epsilon _K|  = \frac{1}{\sqrt{2}}
\left| \frac{{\rm Im}M_{12}}{\Delta m_K} - \xi _0\right| $ 
where $\xi _0$ denotes the phase of the 
$K^0 \ra (\pi \pi )_{I =0}$ isospin zero amplitude; its 
contribution is 
numerically irrelevant.}: 
\be 
|\epsilon _K|  
\simeq 
\frac{1}{\sqrt{2}}
\left| \frac{{\rm Im}M_{12}}{\Delta m_K} \right| 
\ee  
The CP-odd part Im$M_{12}$ is obtained from 
\be 
{\rm Im} M_{12} = 
{\rm Im}\matel{K^0}{{\cal L}_{eff}(\Delta S=2)}{\bar K^0} 
\ee  
whereas for $\Delta m_K$ one inserts the experimental value,   
since the long-distance contributions to $\Delta m_K$ are not 
under 
theoretical control. One then finds 
$$  
|\epsilon _K|_{KM} \simeq |\epsilon _K|_{KM}^{box} \simeq  
$$ 
$$ 
\simeq \frac{G_F^2}{6 \sqrt{2} \pi ^2} 
\frac{M_W^2m_K f_K^2 B_K}{\Delta m_K} 
\left[ {\rm Im}\xi _c^2 E(x_c) \eta _{cc} + 
 {\rm Im}\xi _t^2 E(x_t) \eta _{tt} + 
2{\rm Im}(\xi _c\xi _t) E(x_c,x_t) \eta _{ct} \right] 
$$ 
$$  
\simeq 1.9 \cdot 10^4 B_K \left[ {\rm Im}\xi _c^2 E(x_c) 
\eta _{cc} + 
 {\rm Im}\xi _t^2 E(x_t) \eta _{tt} + 
2{\rm Im}(\xi _c\xi _t) E(x_c,x_t) \eta _{ct} \right] 
\simeq 
$$ 
\be 
\simeq 
7.8 \cdot 10^{-3} \eta B_K (1.4 - \rho ) 
\label{EPSKM} 
\ee  
where I have used the numerical values for the KM parameters 
listed above and 
$x_t \simeq 5$ corresponding to $m_t = 180$ GeV. 

To reproduce the observed value of $|\epsilon _K|$ one needs 
\be 
\eta \simeq \frac{0.3}{B_K} \frac{1}{1.4 - \rho } 
\label{EPSREP} 
\ee 
For a given $B_K$ one thus obtains another $\rho - \eta $ 
constraint. Since $B_K$ is not precisely known 
one has a fairly broad band in the $\rho - \eta $ plane 
rather than a line. Yet I find it quite remarkable and very 
non-trivial that Eq.(\ref{EPSREP}) {\em can} be 
satisfied since 
\be 
\frac{0.3}{B_K} \sim 0.3 \div 0.5 
\ee  
without stretching any of the parameters or bounds, in 
particular 
\be 
\sqrt{\rho ^2 + \eta ^2} \sim 0.38 \pm 0.11 \; .   
\ee  
 While this does of course not amount to a {\em pre}diction, 
one should keep in mind for proper perspective 
that in the 1970's and early 
1980's values like $|V(cb)| \sim 0.04$ and 
$|V(ub)|  \sim 0.004$ would have seemed quite unnatural; claiming 
that the top quark mass had to be 180 GeV would have been 
outright preposterous even in the 1980's! Consider a 
scenario with $|V(cb)| \simeq 0.04$ and $|V(ub)| \simeq 0.003$, 
yet $m_t \simeq 40$ GeV; in the mid 80's this would have appeared 
to be quite natural (and there had even been claims that top quarks 
with a mass of $40\pm 10$ GeV had been discovered). In that 
case one would need 
\be 
\eta \sim \frac{0.75}{B_K} 
\label{EPS40} 
\ee  
to reproduce $|\epsilon _K|$. Such a large value for $\eta$ would hardly 
be compatible with what we know about $|V(ub)|$. 
\begin{center} 
$\spadesuit \; \; \; \spadesuit \; \; \; \spadesuit $ \\ 
{\em  Homework Problem \# 2}: 
\end{center}
Eq.(\ref{EPSKM}) suggests that 
a non-vanishing value for $\epsilon _K$ is  generated from the 
box diagram with internal charm quarks only -- 
Im$\xi _c^2\; E(x_c) = - \eta A^2 \lambda ^6 E(x_c) \neq 0$ -- 
{\em without} top quarks. How does this match up with the 
statement that the intervention of three families 
is needed for a CP asymmetry to arise? 
\begin{center} 
$\spadesuit \; \; \; \spadesuit \; \; \; \spadesuit $
\end{center}

\subsubsection{An Aside on $\Delta m_K$ 
(\& $\Delta \Gamma _K$)}
Having the local operator ${\cal L}^{box}_{\Delta S=2}$ 
generated by the box diagram, one can 
calculate its contribution to $\Delta m_K$, including the 
latter's sign, as expressed in Eq.(\ref{DELTAMM12}). 
One finds 
\be 
\frac{\Delta m_K|_{box}}{\Delta m_K|_{exp}} 
\sim 0.5 - 1  \; ; 
\ee  
i.e., the box diagram yields a major or even the 
dominant contribution to the observed value of 
$\Delta m_K$ and it predicts indeed that the 
CP odd state is heavier, albeit by a whisker only. 
This is due to the following factors: 
\begin{itemize}
\item 
The coefficient of the effective $\Delta S =2$ operator 
is positive.  
\item 
The quantity $B_K$ is reliably estimated to be positive, 
see Eq.(\ref{BAG}). 
\item 
The minus sign in Eq.(\ref{BAGFACT}) 
thus 
cancels against that in Eq.(\ref{DELTAMM12}) to yield $M_B > M_A$; 
the state $K_B$ is the CP odd one in the convention of 
Eq.(\ref{QPPOS}) which was also employed in obtaining 
Eq.(\ref{DELTAMM12}). 
\end{itemize} 
There are sizeable long-distance contributions to 
$\Delta m_K$ which are estimated to be positive, as well. 

There is no reason why the box diagram's contribution 
to $\Delta \Gamma _K$ should have anything to do with 
reality: it is dominated by $K_L \ra \pi \pi$ which cannot be 
described by short-distance dynamics. Nevertheless it is amusing 
to note that one finds $\Delta \Gamma _K^{box} > 0$ -- 
in agreement with reality!  

\subsubsection{$\Delta S=1$ Decays}
At first one might think that no {\em direct} CP asymmetry 
can arise in $K \ra \pi \pi$ decays since it requires the interplay 
of three quark families. Yet upon further reflection one realizes 
that a one-loop diagram produces the so-called Penguin 
operator which changes isopin by half a unit only, 
is {\em local} and contains a CP {\em odd} component since it involves 
virtual charm 
and top quarks. With direct CP violation thus being  
of order $\hbar$, i.e. a pure quantum 
effect, one suspects already at this point that it will be reduced 
in strength.  

The quantity $\epsilon ^{\prime}$ is suppressed relative to 
$\epsilon _K$  due to two other reasons: 
\begin{itemize}
\item 
The GIM factors are actually quite different for $\epsilon _K$ and 
$\epsilon ^{\prime}$: in the former case they are of the type 
$(m_t^2 - m_c^2)/M_W^2$, in the latter log$(m_t^2/m_c^2)$. Both 
of these expressions vanish for $m_t=m_c$, yet for the realistic 
case $m_t \gg m_c$ they behave very differently: $\epsilon _K$ 
is much more enhanced by the large top mass than 
$\epsilon ^{\prime}$. This means of course that 
$|\epsilon ^{\prime}/\epsilon _K|$ is a rather steeply decreasing 
function of $m_t$. 
\item 
There are actually two classes of Penguin operators contributing 
to $\epsilon ^{\prime}$, namely strong as well as electroweak 
Penguins. The latter become relevant  
since they are more enhanced than the former for very heavy top 
masses due to the coupling of the longitudinal virtual $Z$ boson 
(the re-incarnation of one of the original Higgs fields) to 
the internal top line. Yet electroweak and strong Penguins contribute 
with the opposite sign! 
\end{itemize}
CPT invariance together with the measured $\pi \pi$ phase shifts 
tells us that the two complex quantities $\epsilon ^{\prime}$ and 
$\epsilon _K$ are almost completely real to each other; i.e., 
their ratio is practically real: 
\be 
\frac{\epsilon ^{\prime}}{\epsilon _K} \simeq 2 \omega 
\frac{\Phi (\Delta S=1)}{\Phi (\Delta S=2)}
\label{EPSPOVEREPSTH} 
\ee  
where, as defined before,  
\be 
\omega \equiv \frac{|A_2|}{|A_0|} \simeq 0.05 \; \; , \; \; 
\Phi (\Delta S=2) \equiv {\rm arg}\frac{M_{12}}{\Gamma _{12}} 
\; \; , \Phi (\Delta S=1) \equiv {\rm arg} \frac{A_2}{A_0}
\ee 
Eq.(\ref{EPSPOVEREPSTH}) makes two points obvious:
\begin{itemize}
\item 
Direct CP violation -- $\epsilon ^{\prime} \neq 0$ -- 
requires a relative phase between the isospin 0 and 2 amplitudes; 
i.e., $K \ra (\pi \pi )_0$ and $K \ra (\pi \pi )_2$ have to exhibit 
different CP properties. 
\item 
The observable ratio $\epsilon ^{\prime} / \epsilon _K$ is 
{\em artifically reduced} by the 
enhancement of the $\Delta I =1/2$ amplitude, as 
expressed through $\omega$. 
\end{itemize} 
Several $\Delta S=1$ transition operators contribute to 
$\epsilon ^{\prime}$ and their renormalization has to be treated 
quite carefully. Two recent detailed analyses yield 
\cite{BURASPRIME,CIUCHINIPRIME}
\be 
-2.1 \cdot 10^{-4} \leq \frac{\epsilon ^{\prime}}{\epsilon _K} \leq 
13.3 \cdot 10^{-4} 
\label{BURAS1} 
\ee 
\be 
\frac{\epsilon ^{\prime}}{\epsilon _K} =  
(4.6 \pm 3.0 \pm 0.4)\cdot 10^{-4} 
\label{CIUCHINI} 
\ee 
These results are quite consistent with each other and show  
\begin{itemize}
\item  
that the KM ansatz leads to a prediction typically in the range 
{\em below} $10^{-3}$, 
\item 
that the value could happen to be zero or 
even slightly negative and 
\item 
that large theoretical uncertainties persist due to cancellations among 
various contributions. 
\end{itemize} 
This last (unfortunate) point can be illustrated also by comparing 
these predictions with older ones made before top quarks were 
discovered and their mass measured; those old predictions 
\cite{FRANZINI} are 
very similar to Eqs.(\ref{BURAS1},\ref{CIUCHINI}), once the now 
known value of $m_t$ has been inserted. 

Two new experiments running now -- NA 48 at CERN and KTEV at 
FNAL -- and one expected to start up soon -- CLOE at DA$\Phi $NE -- 
expect to measure $\epsilon ^{\prime}/\epsilon _K$ with a 
sensitivity of $\simeq \pm 2\cdot 10^{-4}$. Concerning their future 
results one can distinguish four scenarios: 
\begin{enumerate}
\item 
The `best' scenario: 
$\epsilon ^{\prime}/\epsilon _K \geq 2 \cdot 10^{-3}$. One would 
then 
have established unequivocally direct CP violation of a strength that 
very probably reflects the intervention of new physics beyond the 
KM ansatz. 
\item 
The `tantalizing' scenario: 
$1\cdot 10^{-3}\leq \epsilon ^{\prime}/\epsilon _K 
\leq 2 \cdot 10^{-3}$. It would be tempting to interprete this 
discovery of direct CP violation as a sign for new physics -- yet 
one could not be sure! 
\item 
The `conservative' scenario: 
$\epsilon ^{\prime}/\epsilon _K \simeq 
{\rm few}\cdot10^{-4} > 0$. This strength of direct CP violation could 
easily be accommodated 
within the KM ansatz -- yet no further constraint would 
materialize. 
\item 
The `frustrating' scenario: 
$\epsilon ^{\prime}/\epsilon _K \simeq 0$ within errors! 
No substantial conclusion could be drawn then concerning the 
presence or absence of direct CP violation, and the allowed KM 
parameter space would hardly shrink. 
\end{enumerate}

\section{`Exotica'}
\label{EXOTICA} 
In this section I will discuss important possible manifestations 
of CP and/or T violation that are exotic only in the sense that 
they are unobservably small with the KM ansatz, to be introduced 
in the next section.  

{$K_{3\mu}$ Decays}
In the reaction  
\be 
K^+ \ra \mu ^+ \nu \pi ^0
\ee 
one can search for a transverse polarisation of the emerging 
muons in close analogy to what has just been discussed for 
$K_L \ra \pi ^+ \pi ^- \gamma$: 
\be 
P_{\perp}^{K^+}(\mu ) \equiv 
\langle \vec s (\mu) \cdot 
(\vec p (\mu ) \times \vec p (\pi ^0))\rangle  
\ee
where $\vec s$ and $\vec p$ denote spin and momentum, 
respectively. 
The quantity $P_{\perp}(\mu )$ constitutes a T-odd  
correlation: 
\be 
\left. 
\begin{array}{l}
\vec p \Rightarrow - \vec p \\ 
\vec s \stackrel{\bf T}{\Rightarrow} - \vec s
\end{array} 
\right\} \leadsto 
P_{\perp}(\mu ) \stackrel{T}{\Rightarrow} - P_{\perp}(\mu ) 
\ee   
Once a {\em non-}vanishing value has been observed for a 
parity-odd correlation one has unequivocally found a manifestation 
of parity violation. From $P_{\perp}^{K^+}(\mu ) \neq 0$ one can 
deduce that T is violated -- yet the argument is more subtle as can 
be learnt from the following homework problem. 
\begin{center} 
$\spadesuit \; \; \; \spadesuit \; \; \; \spadesuit $ \\ 
{\em Homework Problem \#3}: 
\end{center}
Consider 
\be 
K_L \ra \mu ^+ \nu \pi ^- 
\ee 
Does $P_{\perp}^{K_L}(\mu ) \equiv 
\langle \vec s(\mu ) \cdot (\vec p(\mu ) \times \vec p(\pi ^-)) 
\rangle \neq 0$ necessarily imply that T invariance does not hold in 
this reaction?
\begin{center}  
$\spadesuit \; \; \; \spadesuit \; \; \; \spadesuit $ \\ 
\end{center}
Data on $P_{\perp}^{K^+}(\mu )$ are still consistent with zero 
\cite{SCHMIDT}: 
\be 
P_{\perp}^{K^+}(\mu ) = ( -1.85 \pm 3.60) \cdot 10^{-3} \; ; 
\label{YALE}
\ee 
yet being published in 1981 they are ancient by the standards of our 
disciplin. 

On general grounds one infers that 
\be 
P_{\perp}^{K^+}(\mu ) \propto {\rm Im} \frac{f_-^*}{f_+} 
\label{POLTH} 
\ee 
holds where $f_-\, [f_+]$ denotes the chirality changing 
[conserving] decay amplitude. Since $f_-$ practically vanishes 
within the Standard Model, one obtains a fortiori 
$P_{\perp}^{K^+}(\mu )|_{KM} \simeq 0$. 

Yet in the presence of charged Higgs fields one has 
$f_- \neq 0$. CPT implies that 
$P_{\perp}^{K^+}(\mu ) \neq 0$ represents CP violation as 
well, and actually one of the {\em direct} variety. A rather 
model independant guestimate on how large such an effect 
could be is obtained from the present bound on 
$\epsilon ^{\prime}/\epsilon _K$: 
\be 
P_{\perp}^{K^+}(\mu ) \leq 20 \cdot 
(\epsilon ^{\prime}/\epsilon _K) \cdot \epsilon _K \leq 
10^{-4} 
\ee 
where the factor 20 allows for the `accidental' reduction of 
$\epsilon ^{\prime}/\epsilon _K$ by the $\Delta I=1/2$ rule: 
$\omega \simeq 1/20$. This bound is a factor of 100 larger 
than what one could obtain within KM.  It could actually be bigger 
still since there is a loophole in this generic 
argument: Higgs couplings to leptons could be strongly 
enhanced through a large ratio of vacuum expectation values $v_1$ 
relative to $v_3$, where $v_1$ controls the couplings to 
up-type quarks and $v_3$ to leptons. 
Then 
\be 
P_{\perp}^{K^+}(\mu )|_{Higgs} \leq {\cal O}(10^{-3}) 
\ee 
becomes conceivable with the Higgs fields as heavy as 
80 - 200 GeV 
\cite{GARISTO}. Such Higgs exchanges would be quite insignificant 
for $K_L \ra \pi \pi $! 

Since $K_{\mu 3}$ studies provide such a unique 
window onto Higgs dynamics, I find it mandatory to probe for 
$P_{\perp}(\mu ) \neq 0$ in a most determined way. 
It is gratifying to note that an on-going KEK experiment will be 
sensitive to $P_{\perp}(\mu )$ down to the $10^{-3}$ level -- 
yet I strongly feel one should not stop there, but push 
further down to the $10^{-4}$ level. 
\subsection{Electric Dipole Moments}
Consider a system -- such as an elementary particle or  
an atom -- in a weak external electric field $\vec E$. The 
energy shift of this system due to the electric field 
can then be expressed through an expansion in powers 
of $\vec E$ \cite{BERN}: 
\be 
\Delta E = \vec d \cdot \vec E + d_{ij}E_i E_j + 
{\cal O}(|\vec E|^3) 
\label{ESHIFTDIP}
\ee 
where summation over the indices $i,j$ is understood. The 
coefficient $\vec d$ of the term linear in $\vec E$ is called 
electric dipole moment or sometimes permanent 
electric dipole moment (hereafter referred to as EDM) whereas that 
of the quadratic 
term is often named an {\em induced} dipole moment. 

For an elementary object one has 
\be 
\vec d = d \vec j 
\ee 
where $\vec j$ denotes its total angular momentum since that 
is the only available vector. Under time reversal one finds  
\be 
\begin{array}{l}
\vec j \; \stackrel{\bf T}{\Rightarrow} \; - \vec j \\ 
\vec E \; \stackrel{\bf T}{\Rightarrow} \vec E \; . 
\end{array} 
\ee 
Therefore 
\be 
{\rm T \; invariance}  \leadsto d = 0 \; ; 
\ee 
i.e., such an electric dipole moment has to vanish, unless T is 
violated (and likewise for parity). 

The EDM is at times confused with an induced electric dipole 
moment objects can possess due to their internal structure. To 
illustrate that consider an atom with two {\em nearly degenerate} 
states of opposite parity: 
\be 
{\bf P}|\pm \rangle = \pm |\pm \rangle \; , \; 
{\bf H}|\pm \rangle = E_{\pm} |\pm \rangle \; , \; E_+ < E_- \; , \; 
\frac{E_- - E_+}{E_+} \ll 1 
\ee 
Placed in a constant external electric field $\vec E$ the states 
$|\pm \rangle $ will mix to produce new energy eigenstates; 
those can be found by diagonalising the matrix of the 
Hamilton operator: 
\be 
H = 
\left( 
\begin{array}{ll}
E_+ & \Delta \\ 
\Delta & E_- 
\end{array} 
\right) 
\ee 
where $\Delta = \vec d_{ind} \cdot \vec E$ with  
$\vec d_{ind}$ being the transition matrix element between 
the $|+\rangle$ and $|-\rangle$ states induced by the electric 
field. The two new energy eigenvalues are 
\be 
E_{1,2} = \frac{1}{2} (E_+ + E_-) \pm 
\sqrt{\frac{1}{4} (E_+ - E_-)^2 + \Delta ^2} 
\ee 
For $E_+ \simeq E_-$ one has 
\be 
E_{1,2} \simeq \frac{1}{2} (E_+ + E_-) \pm |\Delta | \; ; 
\ee 
i.e., the energy shift appears to be linear in $\vec E$: 
\be 
\Delta E = E_2 - E_1 = 2 |\vec d_{ind} \cdot \vec E| 
\label{FAKELINEAR} 
\ee   
Yet with $\vec E$ being sufficiently small one arrives at 
$4(\vec d_{ind}\cdot \vec E)^2 \ll (E_+ - E_-)^2$ and therefore 
\be 
E_1 \simeq E_- + \frac{(\vec d_{ind}\cdot \vec E)^2}{E_- - E_+} \; , \; 
E_2 \simeq E_+ - \frac{(\vec d_{ind}\cdot \vec E)^2}{E_- - E_+} \; ; 
\ee
i.e., the induced energy shift is {\em quadratic} in $\vec E$ 
rather than  
linear and therefore does {\em not} imply T violation! The distinction 
between an EDM and an induced electric dipole moment is somewhat 
subtle -- yet it can be established in an unequivocal way by 
probing for a linear Stark effect with weak electric fields. A more 
careful look at Eq.(\ref{FAKELINEAR}) already indicates that. 
For the energy shift stated there does not change under 
$\vec E \Rightarrow - \vec E$ as it should for an EDM which also 
violates parity! 

The data for neutrons read: 
\be 
d_n = \left\{   
\begin{array}{l} 
(-3 \pm 5)\cdot 10^{-26} \; ecm \; \; \; \; \; {\rm ILL} \\ 
(2.6 \pm 4 \pm 1.6)\cdot 10^{-26} \; ecm \; \; \; \; \; {\rm LNPI}  
\end{array}
\right. 
\label{EDMNEUT}
\ee 
These numbers and the experiments leading to them are very 
impressive: 
\begin{itemize} 
\item 
One uses neutrons emanating from a reactor and 
subsequently cooled down 
to a temperature of order $10^{-7}$ eV. This is comparable to the 
kinetic energy a neutron gains when dropping 1 m in the 
earth's gravitational field. 
\item 
Extrapolating the ratio between the neutron's radius 
-- $r_N \sim 10^{-13}$ cm -- with its EDM of no more than 
$10^{-25}$ ecm to the earth's case, one would say that it 
corresponds to a situation where one has searched for a displacement 
in the earth's mass distribution of order $10^{-12} \cdot r_{earth} 
\sim 10^{-3} {\rm cm} = 10$ microns!  
\end{itemize} 

A truly dramatic increase in sensitivity for the 
{\em electron's} EDM has 
been achieved over the last few years: 
\be 
d_e = (-0.3 \pm 0.8) \cdot 10^{-26} \; \; e \, cm 
\label{EDMEL}
\ee 
This quantity is searched for through measuring electric dipole 
moments of {\em atoms}. At first this would seem to be a 
losing proposition theoretically: for according to Schiff's 
theorem an atom when placed inside an external electric 
field gets deformed in such a way that the electron's EDM is 
completely shielded; i.e., $d_{atom} = 0$. This theorem 
holds true in the nonrelativistic limit, yet is vitiated by relativistic 
effects. Not surprisingly the latter are particularly large for 
heavy atoms; one would then expect the electron's 
EDM to be only partially shielded: $d_{atom} = S\cdot d_e$ with 
$S < 1$. Yet amazingly -- and highly welcome of 
course -- the electron's EDM can actually get magnified by two to three 
orders of magnitude in the atom's electric dipole moment; for 
Caesium one has \cite{BERN} 
\be 
d_{Cs} \simeq 100 \cdot d_e 
\ee
This enhancement factor is the theoretical reason behind the 
greatly improved sensitivity for $d_e$ as expressed through 
Eq.(\ref{EDMEL}); the other one is experimental, namely the great 
strides made by laser technology applied to atomic physics. 

The quality of the number in Eq.(\ref{EDMEL}) can be illustrated 
through a comparison with the electron's magnetic moment. 
The electromagnetic form factor $\Gamma _{\mu}(q)$ 
of a particle like the electron evaluated at momentum 
transfer $q$ contains two tensor terms: 
\be 
d_{atom} =  
\frac{1}{2m_e}\sigma _{\mu \nu} q^{\nu}\left[ i F_2(q^2) + 
F_3(q^2) \gamma _5 \right]  + ... 
\ee  
In the nonrelativistic limit one finds for the EDM: 
\be 
d_e = - \frac{1}{2m_e} F_3(0) 
\ee 
On the other hand one has 
\be 
\frac{1}{2}(g-2) = \frac{1}{e} F_2(0) 
\ee 
The {\em precision} with which $g-2$ is known for the 
electron -- $\delta [(g-2)/2] \simeq 10^{-11}$ -- 
(and which represents one of the great success stories of 
field theory) corresponds to an {\em uncertainty} in the electron's 
{\em magnetic} moment 
\be 
\delta \left[ \frac{1}{2m_e} F_2(0)\right] \simeq 
2\cdot 10^{-22} \; \; e\, cm
\ee 
that is several orders of magnitude larger than the bound on its 
EDM! 

Since the EDM is, as already indicated above, described by a 
dimension-five operator in the Lagrangian 
\be 
{\cal L}_{EDM} = - \frac{i}{2} d 
\bar \psi \sigma _{\mu \nu}\gamma _5 \psi F^{\mu \nu} 
\ee
with $F^{\mu \nu}$ denoting the electromagnetic field strength 
tensor, one can calculate $d$ within a given theory of CP violation 
as a finite quantity.  Within the KM ansatz one finds that the 
neutron's EDM is zero for all practical purposes 
\footnote{I ignore here the Strong CP Problem.}:  
\be 
\left. d_N\right| _{KM} < 10^{-30} \; \; e\, cm 
\ee
and likewise for $d_e$. Yet again that is due to very specific features 
of the KM mechanism and the chirality structure of the Standard 
Model. In alternative models -- where CP violation enters 
through {\em right}-handed currents or a non-minimal 
Higgs sector (with or without involving SUSY) -- one finds 
\be 
\left. d_N\right| _{New\; Physics} \sim 
10^{-27} - 10^{-28} \; \; e\, cm 
\ee  
as reasonable benchmark figures.

\section{Summary on the CP Phenomenology with Light Degrees of 
Freedom}
\label{SUMMARYLIGHT} 
To summarize our discussion up to this point: 
\begin{itemize} 
\item 
The following data represent the most sensitive probes:  
\be 
{\rm BR}(K_L \ra \pi ^+ \pi ^-) = 2.3 \cdot 10^{-3} \neq 0 
\ee  
\be 
\frac{{\rm BR}(K_L \ra l^+ \nu \pi ^-)}
{{\rm BR}(K_L \ra l^- \nu \pi ^+)} \simeq 1.006 \neq 1 
\ee 
\be 
{\rm Re} \frac{\epsilon ^{\prime}}{\epsilon _K} = 
\left\{ 
\begin{array}{l} 
(2.3 \pm 0.7) \cdot 10^{-3} \; \; NA\, 31 \\ 
(0.6 \pm 0.58 \pm 0.32 \pm 0.18) \cdot 10^{-3} \; \; 
E\, 731 
\end{array}  
\right. 
\ee
\be 
{\rm Pol}_{\perp}^{K^+}(\mu ) = (-1.85\pm 3.60) \cdot 10^{-3} 
\ee
\be 
d_N < 12 \cdot 10^{-26} \; \; e\, cm 
\ee   
\be 
d_{Tl} = (1.6 \pm 5.0) \cdot 10^{-24} \; \; e\, cm 
\stackrel{theor.}{\leadsto} 
d_e = (-2.7 \pm 8.3) \cdot 10^{-27} \; \; e\, cm
\ee 
\item 
An impressive amount of experimental ingenuity, acumen and 
commitment 
went into producing this list. We know that CP violation 
unequivocally exists in nature; it can be 
characterized by a {\em single} non-vanishing quantity: 
\be 
{\rm Im} M_{12} \simeq 1.1 \cdot 10^{-8} \; eV \neq 0 
\ee  
\item 
The `Superweak Model' states that there just happens to 
exist a $\Delta S=2$ interaction that is fundamental or effective -- 
whatever the case may be -- generating Im$M_{12} =$ 
Im$M_{12}|_{exp}$ while $\epsilon ^{\prime} =0$. It 
provides merely  
a {\em classification} for possible dynamical implementations 
rather than such a dynamical implementation itself.  

\item 
The KM ansatz allows us to incorporate CP violation into the 
Standard Model. Yet it does not regale us with an understanding. 
Instead it relates the origins of CP violation to central 
mysteries of the Standard Model: Why are there families? 
Why are there three of those?  What is underlying 
the observed pattern in the fermion masses? 
\item 
Still the KM ansatz succeeds in {\em accommodating} the data in an 
unforced way: $\epsilon _K$ emerges to be naturally 
small, $\epsilon ^{\prime}$ naturally tiny (once the huge 
top mass is built in), the EDM's for neutrons [electrons] 
naturally (tiny)$^2$ [(tiny)$^3$] etc. 

\end {itemize}

\section{CP Violation in Beauty Decays -- The KM Perspective}
\label{BEAUTY} 
The KM predictions for strange decays and electric dipole 
moments given above will be subjected to sensitive tests in the 
foreseeable future. Yet there is one question that most naturally 
will come up in this context: "Where else to look?" 
I will show below that on very general grounds one has to 
conclude that the decays of beauty hadrons provide by far the 
optimal lab. Yet first I want to make some historical remarks. 

\subsection{The Emerging Beauty of $B$ Hadrons}
\subsubsection{Persistence Awarded!}
In 1970 Lederman's group studying the Drell-Yan process 
\be 
p p \ra \mu ^+ \mu ^- X
\ee 
at Brookhaven observed a shoulder in the di-muon mass 
distribution around 3 GeV. 1974 saw the `Octobre Revolution' when 
Ting et al. and Richter et al. found a narrow resonance -- 
the $J/\psi$ 
-- with a mass of 3.1 GeV at Brookhaven and SLAC, respectively, and 
announced it. In 1977 Lederman's group working at 
Fermilab discovered 
three resonances in the mass range of 9.5 - 10.3 GeV, the 
$\Upsilon$, $\Upsilon ^{\prime}$ and $\Upsilon ^{\prime \prime}$! 
That shows that persistence can pay off -- at least sometimes and for  
some people. 

\subsubsection{Longevity of Beauty}
The lifetime of weakly decaying beauty quarks can be related 
to the muon lifetime 
\be  
\tau (b) \sim \tau (\mu ) \left( \frac{m_{\mu}}{m_b} \right) ^5 
\frac{1}{9} \frac{1}{|V(cb)|^2} \sim 
3 \cdot 10^{-14} \left|  \frac{{\rm sin}\theta _C}{V(cb)}\right| ^2  
\; {\rm sec} 
\ee  
for a $b$ quark mass of around  5 GeV; the factor 1/9 reflects 
the fact that the virtual $W ^-$ boson in $b$ quark decays can 
materialize as a $d \bar u$ or $s \bar c$ in three colours each and 
as three lepton pairs. I have ignored phase space corrections here. 
Since the $b$ quark has to decay outside its own family one would 
expect $|V(cb)| \sim {\cal O}({\rm sin}\theta _C) = 
|V(us)|$. Yet starting in 1982 data showed a considerably longer 
lifetime 
\be 
\tau ({\rm beauty}) \sim 10^{-12} \; {\rm sec} 
\ee 
implying 
\be 
|V(cb)| \sim {\cal O}({\rm sin}^2\theta _C) \sim 0.05 
\ee 
The technology to resolve decay vertices for objects of such 
lifetimes happened to have just been developed -- for charm 
studies! 

\subsubsection{The Changing Identity of Neutral $B$ Mesons}
Speedy $B_d - \bar B_d$ oscillations were discovered by ARGUS in 
1986: 
\be 
x_d \equiv \frac{\Delta m(B_d)}{\Gamma (B_d)} \simeq 
{\cal O}(1)
\ee 
These oscillations can then be tracked like the decays. This 
observation was also the first evidence that top quarks had to be 
heavier than orginally thought, namely $m_t \geq M_W$. 

\subsubsection{Beauty Goes to Charm (almost always)}
It was soon found that $b$ quarks exhibit a strong preference 
to decay into charm rather than up quarks  
\be 
\left| \frac{V(ub)}{V(cb)} \right| ^2 \ll 1 
\ee 
establishing thus the hierarchy 
\be 
|V(ub)|^2 \ll |V(cb)|^2 \ll |V(us)|^2 \ll 1
\ee 

\subsubsection{Resume}
We will soon see how all these observations form crucial inputs 
to the general message that big CP asymmetries should emerge in 
$B$ decays and that they (together with interesting rare decays)  
are within reach of experiments. It is for this reason that I strongly 
feel that the only appropriate name for this quantum number is 
{\em beauty}! A name like bottom would not do it justice. 

\subsection{The KM Paradigm of Huge CP Asymmetries}
\subsubsection{Large Weak Phases!}
The Wolfenstein representation expresses the CKM matrix as an 
expansion: 
\be 
V_{CKM} = 
\left( 
\begin{array}{ccc} 
1 & {\cal O}(\lambda ) & {\cal O}(\lambda ^3) \\ 
{\cal O}(\lambda ) & 1 & {\cal O}(\lambda ^2) \\ 
{\cal O}(\lambda ^3) & {\cal O}(\lambda ^2) & 1 
\end {array} 
\right) 
\; \; \; , \; \; \; \lambda = {\rm sin}\theta _C 
\ee 
The crucial element in making this expansion meaningful is the 
`long' lifetime of beauty hadrons of around 1 psec. That number 
had to change by an order of magnitude -- which is out of the 
question -- to invalidate the conclusions given below for the 
size of the weak phases. 

The unitarity condition yields 6 triangle relations: 
\be 
\begin{array}{ccc} 
V^*(ud)V(us) + &V^*(cd)V(cs) + &V^*(td) V(ts) = 
\delta _{ds}= 0 \\
{\cal O}(\lambda ) & {\cal O}(\lambda ) & {\cal O}(\lambda ^5) 
\end{array} 
\label{TRI1} 
\ee 
\be 
\begin{array}{ccc} 
V^*(ud)V(cd) + &V^*(us)V(cs) + &V^*(ub) V(cb) = 
\delta _{uc}= 0 \\
{\cal O}(\lambda ) & {\cal O}(\lambda ) & {\cal O}(\lambda ^5) 
\end{array} 
\label{TRI2} 
\ee 
\be 
\begin{array}{ccc} 
V^*(us)V(ub) + &V^*(cs)V(cb) + &V^*(ts) V(tb) = 
\delta _{sb}= 0 \\
{\cal O}(\lambda ^4) & {\cal O}(\lambda ^2) & {\cal O}(\lambda ^2) 
\end{array} 
\label{TRI3} 
\ee 
\be 
\begin{array}{ccc} 
V^*(td)V(cd) + &V^*(ts)V(cs) + &V^*(tb) V(cb) = 
\delta _{ct}=0 \\
{\cal O}(\lambda ^4) & {\cal O}(\lambda ^2) & {\cal O}(\lambda ^2) 
\end{array} 
\label{TRI4} 
\ee 
\be 
\begin{array}{ccc} 
V^*(ud)V(ub) + &V^*(cd)V(cb) + &V^*(td) V(tb) = 
\delta _{db}=0 \\
{\cal O}(\lambda ^3) & {\cal O}(\lambda ^3) & {\cal O}(\lambda ^3) 
\end{array} 
\label{TRI5} 
\ee 
\be 
\begin{array}{ccc} 
V^*(td)V(ud) + &V^*(ts)V(us) + &V^*(tb) V(ub) = 
\delta _{ut}=0 \\
{\cal O}(\lambda ^3) & {\cal O}(\lambda ^3) & {\cal O}(\lambda ^3) 
\end{array}
\label{TRI6}  
\ee 
where below each product of matrix elements I have noted 
their size in powers of $\lambda $. 

We see that the six triangles fall into three categories: 
\begin{enumerate}
\item 
The first two triangles are extremely `squashed': two sides are 
of order $\lambda $, the third one of order $\lambda ^5$ and their 
ratio of order $\lambda ^4 \simeq 2.3 \cdot 10^{-3}$; 
Eq.(\ref{TRI1}) and Eq.(\ref{TRI2}) control the situation in 
strange and charm decays; the relevant weak phases there 
are obviously tiny. 
\item 
The third and fourth triangles are still rather squashed, yet less so: 
two sides are of order $\lambda ^2$ and the third one of order 
$\lambda ^4$. 
\item 
The last two triangles have sides that are all of the same 
order, namely $\lambda ^3$. All their angles are therefore 
naturally large, i.e. $\sim$ several $\times$ $10$ degrees! Since to 
leading order in $\lambda$ one has 
\be 
V(ud) \simeq V(tb) \; , \; V(cd) \simeq - V(us) \; , \; 
V(ts) \simeq - V(cb) 
\ee 
we see that the triangles of Eqs.(\ref{TRI5}, \ref{TRI6}) 
actually coincide to that order. 
\end{enumerate} 
The sides of this triangle having naturally large angles are 
given by $\lambda \cdot V(cb)$, $V(ub)$ and 
$V^*(td)$; these are all quantities that control important 
aspects of $B$ decays, namely CKM favoured and disfavoured 
$B$ decays and $B_d - \bar B_d$ oscillations! 

\subsubsection{Different, Yet Coherent Amplitudes!}
$B^0 - \bar B^0$ oscillations provide us with two different 
amplitudes that by their very nature have to be coherent: 
\be 
B^0 \Rightarrow \bar B^0 \ra f \leftarrow B^0 
\ee 
On general grounds one expects oscillations to be speedy for 
$B^0 - \bar B^0$ (like for $K^0 - \bar K^0$), yet slow for 
$D^0 - \bar D^0$ 
\footnote{$T^0 - \bar T^0$ oscillations cannot 
occur since top quarks decay before they hadronize 
\cite{RAPALLO}.}. 
Experimentally one indeed finds 
\be 
\frac{\Delta m(B_d)}{\Gamma (B_d)} = 0.71 \pm 0.06 
\label{OSCBD}
\ee  
\be 
\frac{\Delta m(B_s)}{\Gamma (B_s)} \geq 10  
\label{OSCBS}
\ee
While Eq.(\ref{OSCBD}) describes an almost optimal 
situation the overly rapid pace of $B_s - \bar B_s$ 
oscillations will presumably cause experimental 
problems. 

The conditions are quite favourable also for {\em direct} 
CP violation to surface. Consider a transition amplitude 
\be 
T(B \ra f) = {\cal M}_1 + {\cal M}_2 = 
e^{i\phi _1}e^{i\alpha _1}|{\cal M}_1| + 
e^{i\phi _2}e^{i\alpha _2}|{\cal M}_2| \; . 
\ee    
The two partial amplitudes ${\cal M}_1$ and ${\cal M}_2$ are 
distinguished by, say, their isospin -- as it was the case for 
$K \ra (\pi \pi )_{I=0,2}$ discussed before; $\phi _1$, $\phi _2$ 
denote the phases in the {\em weak} couplings and 
$\alpha _1$, $\alpha _2$ the phase shifts due to 
{\em strong} final state interactions. For the CP conjugate reaction 
one obtains 
\be 
T(\bar B \ra \bar f) =  
e^{-i\phi _1}e^{i\alpha _1}|{\cal M}_1| + 
e^{-i\phi _2}e^{i\alpha _2}|{\cal M}_2| \; . 
\ee 
since under CP the weak parameters change into their 
complex conjugate values whereas the phase shifts remain 
the same; for the strong forces driving final state 
interactions conserve CP. The rate difference is then given by 
$$ 
\Gamma (B \ra f) - \Gamma (\bar B \ra \bar f) \propto 
|T(B \ra f)|^2 - |T( \bar B \ra \bar f)|^2 = 
$$
\be 
= - 4 {\rm sin}(\phi _1 - \phi _2) \cdot 
{\rm sin}(\alpha _1 - \alpha _2) \cdot 
{\cal M}_1 \otimes {\cal M}_2  
\ee 
For an asymmetry to arise in this way two conditions need to be 
satisfied simultaneously, namely 
\be 
\begin{array}{l} 
\phi _1 \neq \phi _2 \\ 
\alpha _1 \neq \alpha _2
\end{array}
\ee 
I.e., the two amplitudes ${\cal M}_1$ and ${\cal M}_2$ have to 
differ both in their weak and strong forces! The first condition 
implies (within the Standard Model) that the reaction has to 
be KM suppressed, whereas the second one require the intervention 
of nontrivial final state interactions. 

There is a large number of KM suppressed channels in $B$ 
decays that are suitable in this context: they receive significant 
contributions from weak couplings with large phases -- 
like $V(ub)$ in the Wolfenstein representation -- and there 
is no reason why the phase shifts should be small in general 
(although that could happen in some cases).

\subsubsection{Resume}
Let me summarize the discussion just given and anticipate the 
results to be presented below. 
\begin{itemize}
\item 
{\em Large} CP asymmetries are {\em pre}dicted {\em with} 
confidence to occur in $B$ decays. If they are not found, there is 
no plausible deniability for the KM ansatz. 
\item 
Some of these predictions can be made with high 
{\em parametric} reliability. 
\item 
New theoretical technologies have emerged that will allow us to 
translate this {\em parametric} reliability into 
{\em numerical} precision. 
\item 
Some of the observables exhibit a high and unambiguous 
sensitivity to the presence of New Physics since we are 
dealing with coherent processes with observables depending  
{\em linearly} on New Physics amplitudes and where the 
CKM `background' is (or can be brought)  
under theoretical control. 
\end{itemize}

\subsection{General Phenomenology}
Decay rates for CP conjugate channels can be expressed as follows: 
\be  
\begin{array} {l} 
{\rm rate} (B(t) \ra f) = e^{-\Gamma _Bt}G_f(t) \\  
{\rm rate} (\bar B(t) \ra \bar f) = 
e^{-\Gamma _Bt}\bar G_{\bar f}(t)  
\end{array} 
\label{DECGEN}
\ee 
where CPT invariance has been invoked to assign the same lifetime 
$\Gamma _B^{-1}$ to $B$ and $\bar B$ hadrons. Obviously if 
\be
\frac{G_f(t)}{\bar G_{\bar f}(t)} \neq 1 
\ee 
is observed, CP violation has been found. Yet one should 
keep in mind that this can manifest itself in two (or three) 
qualitatively different ways: 
\begin{enumerate} 
\item 
\be 
\frac{G_f(t)}{\bar G_{\bar f}(t)} \neq 1 
\; \; {\rm with} \; \; 
\frac{d}{dt}\frac{G_f(t)}{\bar G_{\bar f}(t)} =0 \; ; 
\label{DIRECTCP1}
\ee   
i.e., the {\em asymmetry} is the same for all times of decay. This 
is true for {\em direct} CP violation; yet, as explained later, it also 
holds for CP violation {\em in the oscillations}.  
\item 
\be 
\frac{G_f(t)}{\bar G_{\bar f}(t)} \neq 1 
\; \; {\rm with} \; \; 
\frac{d}{dt}\frac{G_f(t)}{\bar G_{\bar f}(t)} \neq 0 \; ; 
\label{DIRECTCP2}
\ee   
here the asymmetry varies as a function of the time of decay. 
This can be referred to as CP violation {\em involving 
oscillations}. 
\end{enumerate} 

Quantum mechanics with its linear superposition principle makes 
very specific statements about the possible time dependance of 
$G_f(t)$ and $\bar G_{\bar f}(t)$; yet before going into that 
I want to pose another homework problem: 
\begin{center} 
$\spadesuit \; \; \; \spadesuit \; \; \; \spadesuit $ \\ 
{\em Homework Problem \# 4}: 
\end{center}
Consider the reaction 
\be 
e^+ e^- \ra \phi \ra 
(\pi ^+\pi ^-)_K  (\pi ^+\pi ^-)_K 
\ee 
Its occurrance requires CP violation. For the {\em initial} state -- 
$\phi $ -- carries {\em even} CP parity whereas the 
{\em final} state with the two $(\pi ^+\pi ^-)$ combinations 
forming a P wave must be CP {\em odd}: 
$(+1)^2 (-1)^l = -1$! Yet Bose statistics requiring identical 
states to be in a symmetric configuration would appear to 
veto this reaction; for it places the two $(\pi ^+\pi ^-)$ states 
into a P wave which is antisymmetric. What is the flaw in this 
reasoning? The same puzzle can be formulated in terms of 
\be 
e^+ e^- \ra \Upsilon (4S) \ra B_d \bar B_d \ra 
(\psi K_S)_B (\psi K_S)_B \; . 
\ee 
\begin{center} 
$\spadesuit \; \; \; \spadesuit \; \; \; \spadesuit $
\end{center} 
A straightforward application of quantum mechanics yields  
the general expressions 
\cite{CARTER,BS,CECILIABOOK}:
\be 
\begin{array}{l}
G_f(t) = |T_f|^2 
\left[ 
\left( 1 + \left| \frac{q}{p}\right| ^2|\bar \rho _f|^2 \right) + 
\left( 1 - \left| \frac{q}{p}\right| ^2|\bar \rho _f|^2 \right) 
{\rm cos}\Delta m_Bt 
- 2 ({\rm sin}\Delta m_Bt) {\rm Im}\frac{q}{p} \bar \rho _f 
\right]  \\ 
\bar G_{\bar f}(t) = |\bar T_{\bar f}|^2 
\left[ 
\left( 1 + \left| \frac{p}{q}\right| ^2|\rho _{\bar f}|^2 \right) + 
\left( 1 - \left| \frac{p}{q}\right| ^2|\rho _{\bar f}|^2 \right) 
{\rm cos}\Delta m_Bt 
- 2 ({\rm sin}\Delta m_Bt) {\rm Im}\frac{p}{q} \rho _{\bar f}  
\right]  
\end{array}
\ee 
The amplitudes for the instantaneous $\Delta B=1$ 
transition into a 
final state $f$ are denoted by 
$T_f = T(B \ra f)$ and $\bar T_f = T(\bar B \ra f)$ and  
\be 
\bar \rho _f = \frac{\bar T_f}{T_f} \; \; , 
\rho _{\bar f} = \frac{T_{\bar f}}{\bar T_{\bar f}} \; \; , 
\frac{q}{p} = \sqrt{\frac{M_{12}^* - \frac{i}{2} \Gamma _{12}^*}
{M_{12} - \frac{i}{2} \Gamma _{12}}}
\ee 
Staring at the general expression is not always very illuminating; 
let us therefore consider three very simplified limiting cases: 
\begin{itemize}
\item
$\Delta m_B = 0$, i.e. {\em no} $B^0- \bar B^0$ oscillations: 
\be 
G_f(t) = 2|T_f|^2 \; \; , \; \; 
\bar G_{\bar f}(t) = 2|\bar T_{\bar f}|^2 
\leadsto \frac{\bar G_{\bar f}(t)}{G_{ f}(t)} = 
\left|
\frac{\bar T_{\bar f}}{T_{ f}}
\right|^2 \; \; , \frac{d}{dt}G_f (t) \equiv 0 \equiv 
\frac{d}{dt}\bar G_{\bar f} (t) 
\ee 
This is explicitely what was referred to above as {\em direct} 
CP violation. 
\item 
$\Delta m_B \neq  0$   
and $f$ a flavour-{\em specific} final state with {\em no} 
direct CP violation; i.e., 
$T_{f} = 0 = \bar T_{\bar f}$ and $\bar T_f = T_{\bar f}$   
\footnote{For a flavour-specific mode one has in general 
$T_f \cdot T_{\bar f} =0$; the more intriguing case arises  
when one considers a transition that requires oscillations 
to take place.}: 
\be 
\begin{array} {c} 
G_f (t) = \left| \frac{q}{p}\right| ^2 |\bar T_f|^2 
(1 - {\rm cos}\Delta m_Bt )\; \; , \; \; 
\bar G_{\bar f} (t) = \left| \frac{p}{q}\right| ^2 |T_{\bar f}|^2 
(1 - {\rm cos}\Delta m_Bt) \\ 
\leadsto 
\frac{\bar G_{\bar f}(t)}{G_{ f}(t)} = 
\left| \frac{q}{p}\right| ^4 
\; \; , \; \; \frac{d}{dt} \frac{\bar G_{\bar f}(t)}{G_{ f}(t)} 
\equiv 0  
\; \; , \; \; \frac{d}{dt} \bar G_{\bar f}(t) \neq 0 \neq 
\frac{d}{dt} G_ f(t)
\end{array} 
\ee 
This constitutes CP violation {\em in the 
oscillations}. For the CP conserving decay into the 
flavour-specific 
final state is used merely to track the flavour identity of the 
decaying meson. This situation can therefore be denoted also 
in the following way: 
\be 
\frac{{\rm Prob}(B^0 \Rightarrow  \bar B^0; t) - 
{\rm Prob}(\bar B^0 \Rightarrow  B^0; t)}
{{\rm Prob}(B^0 \Rightarrow \bar B^0; t) + 
{\rm Prob}(\bar B^0 \Rightarrow  B^0; t)} = 
\frac{|q/p|^2 - |p/q|^2}{|q/p|^2 + |p/q|^2} = 
\frac{1- |p/q|^4}{1+ |p/q|^4} 
\ee 

\item 
$\Delta m_B \neq  0$ with $f$ now being a  
flavour-{\em non}specific final state -- a final state 
{\em common} 
to $B^0$ and $\bar B^0$ decays -- of a special nature, namely 
a CP eigenstate -- $|\bar f\rangle = {\bf CP}|f\rangle = 
\pm |f\rangle $ -- {\em without} direct CP violation --  
$|\bar \rho _f| = 1 = |\rho _{\bar f}| $: 
\be 
\begin{array} {c} 
G_f(t) = 2 |T_f|^2 
\left[ 1 - ({\rm sin}\Delta m_Bt) \cdot 
{\rm Im} \frac{q}{p} \bar \rho _f 
\right] \\  
\bar G_f(t) = 2 |T_f|^2 
\left[ 1 + ({\rm sin}\Delta m_Bt )\cdot 
{\rm Im} \frac{q}{p} \bar \rho _f 
\right] \\ 
\leadsto 
\frac{d}{dt} \frac{\bar G_f(t) }{G_f(t)} \neq 0
\end{array} 
\label{GGBAR}
\ee 
is the concrete realization of what was called CP violation 
{\em involving oscillations}. 
\end{itemize} 

\subsubsection{CP Violation in Oscillations}
Using the convention blessed by the PDG 
\be 
B = [\bar b q] \; \; , \; \; \bar B = [\bar q b] 
\eeq
we have 
\be 
\begin{array} {c} 
T(B \ra l^- X) = 0 = T(\bar B \ra l^+ X) \\ 
T_{SL} \equiv  T(B \ra l^+ X) = T(\bar B \ra l^- X) 
\end{array} 
\ee  
with the last equality enforced by CPT invariance. The 
so-called Kabir test can then be realized as follows: 
$$  
\frac
{{\rm Prob}(B^0 \Rightarrow \bar B^0; t) - 
{\rm Prob}(\bar B^0 \Rightarrow B^0; t)} 
{{\rm Prob}(B^0 \Rightarrow \bar B^0; t) + 
{\rm Prob}(\bar B^0 \Rightarrow B^0; t)} = 
$$ 
\be 
= \frac
{{\rm Prob}(B^0 \Rightarrow \bar B^0 \ra l^-X; t) - 
{\rm Prob}(\bar B^0 \Rightarrow B^0 \ra l^+X; t)} 
{{\rm Prob}(B^0 \Rightarrow \bar B^0 \ra l^-X; t) + 
{\rm Prob}(\bar B^0 \Rightarrow B^0 \ra l^+X; t)} = 
\frac{1 - |q/p|^4}{1+|q/p|^4} 
\ee  
Without going into details I merely state the results here 
\cite{CECILIABOOK}: 
\be 
1 - \left|  \frac{q}{p} \right| \simeq 
\frac{1}{2} {\rm Im}\left( \frac{\Gamma _{12}}
{M_{12}} \right) \sim 
\left\{ 
\begin{array}{ccc} 
10^{-3} & \; {\rm for} \; & B_d=(\bar bd) \\ 
10^{-4} & \; {\rm for} \; & B_s =(\bar bs) \\ 
\end{array} 
\right. 
\ee
i.e.,  
\be 
a_{SL} (B^0) \equiv \frac{\Gamma (\bar B^0(t) \ra l^+ \nu X) - 
\Gamma ( B^0(t) \ra l^- \bar \nu X)}
{\Gamma (\bar B^0(t) \ra l^+ \nu X) +  
\Gamma ( B^0(t) \ra l^- \bar \nu X)} \simeq 
\left\{ 
\begin{array}{ccc} 
{\cal O}(10^{-3}) & \; {\rm for} \; & B_d \\ 
{\cal O}(10^{-4}) & \; {\rm for} \; & B_s \\ 
\end{array} 
\right. 
\ee  
The smallness of the quantity $1-|q/p|$ is primarily due to 
$|\Gamma _{12}| \ll |M_{12}|$ or $\Delta \Gamma _B \ll 
\Delta m_B$. Within the Standard Model this hierarchy 
is understood (semi-quantitatively at leaast) as due to the 
hierarchy in the GIM factors of the box diagram 
expressions for $\Gamma _{12}$ and $M_{12}$, namely 
$m_c^2/M_W^2 \ll m_t^2/M_W^2$. 

For $B_s$ mesons the phase between $\Gamma _{12}$ and 
$M_{12}$ is further (Cabibbo) suppressed for reasons that 
are peculiar to the KM ansatz: for to leading order in the 
KM parameters quarks of the second and third family only 
contribute and therefore 
arg$(\Gamma _{12}/M_{12}) = 0$ to that order. If New 
Physics intervenes in $B^0 - \bar B^0$ oscillations, it would 
quite naturally generate a new phase between 
$\Gamma _{12}$ and $M_{12}$; it could also reduce 
$M_{12}$. Altogether this CP asymmetry could get 
enhanced very considerably: 
\be 
a_{SL}^{New \; Physics} (B^0) \sim 1 \% 
\ee  
Therefore one would be ill-advised to accept the somewhat 
pessimistic KM predictions as gospel. 

Since this CP asymmetry does not vary with the time of decay, 
a signal is not diluted by integrating over all times. It is, 
however, essential to `flavour tag' the decaying meson; i.e., 
determine whether it was {\em produced} as  a 
$B^0$ or $\bar B^0$. This can be achieved in several ways 
as discussed later.

\subsubsection{Direct CP Violation}

Sizeable direct CP asymmetries arise rather naturally in 
$B$ decays. Consider 
\be 
b \ra s \bar u u 
\ee 
Three different processes contribute to it, namely 
\begin{itemize}
\item 
the tree process 
\be 
b \ra u W^* \ra u (\bar u s)_W  \; , 
\ee 
\item 
the penguin process with an internal top quark which is 
purely local (since $m_t > m_b$) 
\be 
b \ra s g^* \ra s u \bar u \; , 
\ee 
\item 
the penguin reaction with an internal charm quark. Since 
$m_b > 2m_c + m_s$, this last operator is {\em not} 
local: it contains an absorptive part that amounts to a 
final state interaction including a phase shift. 
\end{itemize}
One then arrives at a guestimate \cite{SONI,CECILIABOOK} 
\be 
\frac{\Gamma (b \ra s u \bar u) - 
\Gamma (\bar b \ra \bar s u \bar u)}
{\Gamma (b \ra s u \bar u) + 
\Gamma (\bar b \ra \bar s u \bar u)} 
\sim {\cal O}(\% ) 
\ee  
Invoking quark-hadron duality one can expect  
(or at least hope) that this quark level analysis -- rather 
than being washed out by hadronisation -- yields 
some average asymmetry or describes the 
asymmetry for some inclusive subclass of nonleptonic 
channels. I would like to draw the following lessons 
from these considerations: 
\begin{itemize}
\item 
According to the KM ansatz the natural scale for direct 
CP asymmetries in the decays of beauty hadrons 
(neutral or charged mesons or baryons) is the 
$10^{-2}$ level -- not $10^{-6} \div 10^{-5}$ as in 
strange decays! 
\item 
The size of the asymmetry in {\em individual} channels -- 
like $B\ra K \pi$ -- is shaped by the strong final state 
interactions operating there. Those are likely to differ 
considerably from channel to channel, and at present we 
are unable to predict them since they reflect long-distance 
dynamics. 
\item 
Observation of such an asymmetry (or lack thereof) will 
not provide 
us with reliable numerical information on the parameters of the 
microscopic theory, like the KM ansatz. 
\item 
Nevertheless comprehensive and detailed studies are an 
absolute must!
\end{itemize} 
Later I will describe examples where the relevant 
long-distance 
parameters -- phase shifts etc. -- can be {\em measured} 
independantly. 

\subsubsection{CP Violation Involving Oscillations}
The essential feature that a final state in this category has to 
satisfy is that it can be fed both by $B^0$ and $\bar B^0$ decays 
\footnote{Obviously no such common channels can exist for 
charged mesons or for baryons.}. However for convenience reasons 
I will concentrate on a special subclass of such modes, namely 
when the final state is a CP eigenstate. A more comprehensive 
discussion can be found in \cite{CECILIABOOK,BOOK}. 

Three qualitative observations have to be made here: 
\begin{itemize}
\item 
Since the final state is shared by $B^0$ and $\bar B^0$ decays 
one cannot even define a CP asymmetry unless one acquires 
{\em independant} information on the decaying meson: was it 
a $B^0$ or $\bar B^0$ or -- more to the point -- was it 
originally 
produced as a $B^0$ or $\bar B^0$? There are several 
scenarios for achieving such {\em flavour tagging}: 
\begin{itemize} 
\item 
Nature could do the trick for us by providing us with 
$B^0$ - $\bar B^0$ production asymmetries through, say, 
associated production in hadronic collisions or the use of 
polarized beams in $e^+ e^-$ annihilation. Those production 
asymmetries could be tracked through decays that are 
necessarily CP conserving -- like $\bar B_d \ra \psi K^- \pi ^+$ vs.  
$B_d \ra \psi K^+ \pi ^-$. It seems unlikely, though, that such 
a scenario could ever be realized with sufficient statistics. 

\item 
{\em Same Side Tagging}: One undertakes to repeat the success 
of the $D^*$ tag for charm mesons -- $D^{+*} \ra D^0 \pi ^+$ 
vs. $D^{-*} \ra \bar D^0 \pi ^-$ -- through finding a conveniently 
placed nearby resonance -- $B^{-**} \ra \bar B_d \pi ^-$  vs. 
$B^{+**} \ra B_d \pi ^+$ -- or through employing correlations 
between the beauty mesons and a `nearby' pion (or kaon 
for $B_s$) as pioneered by the CDF collaboration. This method 
can be calibrated by 
analysing how well $B^0 - \bar B^0$ oscillations are reproduced.

\item 
{\em Opposite Side Tagging}: With electromagnetic and strong 
forces conserving the beauty quantum number, one can employ 
charge correlations between the decay products (leptons and kaons) 
of the two beauty hadrons originally produced together.  

\item 
If the lifetimes of the two mass eigenstates of the neutral $B$ 
meson differ sufficiently from each other, then one can wait 
for the short-lived  component to fade away relative to the 
long-lived one and proceed in qualitative analogy to the $K_L$ 
case. Conceivably this could become feasible -- or even 
essential -- for overly fast oscillating $B_s$ mesons 
\cite{DUNIETZ}. 

\end{itemize} 
The degree to which this flavour tagging can be achieved is a crucial 
challenge each experiment has to face. 

\item 
The CP asymmetry is largest when the two interfering amplitudes 
are comparable in magnitude. With oscillations having to provide 
the second amplitude that is absent initially at time of production, 
the CP asymmetry starts out at zero for decays that occur right after 
production and builds up for later decays. The (first) maximum 
of the asymmetry 
\be 
\left|  1 - \frac{1 - {\rm Im}\frac{q}{p}\bar \rho _f 
{\rm sin}\Delta m_Bt}
{1 + {\rm Im}\frac{q}{p}\bar \rho _f 
{\rm sin}\Delta m_Bt}
\right| 
\ee 
is reached for 
\be 
\frac{t}{\tau _B} = \frac{\pi }{2} \frac{\Gamma _B}{\Delta m_B} 
\simeq 2 
\ee 
in the case of $B_d$ mesons. 

\item 
The other side of the coin is that very rapid oscillations -- 
$\Delta m_B \gg \Gamma _B$ as is the case for $B_s$ mesons -- 
will tend to wash out the asymmetry or at least will severely 
tax the experimental resolution.  

\end{itemize} 

\subsubsection{On the Sign of CP Asymmetries involving 
Oscillations}
Let us consider the asymmetry derived from Eq.(\ref{GGBAR}) 
for decays into a final state $f$ that is a CP eigenstate: 
\be 
A_f \equiv \frac{\Gamma (\bar B_d(t) \ra f) - 
\Gamma ( B_d(t) \ra f)}{\Gamma (\bar B_d(t) \ra f) + 
\Gamma ( B_d(t) \ra f)} = 
({\rm sin}\Delta m_B t) \cdot 
{\rm Im}\frac{q}{p}\bar \rho _f 
\ee 
Obviously one can measure both the size and the sign 
of $A_f$. Yet 
at first sight it would appear that no useful information 
can be extracted from the observed sign since it depends 
on $\Delta m_B$ and the sign of the latter 
cannot be defined nor 
determined {\em experimentally} in a feasible way. 
For the two mass eigenstates in the 
$B_d/\bar B_d$ complex can be distinguished neither 
by an observable difference in lifetimes nor by their 
CP parities -- unlike for kaons: 
(i) One confidently predicts 
$\Delta \Gamma _{B_d}/\Gamma _{B_d}$ 
to not exceed the percent level. Such a small difference 
cannot be observed in the foreseeable future. Remember that 
$\Gamma _{K_S} \gg \Gamma _{K_L}$ represents a 
kinematical accident.  
(ii) As emphasized before the KM ansatz predicts 
large CP violation 
in the $B_d/\bar B_d$ complex. 
 
Yet the essential point is that within a given theory for 
$\Delta B=2$ dynamics one can nevertheless predict 
the overall sign of $A_f$!
 
Let us start from the general discussion in 
Sect.\ref{GENFORM}. Using the conventions 
$q/p = + \sqrt{(M_{12}^* - \frac{i}{2}\Gamma _{12}^*)/
(M_{12} - \frac{i}{2}\Gamma _{12})}$, ${\bf CP}|B_d\rangle 
= |\bar B_d \rangle$ one finds 
\be 
\Delta m_{B_d} \simeq - 2 \bar M_{12}^{B_d} 
\ee 
like in the kaon case, albeit for a different reason, 
namely $|\bar \Gamma _{12}/\bar M_{12}| \ll 1$. One 
relies on the box diagram to derive 
${\cal L}_{eff}(\Delta B=2)$ and evaluate $\Delta m_{B_d}$ 
with it. This is similar to the procedure indicated above for 
$\Delta m_K$. There is actually better justification 
for $\Delta m_B$ being produced by short-distance 
physics than $\Delta m_K$. Again we find 
\be 
\Delta m_{B_d} \equiv m_B - m_A > 0\; , 
\ee 
where the subscripts $A$ and $B$ label the 
two mass eigenstates. 
As stated above we are not able to characterize in an  
{\em empirical} way which of the two states is heavier. 
By that I mean we can only say that one state is heavier 
than the other by an amount $\Delta m_B$; yet for practical 
limitations we cannot relate this finding to an 
{\em observable}  
difference in lifetimes or to 
{\em observable} CP parities. 
Next we consider 
the ratio of decay amplitude into a final state $f$ that 
is a CP eigenstate: ${\bf CP}|f_{\pm} \rangle = 
\pm |f_{\pm} \rangle$: 
\be 
\bar \rho _{f_{\pm}} = 
\frac{\matel{f_{\pm}}{H_{\Delta B=1}}{\bar B_d}}
{\matel{f_{\pm}}{H_{\Delta B=-1}}{ B_d}} \; , \; \; 
{\bf CP}H_{\Delta B=1} ({\bf CP})^{\dagger} = 
H^*_{\Delta B=-1} 
\ee 
Since 
\be 
\matel{f_{\pm}}{H_{\Delta B=1}}{\bar B_d}= 
\matel{f_{\pm}}{({\bf CP})^{\dagger}{\bf CP}
H_{\Delta B=1}({\bf CP})^{\dagger}{\bf CP}}{\bar B_d}= 
\pm \matel{f_{\pm}}{H^*_{\Delta B=-1}}{B_d} 
\ee 
we find 
\be 
\bar \rho _{f_{\pm}} = \pm 
\frac{\matel{f_{\pm}}{H^*_{\Delta B=-1}}{B_d}}
{\matel{f_{\pm}}{H_{\Delta B=-1}}{B_d}}
\ee 
where it is obviously essential to adopt the same convention 
${\bf CP}|B_d\rangle = |\bar B_d \rangle$ used in 
evaluating $\Delta m_{B_d}$. Putting the pieces together 
we arrive at 
\be 
A_{f_{\pm}} \simeq \pm 
\left( {\rm sin}|\Delta m_{B_d}| t \right) \cdot 
{\rm Im}\left( \sqrt{\frac{M_{12}^*}{M_{12}}}
\frac{\matel{f_{\pm}}{H^*_{\Delta B=-1}}{B_d}}
{\matel{f_{\pm}}{H_{\Delta B=-1}}{B_d}} 
\right) 
\label{AFPM}
\ee 
where we have used $q/p \simeq 
\sqrt{\frac{M_{12}^*}{M_{12}}}$. We read off from 
Eq.(\ref{AFPM}) that knowing the CP parity of the 
final state $f_{\pm}$ we can deduce the sign of 
Im$\frac{q}{p}\bar \rho _f$ from the {\em observed} 
sign of $A_{f_{\pm}}$. The ambiguity we have in the 
sign of $\Delta m_B$ is thus compensated by a 
corresponding ambiguity in the sign of 
$\frac{q}{p}\bar \rho _f$. 

This at first sight surprising result can be seen also 
in the following down-to-earth way: 
\begin{itemize}
\item 
Going to a different phase convention by 
changing $q/p \ra - q/p$ maintains 
the defining property $(q/p)^2 = 
(M_{12}^* - \frac{i}{2}\Gamma _{12}^*)/
(M_{12} - \frac{i}{2}\Gamma _{12})$, see 
Eq.(\ref{Q/PSQ}). Observables thus cannot be 
affected. 
\item 
Yet the two mass eigenstates labeled by subscripts 
$A$ and $B$ exchange places, see Eq.(\ref{P1P2}). 
\item 
The difference $\Delta M = - 2 {\rm Re}\left[ 
\frac{q}{p}(M_{12} - \frac{i}{2} \Gamma _{12})\right] $ 
then flips its sign -- yet so does Im$\frac{q}{p}\bar \rho _f$! 
\item 
The product $({\rm sin}\Delta m_B t)\cdot 
{\rm Im}\frac{q}{p}\bar \rho _f$ therefore 
remains invariant. 
\end{itemize}

\subsubsection{Resume}
Three classes of quantities each describe the three types of CP 
violation: 
\begin{enumerate} 
\item 
\be 
\left| \frac{q}{p} \right| \neq 1 
\ee 
\item 
\be 
\left| 
\frac{T(\bar B \ra \bar f)}{T( B \ra f)} 
\right| \neq 1
\ee 

\item 
\be 
{\rm Im} \frac{q}{p} \frac{T(\bar B \ra \bar f)}{T( B \ra f)} 
\neq 0
\ee 
\end{enumerate}
These quantities obviously satisfy one necessary condition 
for being observables: they are insensitive to the phase convention 
adopted for the anti-state. 

\subsection{Parametric KM Predictions}
The triangle defined by 
\be 
\lambda V(cb) - V(ub) + V^*(td) = 0 
\ee
to leading order controls basic features of $B$ transitions. As 
discussed before, it has naturally large angles; it usually is called 
{\em the} KM triangle. Its angles are given by KM matrix elements 
which are most concisely expressed in the Wolfenstein 
representation: 
\be 
e^{i\phi _1} = - \frac{V(td)}{|V(td)|} \; \; ,  \; \; 
e^{i\phi _2} =  \frac{V^*(td)}{|V(td)|} 
\frac{|V(ub)|}{V(ub)}\; \; , \; \; 
e^{i\phi _3} =  \frac{V(ub)}{|V(ub)|} 
\ee 
The various CP asymmetries in beauty decays are expressed in 
terms of these three angles. I will describe `typical' examples now. 

\subsubsection{Angle $\phi _1$}
Consider 
\be 
\bar B_d \ra \psi K_S \leftarrow B_d
\ee 
where the final state is an almost pure odd CP eigenstate. On the 
quark level one has two different reactions, namely one 
describing the direct decay process 
\be 
\bar B_d = [b \bar d] \ra [c\bar c] [s \bar d] 
\label{BDTREE}
\ee 
and the other one involving a $B_d - \bar B_d$ oscillation: 
\be 
\bar B_d = [b \bar d] \Rightarrow B_d = [\bar b d] 
\ra [c \bar c] [ \bar sd] 
\label{BDOSC} 
\ee 
\begin{center} 
$\spadesuit \; \; \; \spadesuit \; \; \; \spadesuit $ \\ 
{\em Homework Problem \# 5}: 
\end{center}
How can the $[s \bar d]$ combination in 
Eq.(\ref{BDTREE}) interfere with 
$[\bar sd]$ in Eq.(\ref{BDOSC})? 
\begin{center} 
$\spadesuit \; \; \; \spadesuit \; \; \; \spadesuit $
\end{center} 
Since the final state in $B/\bar B \ra \psi K_S$ can carry 
isospin 1/2 only, we have for the {\em direct} 
transition amplitudes: 
\be 
\begin{array}{c}
T(\bar B_d \ra \psi K_S) = V(cb)V^*(cs) 
e^{i\alpha _{1/2}} |{\cal M}_{1/2}| \\ 
T(B_d \ra \psi K_S) = V^*(cb)V(cs) 
e^{i\alpha _{1/2}} |{\cal M}_{1/2}| 
\end{array}
\ee 
and thus 
\be 
\bar \rho _{\psi K_S} = \frac{V(cb)V^*(cs)}{V^*(cb)V(cs)} 
\ee 
from which the hadronic quantities, namely the 
phase shift $\alpha _{1/2}$ and the hadronic matrix 
element $|{\cal M}_{1/2}|$ -- both of which {\em cannot} 
be calculated in a reliable manner -- have dropped out. 
Therefore 
\be 
\left| \bar \rho _{\psi K_S} \right| = 
\left|  
\frac{T(\bar B_d \ra \psi K_S}{T(B_d \ra \psi K_S}
\right|  = 1 \; ; 
\ee 
i.e., there can be {\em no direct} CP violation in this channel. 

Since $|\Gamma _{12}| \ll |M_{12}|$ one has 
\be 
\frac{q}{p} \simeq \sqrt{\frac{M^*_{12}}{M_{12}}} = 
\frac{M^*_{12}}{|M_{12}|} \simeq 
\frac{V^*(tb)V(td)}{V(tb)V^*(td)}
\ee  
which is a pure phase. Altogether one obtains 
\footnote{The next-to-last (approximate) equality 
in Eq.(\ref{SIN2PHI1}) holds in the Wolfenstein 
representation, although the overall result is general.}
\be 
{\rm Im} \frac{q}{p} \bar \rho _{\psi K_S} = 
{\rm Im} \left( 
\frac{V^*(tb)V(td)}{V(tb)V^*(td)} 
\frac{V(cb)V^*(cs)}{V^*(cb)V(cs)}  
 \right) 
\simeq {\rm Im} \frac{V^2(td)}{|V(td)|^2} = 
{\rm sin}2\phi _1 
\label{SIN2PHI1} 
\ee 
That means 
that to a very good approximation the observable 
Im $\frac{q}{p} \bar \rho _{\psi K_S}$, which is the amplitude 
of the oscillating CP asymmetry, is in general given by 
{\em microscopic} parameters of the theory; within 
the KM ansatz they combine to yield the angle $\phi _1$ 
\cite{BS}. 
Within the Wolfenstein representation one has 
\be 
{\rm Im} \frac{q}{p}\bar \rho _{\psi K_S} \simeq 
 \frac{2\eta (1- \rho )}{(1-\rho )^2 + \eta ^2} > 0  
\ee 
since the analysis of $K_L \ra \pi \pi$ yields 
$\eta >0$, $|\rho | < 1$, see 
Sect. \ref{THEPS}. The first pilot studies yield 
\cite{OPAL,CDF}: 
\be 
{\rm Im} \left( \frac{q}{p}\bar \rho _{\psi K_S}
\right) = 
\left\{ 
\begin{array}{ll} 
3.2{+1.8 \atop - 2.0}\pm 0.5 \; , &{\rm ~~OPAL \; 
Collaboration} 
\\ 
1.8 \pm 1.1 \pm 0.3 \; , &{\rm ~~ CDF \; 
Collaboration} 
\end{array} 
\right. 
\ee 

Several other channels are predicted to exhibit a CP asymmetry 
expressed by sin$2\phi _1$, like $B_d \ra \psi K_L$ 
\footnote{Keep in mind that 
Im$\frac{q}{p}\bar \rho _{\psi K_L} =- 
{\rm Im}\frac{q}{p}\bar \rho _{\psi K_S}$ holds because 
$K_L$ is mainly CP odd and $K_S$ mainly CP even.}, 
$B_d \ra D \bar D$ etc. 

\begin{center} 
$\spadesuit \; \; \; \spadesuit \; \; \; \spadesuit $ \\ 
{\em Homework Problem \# 6}: 
\end{center}
At first sight it would seem that Eq.(\ref{SIN2PHI1} 
cannot be correct since the quantity 
$\frac{V^*(tb)V(td)}{V(tb)V^*(td)} 
\frac{V(cb)V^*(cs)}{V^*(cb)V(cs)}$ is not invariant under 
changes in the phase conventions adopted for the $d$ and $s$ 
quark fields whereas observables like sin$2\phi _1$ have to be. 
There is a spurious factor not listed explicitely that 
takes care of this problem. Explain what it is. 
({\em Hint}: Remember Homework Problem \# 5.)
\begin{center} 
$\spadesuit \; \; \; \spadesuit \; \; \; \spadesuit $
\end{center} 

\subsubsection{Angle $\phi _2$}
The situation is not quite as clean for the angle $\phi _2$. 
The asymmetry in $\bar B_d \ra \pi ^+ \pi ^-$ vs. 
$B_d \ra \pi ^+ \pi ^-$ is certainly sensitive to $\phi _2$, 
yet there are two complications:  
\begin{itemize}
\item 
The final state is described by a superposition of {\em two} 
different isospin states, namely $I = 0$ and $2$. The 
spectator process contributes to both of them. 
\item 
The Cabibbo suppressed Penguin operator 
\be 
b \ra d g^* \ra d u \bar u 
\ee 
will also contribute, albeit only to the $I=0$ amplitude.  
\end{itemize} 

The direct transition amplitudes are then expressed as follows: 
$$  
T(\bar B_d \ra \pi ^+ \pi ^-) = 
V(ub)V^*(ud)e^{i \alpha _2}|{\cal M}_2^{spect}| + 
$$ 
\be 
+ e^{i \alpha _0}\left( V(ub)V^*(ud) |{\cal M}_0^{spect}| + 
V(tb)V^*(td) |{\cal M}_0^{Peng}| 
\right)  
\ee   
$$  
T(B_d \ra \pi ^+ \pi ^-) = 
V^*(ub)V(ud)e^{i \alpha _2}|{\cal M}_2^{spect}| + 
$$ 
\be 
+ e^{i \alpha _0}\left( V^*(ub)V(ud) |{\cal M}_0^{spect}| + 
V^*(tb)V(td) |{\cal M}_0^{Peng}| 
\right)  
\ee  
where the phase shifts for the $I=0,2$ states have been factored 
off. 

{\em If} there were no Penguin contributions, we would have 
\be 
{\rm Im} \frac{q}{p} \bar \rho _{\pi \pi} = 
{\rm Im} \frac{V(td)V^*(tb)V(ub)V^*(ud)}
{V^*(td)V(tb)V^*(ub)V(ud)} = 
- {\rm sin}2\phi _2 
\ee 
without direct CP violation -- $|\bar \rho _{\pi \pi }| =1$ --  
since the two isospin amplitudes still contain the same weak 
parameters. The Penguin contribution changes the picture in 
two basic ways: 
\begin{enumerate}
\item 
The CP asymmetry no longer depends on $\phi _2$ alone: 
$$  
{\rm Im} \frac{q}{p} \bar \rho _{\pi \pi } \simeq  
- {\rm sin} 2\phi _2  + \left|  \frac{V(td)}{V(ub)} \right|  
\left[  {\rm Im}\left( e^{-i\phi _2}
\frac{{\cal M}^{Peng}}{{\cal M}^{spect}}\right)  
- {\rm Im}\left( e^{-3i\phi _2}
\frac{{\cal M}^{Peng}}{{\cal M}^{spect}}\right) 
\right] + 
$$ 
\be 
+ {\cal O}(|{\cal M}^{Peng}|^2/|{\cal M}^{spect}|^2) 
\label{PENGPOLL} 
\ee 
where 
\be 
{\cal M}^{spect} = e^{i \alpha _0}|{\cal M}^{spect}_0|  + 
e^{i \alpha _2}|{\cal M}^{spect}_2| \; \; , \;  \; 
{\cal M}^{Peng} = e^{i \alpha _0}|{\cal M}^{Peng}_0| 
\ee 
\item 
A direct CP asymmetry emerges:
\be 
|\bar \rho _{\pi \pi}| \neq 1 
\ee 
\end{enumerate}
Since we are dealing with a Cabibbo suppressed Penguin operator, 
we expect that its contribution is reduced relative to the 
spectator term: 
\be 
\left| \frac{{\cal M}^{Peng}}{{\cal M}^{spect}}  
\right| 
< 1 \; , 
\ee 
which was already used in Eq.(\ref{PENGPOLL}). 
Unfortunately this reduction might 
not be very large. This concern is based on the observation 
that the branching ratio for $\bar B_d \ra K^- \pi ^+$ appears to 
be somewhat larger than for $\bar B_d \ra \pi ^+ \pi ^-$ 
implying that the Cabibbo favoured Penguin amplitude  
is at least not smaller than the spectator amplitude. 

Various strategies have been suggested to unfold the 
Penguin contribution through a combination of additional 
or other  
measurements (of other $B \ra \pi \pi $ channels 
or of $B\ra \pi \rho$, $B \ra K\pi$ etc.) and supplemented by 
theoretical considerations like $SU(3)_{Fl}$ symmetry 
\cite{PENGTRAP}.  
I am actually hopeful that the multitude of exclusive 
nonleptonic decays (which is the other side of the 
coin of small branching ratios!) can be harnessed to 
extract a wealth of information on the strong dynamics that 
in turn will enable us to extract 
sin$2\phi _2$ with decent accuracy. 

\subsubsection{The $\phi _3$ Saga}
Of course it is important to determine $\phi _3$ as accurately  
as possible. This will not be easy, and one better keep 
a proper perspective. I am going to tell this saga now in 
two installments. 

{\bf (I)} {\em CP asymmetries involving $B_s - \bar B_s$ 
Oscillations}: In principle one can extract $\phi _3$ from 
KM suppressed $B_s$ decays like one does $\phi _2$ from 
$B_d$ decays, namely by measuring and analyzing the difference 
between the rates for, say, $\bar B_s(t) \ra K_S \rho ^0$ and 
$B_s(t) \ra K_S \rho ^0$: 
Im$\frac{q}{p}\bar \rho _{K_S\rho ^0} \sim {\rm sin}2\phi _3$. 
One has to face the same complication, namely that in 
addition to the spectator term a 
(Cabibbo suppressed) Penguin amplitude contributes to $\bar \rho 
_{K_S\rho ^0}$ with different weak parameters. Yet the situation 
is much more challenging due to the rapid 
pace of the $B_s - \bar B_s$ oscillations. 

\noindent A more promising way might be to compare the rates for 
$\bar B_s (t) \ra D_s^+ K^-$ with $B_s (t) \ra D_s^- K^+$ as a function 
of the time of decay $t$ since there is no Penguin contribution. 
The asymmetry depends on sin$\phi _3$ rather than 
sin$2\phi _3$ 
\footnote{Both $D_s^+K^-$ and $D_s^- K^+$ are final states common to 
$B_s$ and $\bar B_s$ decays although they are not CP 
eigenstates.}.  

{\bf (II)} {\em Direct CP Asymmetries}: The largish direct CP 
asymmetries sketched above for $B \ra K \pi$ depend on 
sin$\phi _3$ -- and on the phase shifts which in general are neither  
known nor calculable. Yet in some cases they can be determined 
experimentally -- as first described for 
$B^{\pm} \ra D_{neutral} K^{\pm}$ 
\cite{WYLER}. There are 
{\em four independant} rates that can be measured, namely 
\be 
\Gamma (B^- \ra D^0 K^-) \; , \; 
\Gamma (B^- \ra \bar D^0 K^-) \;  , \;  
\Gamma (B^- \ra D_{\pm} K^-) \; , \; 
\Gamma (B^+ \ra D_{\pm} K^+) 
\ee  
The {\em flavour eigenstates} $D^0$ and $\bar D^0$ are defined 
through flavour specific modes, namely 
$D^0 \ra l^+ X$ and $\bar D^0 \ra l^- X$, respectively; 
$D_{\pm}$ denote the even/odd CP eigenstates 
$D_{\pm} = (D^0 \pm \bar D^0)/\sqrt{2}$ defined by 
$D_+ \ra K^+K^-, \, \pi ^+ \pi ^-,$ etc., 
$D_- \ra K_S\pi ^0, \, K_S \eta ,$ etc. \cite{PAISSB}. 

From these four observables one can (up to a binary ambiguity) 
extract the four basic quantities, namely the moduli of the two 
independant amplitudes ($|T(B^- \ra D^0 K^-)|$, 
$|T(B^- \ra \bar D^0 K^-)|$), their strong phaseshift -- and 
sin$\phi _3$, the goal of the enterprise! 

\subsubsection{A Zero-Background Search for New 
Physics: $B_s \ra \psi \phi , \, D_s^+ D_s^-$}
The two angles $\phi _1$ and $\phi _2$ will be measured in 
the next several years with decent or even good accuracy. I 
find it unlikely that any of the direct measurements 
of $\phi _3$ sketched above will yield a more precise 
value than inferred from simple trigonometry: 
\be 
\phi _3 = 180^o - \phi _1 - \phi _2 
\label{180}
\ee 
Eq.(\ref{180}) holds 
within the KM ansatz; of course the real goal is to uncover 
the intervention of New Physics in $B_s$ transitions. It then 
makes eminent sense to search for it in a reaction where 
Known Physics predicts a practically zero result. 
$B_s \ra \psi \phi , \, \psi \eta , \, D_s \bar D_s$ fit this bill 
\cite{BS}: 
to leading order in the KM parameters the CP asymmetry has 
to vanish since on that level quarks of the second and third 
family only participate in $B_s - \bar B_s$ oscillations -- 
$[s \bar b] \Rightarrow t^* \bar t^* \Rightarrow [b \bar s]$ -- and 
in these direct decays -- $[b \bar s] \ra c \bar c s \bar s$. Any 
CP asymmetry is therefore Cabibbo suppressed, i.e. $\leq 4$\% . 
More specifically 
\be 
\left. {\rm Im} \frac{q}{p}\bar \rho _{B_s \ra \psi \eta , 
\psi \phi , D_s \bar D_s} \right| _{KM} \sim 2\% 
\ee  
Yet New Physics has a good chance to contribute to 
$B_s - \bar B_s$ oscillations; if so, there is no reason for 
it to conserve CP and asymmetries can emerge that are easily 
well in excess of 2\% . New Physics scenarios with non-minimal 
SUSY or flavour-changing neutral currents could actually 
yield asymmetries of $\sim 10 \div 30 \%$ 
\cite{GABB} -- completely 
beyond the KM reach!

\subsubsection{A Menu for Gourmets}
Quite often people in the US tend to believe that a restaurant that 
presents them with a long menu must be a very good one. The 
real experts -- like the French and Italians -- of course know 
better: it is the hallmark of a top cuisine to concentrate on a 
few very special dishes and prepare them in a spectacular fashion 
rather than spread one's capabilities too thinly.There is a 
first class menu consisting of three main dishes and one side 
dish, namely 
\begin{enumerate} 
\item 
measure $\Delta m(B_s)$ which within the Standard Model 
allows to extract $|V(td)|$ through 
\be 
\frac{\Delta m(B_d)}{\Delta m(B_s)} \simeq 
\frac{Bf_{B_d}^2} {Bf_{B_s}^2} \left|  \frac{V(td)}{V(ts)} 
\right| ^2 \; ; 
\ee  

\item 
determine the rates for $\bar B_d \ra \psi K_S$ and 
$B_d \ra \psi K_S$ to obtain the value of 
sin$2\phi _1$; 

\item 
compare $\bar B_s \ra \psi \phi , \, D_s \bar D_s$ with 
$B_s \ra \psi \phi , \, D_s \bar D_s$ 
as a clean search for New Physics and 

\item 
as a side dish: measure the $B_s$ lifetime separately in 
$B_s \ra l \nu D_s^{(*)}$ and $B_s \ra \psi \phi , \, D_s 
\bar D_s$ where 
the former yields the algebraic average of the $B_{s, short}$ 
and $B_{s,long}$ lifetimes and the latter the $B_{s, short}$ 
lifetime. One predicts for them \cite{URALTSEV}:
\be 
\frac{\tau (B_s \ra l \nu D_s^{(*)}) - \tau (B_s \ra \psi \phi 
,\, D_s \bar D_s)}
{\tau (B_s \ra l \nu D_s^{(*)})} \simeq 0.1 \cdot 
\left(  \frac{f_{B_s}}{200\; {\rm MeV}} \right) ^2
\ee  

\end{enumerate} 
This menu featuring $B_s$ decays so prominently can be 
prepared at hadronic machines only. Thus it represents 
a task for HERA-B, CDF/D0 and later LHCB and BTeV. 

While the experiments have to exhibit the voracious 
appetite of a gourmand to gobble up 
enough statistics, they have to demonstrate the 
highly discriminating taste of a gourmet to 
succeed!

\subsection{KM Trigonometry}
One side of the triangle is exactly known since the base line can be 
normalized to unity without affecting the angles: 
\be 
1 - \frac{V(ub)}{\lambda V(cb)} + 
\frac{V^*(td)}{\lambda V(cb)} = 0 
\ee 
The second side is known to some degree from semileptonic $B$ 
decays: 
\be 
\left| 
\frac{V(ub)}{V(cb)} 
\right| 
\simeq 0.08 \pm 0.03
\ee
where the quoted uncertainty is mainly theoretical and amounts 
to little more than a guestimate. In the Wolfenstein representation 
this reads as 
\be 
\sqrt{\rho ^2 + \eta ^2} \simeq 0.38 \pm 0.11
\ee  
The area cannot vanish since $\epsilon _K \neq 0$. Yet at present 
not much more can be said for certain. 

In principle one would have enough observables -- namely 
$\epsilon _K$ and $\Delta m(B_d)$ in addition to 
$|V(ub)/V(cb)|$ -- to determine the two KM parameters 
$\rho$ and $\eta$ in a {\em redundant} way. In practise, though, 
there are two further unknowns, namely the size of the 
$\Delta S=2$ and $\Delta B=2$ matrix elements, as expressed through 
$B_K$ and $B_B f_B^2$. For $m_t$ sufficiently large $\epsilon _K$ 
is dominated by the top contribution:   
$d \bar s \Rightarrow  t^* \bar t^* \Rightarrow s \bar d$. The same holds 
always for $\Delta m(B_d)$. In that case things are simpler: 
\be 
\frac{|\epsilon _K|}{\Delta m(B_d)} \propto 
{\rm sin}2\phi _1 \simeq 
0.42 \cdot UNC 
\label{SIN2BETA} 
\ee 
with the factor $UNC$ parametrising the uncertainties 
\be 
UNC \simeq \left( \frac{0.04}{|V(cb)|}\right) 
\left( \frac{0.72}{x_d}\right) \cdot 
\left( \frac{\eta _{QCD}^{(B)}}{0.55}\right) \cdot 
\left( \frac{0.62}{\eta _{QCD}^{(K)}}\right) \cdot 
\left( \frac{2B_B}{3B_K}\right) \cdot 
\left( \frac{f_B}{160\, {\rm MeV}}\right) ^2  
\ee
where   
$x_d \equiv \Delta m(B_d)/\Gamma (B_d)$; 
$\eta _{QCD}^{(B)}$ and $\eta _{QCD}^{(K)}$ denote the 
QCD radiative corrections for ${\cal H}(\Delta B=2)$ and 
${\cal H}(\Delta S=2)$, respectively; $B_B$ and $B_K$ 
express the expectation value of ${\cal H}(\Delta B=2)$ or 
${\cal H}(\Delta S=2)$ in units of the `vacuum saturation' 
result which is given in terms of the decay constants 
$f_B$ and $f_K$ (where the latter is known). The main 
uncertainty is obviously of a theoretical nature related to 
the hadronic parameters $B_B$, $B_K$ and $f_B$; as discussed 
before, state-of-the-art theoretical technologies yield 
$B_B \simeq  1$, $B_K \simeq 0.8 \pm 0.2$ and 
$f_B \simeq 180 \pm 30 \, {\rm MeV}$ where the latter range 
might turn out to be anything but conservative! 
Eq.(\ref{SIN2BETA}) represents an explicit illustration that some CP 
asymmetries in $B^0$ decays are huge. 

For $m_t \simeq 180$ GeV the $c \bar c$ and $c\bar t + t \bar c$ 
contributions to $\epsilon _K$ are still sizeable; nevertheless 
Eq.(\ref{SIN2BETA}) provides a good approximation. Furthermore 
sin$2\phi _1$ can still be expressed reliably as a function of the 
hadronic matrix elements: 
\be 
{\rm sin}2\phi _1 = f(B_Bf_B^2/B_K) 
\ee  
It will become obvious why this is relevant. 

The general idea is, of course, to construct the triangle as accurately 
as possible and then probe it; i.e. search for inconsistencies that 
would signal the intervention of New Physics. A few remarks on that 
will have to suffice here. 

As indicated before we can expect the value of 
$|V(ub)/V(cb)|$ to be known to better than 10\% and hope for 
$|V(td)|$ to be determined with decent accuracy as well. 
The triangle 
will then be well determined or even overdetermined. Once the 
first asymmetry in $B$ decays that can be interpreted reliably -- 
say in $B_d \ra \psi K_S$ -- has been measured and $\phi _1$ 
been determined, the triangle is fully constructed from 
$B$ decays alone. Furthermore one has arrived at the first 
sensitive consistency check of the triangle: one  
compares the measured value of sin$2\phi _1$ with 
Eq.(\ref{SIN2BETA}) to infer which value of 
$B_Bf_B^2$ is thus required; this value is inserted into the 
Standard Model expression for $\Delta m(B_d)$ together 
with $m_t$ to see whether the experimental result is 
reproduced. 

A host of other tests can be performed that are highly sensitive to 
\begin{itemize}
\item 
the presence of New Physics and 
\item 
to some of their salient dynamical features. 
\end{itemize}
Details can be found in the ample literature on that subject.

\section{Oscillations and CP Violation in Charm Decays -- 
The Underdog's Chance for Fame}
\label{CHARM} 
It is certainly true that 
\begin{itemize}
\item 
$D^0-\bar D^0$ oscillations proceed very slowly in the 
Standard Model and 
\item 
CP asymmetries in $D$ decays are small or even tiny within 
the KM ansatz. 
\end{itemize}
Yet the relevant question quantitatively is: how slow and how small? 

\subsection{$D^0-\bar D^0$ Oscillations}
Bounds on $D^0 - \bar D^0$ oscillations are most cleanly   
expressed through `wrong-sign' semileptonic decays: 
\be 
r_D = \frac{\Gamma (D^0 \ra l^-X)}{\Gamma (D^0 \ra l^+X)} 
\simeq \frac{1}{2} \left( x_D^2 + y_D^2\right)  
\ee 
with $x_D = \Delta m_D/\Gamma _D$, 
$y_D = \Delta \Gamma _D/2\Gamma _D$. It is often stated that  
the Standard Model predicts 
\be 
r_D \leq 10^{-7} \; \; \hat = \; \; 
x_D, \, y_D \leq 3 \cdot 10^{-4} 
\ee 
I myself am somewhat flabbergasted by the boldness of such 
predictions. For one should keep the following in mind for 
proper perspective: there are quite a few channels that can drive 
$D^0 - \bar D^0$ 
oscillations -- like $D^0 \Rightarrow K\bar K , \; \pi \pi \Rightarrow  
\bar D^0$ or $D^0 \Rightarrow K^- \pi ^+ \Rightarrow 
\bar D^0$ -- and 
they branching ratios on the $({\rm few})\times 10^{-3}$ 
level 
\footnote{For the $K^- \pi ^+$ mode this represents the average of 
its Cabibbo allowed and doubly Cabibbo suppressed incarnations.}. 
In the limit of $SU(3)_{Fl}$ symmetry all these contributions have 
to cancel of course. Yet there are sizeable violations of $SU(3)_{Fl}$ 
invariance in $D$ decays, and one should have little confidence in an 
imperfect symmetry to ensure that a host of channels with branching 
ratios of order few$\times 10^{-3}$ will cancel as to render 
$x_D, \, y_D \leq 3\cdot 10^{-4}$. To say it differently: 
The relevant question in this context is {\em not} whether 
$r_D \sim 10^{-7} \div 10^{-6}$ is a possible or even reasonable 
Standard Model estimate, but whether 
$10^{-6} \leq r_D \leq 10^{-4}$ can {\em reliably be ruled out}! I 
cannot see how anyone could make such a claim with the 
required confidence. 

The present experimental bound is 
\be 
r_D |_{exp} \leq 3.4 \cdot 10^{-3} \; \; \hat = \; \; 
x_D, \, y_D \leq 0.1 
\ee 
to be compared with a {\em conservative} Standard Model bound 
\be 
r_D |_{SM} < 10^{-4} \; \; \hat = \; \; y_D, \, x_D|_{SM}
\leq 10^{-2} \; 
\ee 
New Physics on the other hand can enhance $\Delta m_D$ 
(though not 
$\Delta \Gamma _D$) very considerably up to 
\be 
x_D |_{NP} \sim 0.1 \; , 
\ee 
i.e. the present experimental bound. 

\subsection{CP Violation involving $D^0 - \bar D^0$ Oscillations}
One can discuss this topic in close qualitative analogy to 
$B$ decays. First one considers final states that are CP 
eigenstates like $K^+K^-$ or $\pi ^+ \pi ^-$ 
\cite{BSDDBAR}: 
$$  
{\rm rate}(D^0(t) \ra K^+ K^-) \propto e^{-\Gamma _D t} 
\left(  
1+ ({\rm sin}\Delta m_Dt) \cdot {\rm Im}\frac{q}{p}
\bar \rho _{K^+K^-}   
\right) 
\simeq 
$$
\be 
\simeq e^{-\Gamma _D t} 
\left(  
1+ \frac{\Delta m_Dt}{\Gamma _D}\cdot 
\frac{t}{\tau _D} 
\cdot {\rm Im}\frac{q}{p} \bar \rho _{K^+K^-}   
\right) 
\ee 
With $x_D|_{SM} \leq 10^{-2}$ and 
Im$\frac{q}{p} \bar \rho _{K^+K^-}|_{KM} \sim {\cal O}(10^{-3})$ one 
arrives at an asymmetry of around $10^{-5}$, i.e. for all practical 
purposes zero, since it presents the product of two 
very small numbers. 
Yet with New Physics one conceivably has $x_D|_{NP} \leq 0.1$, 
Im$\frac{q}{p} \bar \rho _{K^+K^-}|_{NP} \sim {\cal O}(10^{-1})$ 
leading to an asymmetry that could be as large as of order 1\%. 
Likewise one should compare the doubly Cabibbo suppressed 
transitions \cite{BIGIBERK,NIR}
$$ 
{\rm rate}(D^0(t) \ra K^+ \pi ^-) \propto 
e^{-\Gamma _{D^0} t} {\rm tg}^4\theta _C|\hat \rho _{K\pi }|^2 \cdot 
$$ 
$$ 
\cdot \left[ 1 - \frac{1}{2}\Delta \Gamma _D t + 
\frac{(\Delta m_Dt)^2}{4{\rm tg}^4\theta _C|\hat \rho _{K\pi }|^2} 
+ \frac{\Delta \Gamma _Dt}
{2{\rm tg}^2\theta _C|\hat \rho _{K\pi }|}
{\rm Re}\left( 
\frac{p}{q}\frac{\hat \rho _{K\pi }}{|\hat \rho _{K\pi }|}   
\right)  
- \right.  
$$ 
\be
\left. - \frac{\Delta m_Dt}{{\rm tg}^2\theta _C|\hat \rho _{K\pi }|} 
{\rm Im}\left( 
\frac{p}{q}\frac{\hat \rho _{K\pi }}{|\hat \rho _{K\pi }|}   
\right)  
\right] 
\ee 
$$ 
{\rm rate}(\bar D^0(t) \ra K^- \pi ^+) \propto 
e^{-\Gamma _{D^0} t} {\rm tg}^4\theta _C|
\hat{\bar \rho }_{K\pi }|^2 \cdot 
$$ 
$$ 
\cdot \left[ 1 - \frac{1}{2}\Delta \Gamma _D t + 
\frac{(\Delta m_Dt)^2}{4{\rm tg}^4\theta _C
|\hat{\bar \rho}_{K\pi }|^2} 
+ \frac{\Delta \Gamma _Dt}
{2{\rm tg}^2\theta _C|\hat{\bar \rho }_{K\pi }|}
{\rm Re}\left( 
\frac{p}{q}\frac{\hat{\bar \rho }_{K\pi }}
{|\hat{\bar \rho }_{K\pi }|}   
\right)  
+\right. 
$$ 
\be
\left. + \frac{\Delta m_Dt}{{\rm tg}^2\theta _C
|\hat{\bar \rho }_{K\pi }|} 
{\rm Im}\left( 
\frac{p}{q}\frac{\hat{\bar \rho }_{K\pi }}
{|\hat{\bar \rho }_{K\pi }|}   
\right)  
\right] 
\ee
where 
\be 
{\rm tg}^2\theta _C \cdot \hat \rho _{K\pi } \equiv 
\frac{T(D^0 \ra K^+ \pi ^-)}{T(D^0 \ra K^- \pi ^+)}\; \; , \; \; 
{\rm tg}^2\theta _C \cdot \hat{\bar \rho }_{K\pi } \equiv 
\frac{T(\bar D^0 \ra K^- \pi ^+)}{T(\bar D^0 \ra K^+ \pi ^-)}\; ; 
\ee  
in such New Physics scenarios 
one would expect a considerably enhanced asymmetry 
of order $1\%/{\rm tg}^2 \theta _C \sim 20\%$ -- at the cost 
of  smaller statistics. 

Effects of that size would unequivocally signal the intervention 
of New Physics!

\subsection{Direct CP Violation}
As explained before a direct CP asymmetry requires the presence 
of two coherent amplitudes with different weak and different 
strong phases. Within the Standard Model (and the 
KM ansatz) such effects can occur in Cabibbo suppressed 
\footnote{The effect could well reach the $10^{-3}$ 
and exceptionally the $10^{-2}$ level.}, yet not 
in Cabibbo allowed or doubly Cabibbo suppressed modes. There 
is a subtlety involved in this statement. Consider for example  
$D^+ \ra K_S \pi ^+$. At first sight it appears to be 
a Cabibbo allowed mode described by a single amplitude without 
the possibility of an asymmetry. However \cite{YAMAMOTO}  
\begin{itemize} 
\item 
due to $K^0 - \bar K^0$ mixing the final state 
can be reached also through a doubly Cabibbo suppressed 
reaction, and the two amplitudes necessarily interfere; 
\item 
because of the CP violation in the $K^0 - \bar K^0$ complex 
there is an asymmetry that can be predicted on general grounds 
$$  
\frac{\Gamma (D^+ \ra K_S \pi ^+) - \Gamma (D^- \ra K_S \pi ^-)} 
{\Gamma (D^+ \ra K_S \pi ^+) + \Gamma (D^- \ra K_S \pi ^-)} 
\simeq - 2 {\rm Re}\, \epsilon _K  \simeq - 3.3 \cdot 10^{-3} 
\simeq 
$$  
\be 
\simeq 
\frac{\Gamma (D^+ \ra K_L \pi ^+) - \Gamma (D^- \ra K_L \pi ^-)} 
{\Gamma (D^+ \ra K_L \pi ^+) + \Gamma (D^- \ra K_L \pi ^-)} 
 \; ; 
\label{DIRECTCPEPS}
\ee   
\item
 If New Physics contributes to the doubly Cabibbo suppressed 
amplitude $D^+ \ra K^0 \pi ^+$ (or $D^- \ra \bar K^0 \pi ^-$) then 
an asymmetry could occur quite conceivably on the few percent 
scale; 
\item 
such a manifestation of New Physics would be unequivocal; against 
the impact of $\epsilon _K$, Eq.(\ref{DIRECTCPEPS}) it 
could be distinguished not only through the size of the asymmetry, 
but also how it surfaces in $D^+ \ra K_L \pi ^+$ vs. 
$D^- \ra K_L \pi ^-$: if it is New Physics one has 
$$  
\frac{\Gamma (D^+ \ra K_S \pi ^+) - \Gamma (D^- \ra K_S \pi ^-)} 
{\Gamma (D^+ \ra K_S \pi ^+) + \Gamma (D^- \ra K_S \pi ^-)} 
= 
$$ 
\be = - 
\frac{\Gamma (D^+ \ra K_L \pi ^+) - \Gamma (D^- \ra K_L \pi ^-)} 
{\Gamma (D^+ \ra K_L \pi ^+) + \Gamma (D^- \ra K_L \pi ^-)} 
\ee 
i.e.,  the CP asymmetries in $D \ra K_S \pi$ and $D \ra K_L \pi$ 
differ in sign -- in contrast to Eq.(\ref{DIRECTCPEPS}). 
\end{itemize} 

\section{Heavy Quark Expansions (HQE)}
\label{HQE}
\subsection{Overview}

One other intriguing and gratifying aspect of heavy flavour 
decays has become understood just over the last several 
years. It concerns primarily the strong interactions rather 
than the weak interactions: the decays in particular of beauty 
{\em hadrons}  
can be treated with a reliability and accuracy that before would 
have seemed unattainable. These new theoretical 
technologies can be referred to as {\em Heavy Quark Theory};    
it combines two basic elements, namely an asymptotic 
symmetry principle and a dynamical treatment 
telling us how the asymptotic limit is approached: 
\begin{itemize}
\item 
The symmetry principle is {\em Heavy Quark Symmetry} 
stating that all sufficiently heavy quarks behave identically 
under the strong interactions. Its origin can be understood 
in an intuitive way: consider a hadron $H_Q$ containing a 
heavy quark $Q$ with mass $m_Q \gg \Lambda _{QCD}$ surrounded 
by a "cloud" of light degrees of freedom carrying quantum 
numbers of an antiquark $\bar q$ or diquark $qq$ 
\footnote{This cloud is often referred to -- 
somewhat disrespectfully -- as `brown muck'.}. 
This cloud has 
a rather complex structure: in addition to $\bar q$ 
(for mesons) or $qq$ (for baryons) it 
contains an indefinite number of $q \bar q$ pairs and gluons 
that are strongly coupled to and constantly 
fluctuate into each other. There is, however, 
one thing we know: since typical frequencies 
of these fluctuations are $\sim {\cal O}({\rm few} \times 
\Lambda _{QCD}$, the normally dominant {\em soft} dynamics 
allow the heavy quark to exchange momenta of order few times 
$\Lambda _{QCD}$ only with its surrounding medium. 
$Q \bar Q$ pairs then cannot play a significant role, 
and the heavy quark can be treated as a quantum mechanical 
object rather than a field theoretic entity 
requiring second quantization. This provides a 
tremendous computational simplification even while 
maintaining a field theoretic description for the light 
degrees of freedom. Furthermore techniques developed 
long ago in QED can profitably be adapted here. 
\item 
We can go further and describe the interactions between 
$Q$ and its surrounding light degrees of freedom through 
an expansion in powers of $1/m_Q$. This allows us to 
analyze {\em pre}-asymptotic effects, i.e. effects 
that fade away like a power of $1/m_Q$ as 
$m_Q \ra \infty$.
\end{itemize}   
This situation is qualitatively similar 
to chiral considerations which start from the limit of chiral 
invariance and describe the deviations from it through 
chiral perturbation theory. In both cases one has succeeded in 
describing nonperturbative dynamics in special cases. 

The lessons we have learnt can be summarized as follows 
\cite{URIVAR,HQEREV}: we have 
\begin{itemize}
\item 
identified the sources of the non-perturbative corrections;  
\item 
found them to be smaller than they could have been; 
\item 
succeeded in relating the basic quantities of the Heavy Quark 
Theory -- KM paramters, masses and kinetic energy of 
heavy quarks, 
etc. -- to various a priori independant observables with a fair 
amount of  redundancy; 
\item 
developed a better understanding of incorporating perturbative 
and nonperturbative corrections without double-counting.  
\end{itemize} 
In the following I will sketch the concepts 
on which the Heavy Quark Expansions are based, the 
techniques employed, the results obtained and 
the problems encountered. It will not constitute 
a self-sufficient introduction into this vast 
and ever expanding field. My intent is to provide 
a vademecum that 
\begin{itemize}
\item 
creates an interest in the uninitiated for further 
reading and can serve as a guide for such a 
journey or 
\item 
refreshes the memory of readers who have heard it before 
while also pointing out the present frontline. 

\end{itemize}

\subsection{Theoretical Tools and Concepts}

In describing weak decays of heavy flavour {\em hadrons} 
one has to incorporate perturbative as well as 
nonperturbative contributions in a self-consistent 
and complete  
way. The only known way to 
tackle such a task invokes the 
{\em Operator Product Expansion a la Wilson} 
involving an {\em effective} Lagrangian. Further 
conceptual insights as well as practical results can be 
gained by analysing {\em sum rules}; in particular they 
shed light on various aspects and formulations of 
{\em quark-hadron duality}. 

\subsubsection{Effective Lagrangians}

The Standard Model defines the interaction driving 
beauty decays at an ultraviolet scale at or above 
the $W$ boson mass $M_W$ in a world with quarks, gluons 
and weak bosons. Yet in describing the decays of 
{\em hadrons}, we have to evaluate matrix elements 
of transition operators at ordinary hadronic scales.  
Those are much lower, namely 
around 1 GeV or so, at which point the weak bosons and 
heavy quarks have long ceased  
to represent dynamical entities. Such changes arise 
naturally through {\em effective} Lagrangians. 
In a quantum field theory one has to  
define  a 
(renormalizable) Lagrangian 
in terms of certain fields at an ultraviolet scale $M_{UV}$. 
When considering this Lagrangian at a lower 
energy scale $\mu$, 
all modes with characteristic frequencies 
{\em above} $\mu$ have to be `integrated out' leaving 
only modes with {\em lower} frequencies as 
quantum {\em fields}. Yet the heavy degrees of freedom 
leave their mark in two ways: 
\begin{itemize}
\item 
They induce new nonrenormalizable couplings among 
the light fields which scale like 
$(\mu /M_{UV})^{d-4}$ with $d$ denoting the dimension 
of the new interaction operator. In particular 
removing the $W$ boson fields gives rise to 
dimension six current-current operators. 
\item 
They affect the coefficients of the operators in the 
emerging Lagrangian. For example integrating out 
$t$, $b$, ... quark fields generates an imaginary 
part in the coefficients of the 
$\Delta B=1,2$ (and $\Delta S=1,2$) operators. 
\end{itemize} 
The scale $\mu$ separates short and long distance 
dynamics 
\be 
{\rm short \; distance} \; \; < \; \; 
\mu ^{-1} \; \; < \; \; 
{\rm long \; distance} 
\ee 
with the former entering through the coefficients and 
the latter through the effective operators; their 
matrix elements will thus depend on $\mu$. 

{\em In principle} the value of $\mu$ does not 
matter: it reflects merely our computational procedure 
rather than how nature goes about its business. The 
$\mu$ dependance of the coefficients thus has to cancel 
against that of the corresponding matrix elements. 

{\em In practise} however there are competing 
demands on the choice of $\mu$: 
\begin{itemize}
\item 
On one hand one has to choose 
\be 
\mu \gg \Lambda _{QCD} \; ; 
\label{MULARGE} 
\ee 
otherwise radiative corrections cannot be treated 
within {\em perturbative} QCD. 
\item 
On the other hand many computational techniques 
for evaluating {\em matrix elements}  
-- among them the Heavy Quark Expansions -- 
require 
\be 
\mu \ll m_b 
\label{MUSMALL} 
\ee 
\end{itemize} 

All of this has been well known for a long 
time, of course -- in principle. Yet various 
subtleties that usually had been ignored 
became quite relevant when deriving effective 
Lagrangians for QCD itself with heavy quarks: 
\be 
{\cal L}_{QCD} = {\cal L}_{QCD}^{light} +  
{\cal L}^{heavy}
\ee 
\be 
{\cal L}_{QCD}^{light}= 
- \frac{1}{4} G^a_{\mu \nu} G^a_{\mu \nu} 
+ \sum _q \bar q i \not D q \; \; , \; 
{\cal L}^{heavy} = 
\sum _Q \bar Q (i \not D-m_Q)Q \; . 
\ee
$G^a_{\mu \nu}$ denotes the gluon field-strength 
tensor, $D_{\mu}$ the covariant 
derivative, $q$ the light quark fields $u$, $d$ and 
$s$ (for simplicity assumed to be massless) and 
$Q$ the heavy quark fields, certainly $b$ and 
possibly $c$. The heavy quark Lagrangian 
can be expanded in powers of $1/m_Q$: 
\be 
{\cal L}^{heavy} = \sum _Q 
\left[ \bar Q (i \not D-m_Q)Q + 
\frac{c_G}{2m_Q}\bar Q \frac{i}{2} \sigma \cdot G Q +  
\sum _{q,\Gamma}\frac{d_{Qq}^{(\Gamma )}}{m_Q^2} 
\bar Q \Gamma Q \bar q \Gamma q \right] 
+{\cal O}\left( 1/m_Q^3 \right) 
\label{LHEAVYEXP} 
\ee 
where $c_G$ and $d_{Qq}^{(\Gamma )}$ are coefficient 
functions, the $\Gamma$ denote the possible 
Lorentz covariant fermion bilinears and 
$\sigma \cdot G = \sigma _{\mu \nu} 
G_{\mu \nu}$ with 
$G_{\mu \nu} = g t^a G^a_{\mu \nu}$. Thus a
dimension five operator arises -- usually referred 
to as {\em chromomagnetic} operator -- and 
various dimension six four-fermion operators. In 
expanding expectation values later we will encounter 
also the so-called kinetic energy operator of 
dimension five 
\be 
O_{kin} = \bar Q \vec \pi ^2 Q \; \; , \; 
\vec \pi = - i \vec D \; ; 
\ee   
since it is not Lorentz invariant, it cannot appear 
in the Lagrangian.

\subsubsection{Operator Product Expansion (OPE) 
for Inclusive Weak Decays}

Similar to the well-known case of 
$\sigma (e^+ e^- \to had)$ one invokes the optical 
theorem to describe the decay into a sufficiently 
{\em inclusive} final state $f$ through the 
imaginary part of the forward scattering operator 
evaluated to second order in the weak interactions  
\be 
\hat T( Q \to Q) = 
{\rm Im}  \int d^4x 
\, i\{ {\cal L}_W(x) {\cal L}_W^{\dagger}(0) \} _T
\label{TRANSOP}
\ee 
with the subscript $T$ denoting the time-ordered 
product and ${\cal L}_W$ the relevant weak Lagrangian  
\footnote{There are two qualitative differences to the 
case of $e^+ e^- \to had$: in describing weak decays 
of a hadron $H_Q$ (i) one employs the weak rather than the 
electromagnetic Lagrangian, and (ii) one takes the 
expectation value between the $H_Q$ state rather than the 
vacuum.}.   
The expression in Eq.(\ref{TRANSOP}) represents in general 
a non-local operator with the space-time separation 
$x$ being fixed by the inverse of the energy release. If 
the latter is large compared to typical hadronic 
scales, then the product is dominated by short-distance 
physics, and one can apply an operator 
product expansion a la Wilson on it yielding an infinite 
series of {\em local} operators of increasing dimension 
\footnote{
I will formulate the expansion in powers of 
$1/m_Q$, although it has to be kept in mind that it is 
really controlled by the inverse of the energy release.  
While there is no fundamental difference between 
the two for $b \to c/u l \bar \nu$ or 
$b \to c/u \bar ud$, since $m_b, \, 
m_b - m_{c,u} \gg  
\Lambda _{QCD}$, the expansion becomes of 
somewhat dubious reliability for $b \to c \bar cs$. 
It actually would break down for a scenario 
$Q_2 \to Q_1 l \bar \nu$ with $m_{Q_2} \simeq 
m_{Q_1}$ -- in contrast to HQET! 
}. 
The width for the decay of a hadron $H_Q$ containing 
$Q$ is then obtained by taking the $H_Q$ expectation value 
of the operator $\hat T$:  
$$ 
\frac{\matel{H_Q}{{\rm Im}\hat T(Q \to f \to Q)}{H_Q}}
{2M_{H_Q}} \propto 
\Gamma (H_Q \to f) = \frac{G_F^2m_Q^5}{192 \pi ^3}
|V_{CKM}|^2 \cdot 
$$
$$
\cdot \left[ c_3^{(f)}(\mu )
\frac{\matel{H_Q}{\bar QQ}{H_Q}_{(\mu )}}{2M_{H_Q}} + 
\frac{c_5^{(f)}(\mu )}{m_Q^2}
\frac{\matel{H_Q}{\bar Q\frac{i}{2}\sigma \cdot GQ}
{H_Q}_{(\mu )}}{2M_{H_Q}} + \right. 
$$
\be
\left. + \sum _i \frac{c_{6,i}^{(f)}(\mu )}{m_Q^3} 
\cdot \frac{
\matel{H_Q}{(\bar Q\Gamma _iq)(\bar q\Gamma _iQ)}
{H_Q}_{(\mu )}}{2M_{H_Q}} + {\cal O}(1/m_Q^4) 
\right] 
\label{MASTER}
\ee  
Eq.(\ref{MASTER}) exhibits the following important features:
\begin{itemize}
\item 
As stated before in general terms, 
{\em short-distance} dynamics 
shape the c number coefficients $c_i^{(f)}$. 
{\em In practise} they are evaluated in 
{\em perturbative} QCD. It is 
quite conceivable, though, that also {\em nonperturbative}  
contributions arise there; yet they are believed to be 
fairly small in beauty decays \cite{CHIBISOV}. 
\item  
Nonperturbative contributions on 
the other hand enter through the {\em expectation values} 
of operators of dimension higher than three -- 
$\bar Q\frac{i}{2}\sigma \cdot GQ$ etc. -- and higher order 
corrections to the expectation 
value of the leading operator   
$\bar QQ$, see next. 
\item 
I will describe later what we know about 
these expectation values. Let me just anticipate 
that $\matel{H_Q}{\bar QQ}{H_Q}_{(\mu )}/2M_{H_Q} 
= 1 + {\cal O}(1/m_Q^2)$. One then realize that 
the free quark model expression emerges 
asymptotically for the 
total width, i.e. for $m_Q \to \infty$. 
\item 
These nonperturbative contributions which are 
power suppressed can be described only if considerable 
care is applied in treating the {\em parametrically 
larger} perturbative corrections. 
\item 
The {\em leading} nonperturbative corrections 
arise at order $1/m_Q^2$ only. That means they are 
rather small in beauty decays since 
$(\mu /m_Q)^2 \sim $ few \% for $\mu \leq 1$ GeV. 
\item 
Explicitely flavour dependant effects arise in order 
$1/m_Q^3$. They mainly drive the differences in the 
lifetimes of the various mesons of a given heavy 
flavour.
\end{itemize} 
The absence of corrections of order 
$1/m_Q$ is particularly noteworthy and intriguing since 
such corrections do exist for hadronic masses --  
$M_{H_Q} = m_Q (1+ \bar \Lambda/m_Q + 
{\cal O}(1/m_Q^2) )$ -- and those control 
the phase space. Technically this follows from the 
fact that there is no {\em independant} 
dimension-four operator that could emerge in the OPE 
\footnote{The operator $\bar Q i \not D Q$ can be reduced 
to the leading operator $\bar QQ$ through the equation 
of motion.}. This result can be illuminated in more 
physical terms as follows. Bound-state effects in the 
initial state like mass shifts do generate corrections 
of order $1/m_Q$ to the total width; yet so does 
hadronization in the final state. {\em Local} 
colour symmetry demands that those effects cancel 
each other out. {\em It has to be emphasized that the 
absence of corrections linear in $1/m_Q$ is an 
unambiguous consequence of the OPE description.} 
{\em If} their presence were forced upon us, we would have 
encountered a {\em qualitative} change in our QCD paradigm. 
A discussion of this point has arisen recently phrased 
in the terminology of quark-hadron duality. I will return 
to this point later.

\subsubsection{Sum Rules}

Semileptonic decays of $H_Q$ hadrons can be viewed as the 
crossed version of the deep-inelastic scattering of leptons 
off an $H_Q$ target. Pursuing this well-known analogy 
one expresses the differential decay rate through a 
leptonic tensor $L_{\mu \nu }$ contracted with a 
hadronic tensor $W_{\mu \nu}$; the latter is decomposed 
into five Lorentz covariants with Lorentz invariant 
structure functions $w_i(q_0, q^2)$, where $q_0$ and 
$\sqrt{q^2}$ denote the energy and invariant mass of the 
lepton pair. Only two of those contribute, usually 
labelled $w_1$ and $w_2$, when lepton masses are neglected,  
and the semileptonic width is given by 
\cite{CGG,KOYRAKH} 
\be 
\Gamma _{SL} \equiv  \Gamma (B \to l X_c) = 
\frac{G_F^2}{8\pi ^3} \cdot |V(cb)|^2 \cdot \gamma 
\ee 
\be  
\gamma = \frac{1}{2\pi}\int _0^{q_{max}^2} 
dq^2 \int _{\sqrt{q^2}}^{q_{0,max}} dq_0 
\sqrt{q_0^2 - q^2} 
\left[ 
q^2 w_1(q_0, q^2) + \frac{q_0^2 - q^2}{3}  
w_2(q_0, q^2)
\right]  
\ee
where 
\be 
q^2_{max} = (M_B - M_D)^2 \; , \; \; 
q_{0,max} = 
\frac{M_B^2 + q^2 - M_D^2}{2M_B} 
\ee 

In deep inelastic lepton-nucleon scattering QCD allows us to 
make two types of predictions: (i) If one manages 
to know -- say from the data -- what a given 
{\em moment} of a 
structure function is at a certain scale $q_0^2$, then 
one can predict what it should be at a higher scale 
$q^2$. (ii) Forming linear combinations of some moments 
of structure functions will project out the expectation 
value of a certain operator in the OPE. 
Symmetry considerations 
tell us the value of such matrix elements in certain instances,  
in which case we can predict the absolute magnitude of 
such moments. This is referred to as a {\em sum rule} 
like the Gross-Llewelyn-Smith sum rule or the Adler 
sum rule etc. 

Both types of predictions can be made 
concerning heavy flavour decays as well. Information 
obtained on charm decays can be extrapolated to beauty 
decays to the degree that heavy quark expansions apply already 
at the charm scale 
(a proposition on which reasonable people can 
disagree). More powerful results can be deduced 
from the sum rules approach, in particular since the 
heavy quark symmetry yields much information 
about expectation values of various operators between 
$H_Q$ states. The basic idea is the same underlying 
the QCD sum rules: one equates the integral over 
a transition rate evaluated on the quark-gluon level 
with the corresponding quantity {\em parametrized} in 
terms of hadronic quantities.

Some nontrivial results can be 
stated {\em without} actually evaluating moments. 
Let me illustrate this by citing three examples: 
\begin{itemize}
\item 
Considering a semileptonic transition driven by the 
pseudoscalar weak current 
$J_5 = \int d^3x \bar \{ c i \gamma _5 b\} (x)$ one can 
deduce a sum rule at `zero recoil' 
($\vec q =0$) for a structure function $w^{(5)}$:
\be 
\frac{1}{2\pi} \int _0^{\mu } d\epsilon 
w^{(5)}(\epsilon ) = \left(
\frac{1}{2m_c} - \frac{1}{2m_b}
\right) ^2 
\left( 
\mu ^2_{\pi }(\mu ) - \mu ^2_G(\mu )  
\right)  
\label{SR1}
\ee
where $\epsilon$ denotes the excitation energy above 
threshold, and $\mu ^2_{\pi }(\mu )$ and 
$\mu ^2_G(\mu )$ the expectation values of the kinetic and 
chromomagetic operator, respectively: 
\be 
\mu ^2_{\pi }(\mu ) \equiv \frac{1}{2M_{H_Q}} 
\matel{H_Q}{\bar Q \vec \pi ^2 Q}{H_Q}_{(\mu )} 
\; , \; 
\mu ^2_G (\mu ) \equiv \frac{1}{2M_{H_Q}} 
\matel{H_Q}{\bar Q \frac{i}{2} 
\sigma \cdot G Q}{H_Q}_{(\mu )}
\ee 
With structure functions being nonnegative one deduces 
\be 
\mu ^2_{\pi }(\mu ) \geq \mu ^2_G(\mu )
\label{INEQ}
\ee 
with the normalization point $\mu$ provided by the 
cut-off in the integral on the left-hand side of 
Eq.(\ref{SR1}). Eq.(\ref{INEQ}) forms an important element 
in several applications of the HQE. 
\item 
From a zero-recoil sum rule for the semileptonic 
transitions produced by the axial vector 
current $\bar c \gamma _i \gamma _5 b$ one infers 
in an analogous fashion a constraint on 
$|F_{D^*}(0)|$, the form factor for 
$B \to l  \nu D^*$ at zero recoil. The result will be 
used later. 
\item 
Sum rules shed a considerable amount of light on the 
workings of quark-hadron duality to be discussed next. 
\end{itemize}

\subsubsection{Quark-Hadron Duality}

The general notion that a quark-level based description 
should provide an approximation for a sufficiently 
inclusive hadronic transition goes back to the early days 
of the quark model, and it received a new impetus from 
the parton model. It did not represent a well-defined 
concept, though, and it was applied by different 
authors with considerable latitude. In a 1976 paper 
\cite{POGGIO} 
Poggio et al. presented a more explicit discussion 
of how duality should work around the charm threshold in 
$e^+ e^- \to had$: perturbative QCD allows to 
evaluate the total cross section in the 
Euclidean rather than the Minkowskian 
domain; the result is related to the 
observable cross section through a dispersion relation. 
This means that in general the physical cross 
section can be predicted only averaged 
-- or `smeared' -- over some  
energy interval rather than energy-by-energy; it was 
estimated that this interval had to amount to 0.5 - 1 
GeV. On the other hand, far away from any threshold 
the cross section would be a sufficiently smooth 
function such that smearing was no longer required and 
{\em local} duality applies. 

Another milestone was reached in the mid 1980's 
when Shifman and Voloshin \cite{SV1} first realized 
that under 
the special circumstance of heavy quark symmetry 
quark-hadron duality applies approximately even if 
only two channels dominate a transition -- as 
it emerges for semileptonic $B$ meson decays. This 
program was then pursued further and completed by 
Isgur and Wise \cite{WISGUR} (see also \cite{CGG}). 

Heavy quark expansions have not solved the dynamical 
problems underlying quark-hadron duality, yet 
they have helped us in transforming it into a 
well-defined concept and have provided us with 
novel insights into its inner 
workings -- at least for the 
dynamics of heavy flavours. In \cite{OPTICAL} 
the concept of {\em global} 
duality has been introduced: it states that if 
QCD is to represent the theory of the strong 
interactions, the heavy flavour transition rate 
calculated in the Euclidean region 
-- namely for the energy of the lepton pair $q_0$ purely 
imaginary -- has to equal 
the dispersion integrals over the physical 
heavy flavour transition rates; for otherwise QCD 
had to generate a singularity in, say, the complex 
energy plane for which there were no physical 
counterpart!  

This is however not the end of the 
story. For the dispersion integral extends over 
{\em all} physical heavy flavour transitions: decay 
processes as well as scattering and production 
processes! One has to argue then that the dispersion 
integral over the decay region alone is largely 
insensitive to the other physical singularities; 
those are referred to as the `distant' cuts. 
This shows that duality in general cannot be more than an 
approximation the quality of which can depend 
on the specifics of the transition under study.   

One has to keep in mind also that the OPE does not 
yield an expression that is convergent or even Borel 
summable. Evaluating the power expansion in the 
Euclidean domain does not enable us to determine 
those contributions that turn into {\em oscillating} 
terms in Minkowski space and introduce duality violations. 

Soluble model field theories that exhibit quark 
confinement can be harnessed 
to obtain intriguing and non-trivial insights on the 
realization of duality and its limitations. This holds 
in particular for the t'Hooft model which is 
the $N_C \to \infty$ limit of 1+1 dimensional QCD. 
Based on {\em numerical} studies it has been 
claimed that at least in this model local duality is 
violated in the integrated widths of `nonleptonic' 
decays (i) through the emergence 
of a $1/m_Q$ term {\em absent} in the OPE 
\cite{GL1} and 
(ii) even more massively in explicitely flavour dependant 
processes like Weak Annihilation \cite{GL2}. 

However it has been shown by us \cite{THOOFT1,THOOFT2} 
through a careful {\em analytical} analysis that 
\begin{itemize}
\item 
these claims are erroneous and 
\item 
their analysis flawed. 
\end{itemize} 
Since the spectrum of states in the t'Hooft model is 
known, one can calculate the inclusive width through the 
OPE on one hand and through a sum over the spectrum 
of the model on the other and compare the results. 
They are fully consistent with each other, i.e. 
{\em local duality does indeed hold in this model field 
theory to a high degree of accuracy and even the 
form and strength of duality violations can be 
estimated!} 

The widths of weak decays sensitively depend on the 
phase space which is controlled by the masses of the 
states involved. It actually depends of a high power 
of these masses as exemplified by the simple 
parton model result 
\bea 
\Gamma (b \to c l \bar \nu ) &\propto& 
G_F^2 m_b^5 f(x_c) \; , \; \; 
x_c = \frac{m_c^2}{m_b^2} \\
f(x)&=& 1 - 8x +8 x^3 - x^2 -12 x^2 {\rm log} x 
\label{FIVE} 
\eea 
It would then seem highly unlikely if a decay width 
given by {\em hadronic} masses could be approximated 
by a width expressed in terms of {\em quark} masses. 
Yet that is exactly what happens, and this apparent 
miracle is driven by the sum rules. 

Let us consider 
the simple case of $b \to u l \bar \nu$ with 
$m_u =0$ in a framework where the power of the 
$b$ quark mass in Eq.(\ref{FIVE}) has been made a 
free parameter $n$ in which one can expand transition 
rates \cite{FIVEINFINITY}: 
\be 
\Gamma (B \to l X_u) \propto |V(ub)|^2 G_F^2 m_b^5 
\Longrightarrow 
\Gamma _n(B \to l X_u) \propto |V(ub)|^2 G_F^2 m_b^n 
\ee 
This can be achieved, 
for example, by having 
$l$ extra scalar "leptons" emitted at the weak vertex; 
expressing the semileptonic width in terms of the 
structure functions then reads as follows: 
\be 
\Gamma _n (B \to l X_u) = |V(ub)|^2 
\frac{G_F^2}{8 \pi ^3} \cdot \gamma _n 
\ee 
\be 
\gamma _n = \frac{1}{2\pi} \int _0^{q^2_{max}} 
dq^2 (q^2)^l \int _{\sqrt{q^2}}^{q_{0,max}} 
dq_0 \sqrt{q_0^2 - q^2} 
\left[ 
q^2 w_1(q_0,q^2) + \frac{q_0^2 - q^2}{3} w_2(q_0,q^2)
\right] 
\ee 
with 
\be 
n = 5 + 2l
\ee 
and $l=0$ in the real world. Using a more convenient set 
of kinematical variables 
\be 
\epsilon = M_B - q_0 - \sqrt{M_D^2 + q_0^2 - q^2} \; , 
\; \;  
T=\sqrt{M_D^2 + q_0^2 - q^2} - M_D 
\ee 
one obtains 
$$ 
\gamma _n = \frac{1}{\pi}\int _0^{T_{max}}dT(T+M_D) 
\sqrt{T^2 + 2M_DT} \int _0^{\epsilon _{max}}
d\epsilon 
(\Delta ^2 - 2 M_BT - 2\Delta \epsilon + 2T\epsilon 
+ \epsilon ^2)^l \cdot 
$$
\be 
\cdot \left[ 
(\Delta ^2 - 2M_BT - 2\Delta \epsilon + 2T\epsilon 
+ \epsilon ^2) w_1(T,\epsilon ) + 
\frac{T^2 + 2M_DT}{3} w_2(T,\epsilon ) 
\right] 
\label{gamman} 
\ee
with 
\be 
\Delta = M_B - M_D\, , \; 
T_{max}=\frac{\Delta^2}{2M_B} \, , \; 
\epsilon _{max} = \Delta - T - 
\sqrt{T^2 + 2M_DT} 
\ee 
Expanding the integrand in 
Eq.(\ref{gamman}) in $\epsilon$ 
we find through order $\epsilon$
$$ 
\gamma _n = \frac{1}{\pi} \int _0^{M_B/2} dT T^2 M_B^l 
(M_B - 2T)^l 
\left[ 
M_B(M_B - 2T)\int _0^{M_B - 2T} d\epsilon w_1 - 
\right. 
$$ 
$$ 
\left. 
2(l+1)(M_B - T) \int _0^{M_B - 2T} d\epsilon 
\epsilon w_1 
\right] + 
$$ 
$$  
+ \frac{1}{3\pi} \int _0^{M_B/2} dT T^4 M_B^{l-1}  
(M_B - 2T)^{l-1}  
\left[ 
M_B(M_B - 2T)\int _0^{M_B - 2T} d\epsilon w_2 - 
\right. 
$$ 
\be 
\left. 
2l(M_B - T) \int _0^{M_B - 2T} d\epsilon 
\epsilon w_2 
\right]
\ee 
With the sum rules 
\bea 
\frac{1}{2\pi}\int d\epsilon w_1 &=& 1 \; , \; \; 
\frac{1}{2\pi}\int d\epsilon \epsilon w_1 = M_B - m_b 
=\bar \Lambda 
\\ 
\frac{1}{2\pi}\int d\epsilon w_2 &=& \frac{2m_b}{T} 
\; , \; \; 
\frac{1}{2\pi}\int d\epsilon \epsilon w_2 = 
\frac{2m_b}{T}\bar \Lambda
\eea 
integration over $T$ yields 
$$  
\gamma _n = 
\frac{M_B^{2l+5}}{2(l+4)(l+3)(l+2)}
\left[ 
1 - (2l+5)\frac{\bar \Lambda }{M_B} 
\right] + 
$$ 
\be 
+ \frac{m_bM_B^{2l+4}}{2(l+4)(l+3)(l+2)(l+1)}
\left[ 
1 - (2l+4)\frac{\bar \Lambda }{M_B} 
\right]
\ee 
Since 
\be 
M_B\left( 1-\frac{\bar \Lambda}{M_B}
+{\cal O}(1/M_B^2) \right) = m_b 
\ee  
we see that the meson mass $M_B$ is replaced by 
the quark mass $m_b$ actually for {\em any} value 
of $n= 5 + 2l$  -- due to the imposition of the 
sum rules! That means that the two aspects of 
long-distance dynamics, namely `dressing up'  
\begin{itemize}
\item 
the quark masses into hadron masses and 
\item 
quark operators into hadron operators, 
\end{itemize}  
actually cancel out as far as decay widths 
are concerned, 
and that this is implemented through the sum rules.

\subsection{Basic Elements}

\subsubsection{Heavy Quark Masses}
An internally consistent definition of the 
heavy quark mass is crucial for $1/m_Q$ expansions 
conceptually as well as for quantitative 
studies. While this 
remark is obvious in hindsight, the theoretical 
implications were at first not fully appreciated. 

In QED it is very natural to adopt the 
pole mass for the electron, which is defined as the 
position of the pole in the electron Green function 
(actually the beginning of the cut, to be more precise): 
it is gauge invariant and it can be measured since 
it represents the mass of an isolated electron. 
For quarks the situation is qualitatively different 
because of confinement! 
Yet computational convenience suggested to use the 
pole mass for quarks as well: while not measurable, 
it is still gauge invariant and {\em perturbatively} 
infrared stable {\em order by order}. It thus constitutes 
a useful theoretical construct -- as long as one addresses 
{\em purely perturbative} effects. Yet the pole mass is 
{\em not}  
infrared stable in {\em full} QCD -- it exhibits a 
renormalon ambiguity: 
\be 
\frac{\delta _{IR}m_Q^{pole}}{m_Q} \sim 
{\cal O}\left( \frac{\Lambda _{QCD}}{m_Q} \right) 
\label{RENORMALON} 
\ee
The origin of this {\em irreducible} uncertainty 
can be understood on physical grounds by considering 
the energy stored in the chromomagnetic field in a sphere 
of radius $R \gg 1/m_Q$ around a static colour 
source of mass $m_Q$: 
\be 
\delta {\cal E}_{Coul}(R) \propto 
\int _{1/m_Q \leq |x| < R} d^3x \vec E_{Coul}^2 
\propto {\rm const.} - 
\frac{\alpha _S(R)}{\pi}\frac{1}{R} 
\ee    
The definition of the pole mass amounts to 
setting $R \to \infty$; i.e., in evaluating the 
pole mass one undertakes to integrate the energy density 
associated with the colour source over {\em all space}  
assuming it has a Coulomb form as inferred from 
perturbation theory. Yet in the full theory the colour 
interaction becomes strong at distances approaching 
$R_0 \sim 1/\Lambda _{QCD}$, and 
the colour field can no longer be approximated by 
a $1/R$ field. Thus the long-distance 
or infrared region around and 
beyond 
$R_0$ cannot be included in a meaningful way; its 
contribution has to be viewed as an intrinsic 
uncertainty in the pole mass which is then estimated 
in accordance with Eq.(\ref{RENORMALON}) 
\footnote{The reader can rest assured that 
Eq.(\ref{RENORMALON}) is derived in a more 
rigorous way.}. 

Why the pole mass is in principle inappropriate 
for heavy quark expansions can be read off from 
Eq.(\ref{RENORMALON}) directly: its uncertainty 
of order $1/m_Q$ would dominate the leading 
nonperturbative contributions of order $1/m_Q^2$ 
one works so hard to incorporate -- in particular 
when entering through the high power $m_Q^5$! 
Instead one needs a {\em running} mass $m_Q(\mu )$ 
defined at a scale $\mu$ that shields it against 
the infrared dynamics. {\em In principle} any scale 
$\mu \gg \Lambda _{QCD}$ will do; yet {\em in practise} 
some provide a significantly more favourable 
computational environment than others.  
It can be shown that 
$\mu \sim 1$ GeV is an appropriate scale for these 
purposes whereas $\mu \simeq m_Q$ leads to 
higher order perturbative contributions 
that are artificially large 
\cite{FIVEINFINITY}.  

{}From the measured masses of the charm and beauty 
hadrons one infers 
\be 
m_b - m_c \simeq 3.50 + 40 \, {\rm MeV} \cdot 
\frac{\mu _{\pi}^2 - 0.5 \, {\rm GeV^2}}{0.1\, {\rm GeV^2}}  
\pm 0.015 \, {\rm GeV}
\label{MBMC} 
\ee 
An analysis of $e^+e^- \to \,  had$ just above the 
threshold for open beauty production yields 
\be 
m_b(1\, {\rm GeV}) = 4.64 \, {\rm GeV} 
\pm 0.05 \, {\rm GeV}  
\label{mb} 
\ee 
These results could suffer from some potential 
theoretical uncertainties not stated in 
Eqs.(\ref{MBMC}, \ref{mb}). In the future we will be 
able to extract $m_b$ and $m_c$ in a systematically 
independant way, namely from the {\em spectra} 
of semileptonic and radiative $B$ decays.

\subsubsection{Matrix Elements}

Expanding the expectation value of the leading operator 
$\bar QQ$ in powers of $1/m_Q$ yields 
\be 
\frac{1}{2M_{P_Q}} 
\matel{P_Q}{\bar QQ}{P_Q} = 
1 - \frac{\mu _{\pi}^2}{2m_Q^2} + 
\frac{\mu _G^2}{2m_Q^2} 
+{\cal O}(1/m_Q^3) 
\label{QQ}
\ee
where $P_Q$ [$V_Q$] denotes a pseudoscalar [vector] 
meson with quantum number $Q$. 

The value of 
$\mu _G^2$ for mesons is deduced from their hyperfine splitting: 
\bea 
\mu _G^2 &=& \frac{1}{2M_{P_Q}}
\matel{P_Q}{\bar Q \frac{i}{2} \sigma \cdot G Q}{P_Q} 
= \frac{3}{4} (M^2_{V_Q} - M^2_{P_Q}) \simeq \\ 
&\simeq& \frac{3}{4} \left( M^2_{B^*} - M^2_B \right) 
\simeq 0.36 \, {\rm GeV} \; .  
\eea 
For baryons one finds 
\be 
\matel{\Lambda _Q}{\bar Q \frac{i}{2} \sigma \cdot G Q}
{\Lambda _Q} \simeq 
\matel{\Xi _Q}{\bar Q \frac{i}{2} \sigma \cdot G Q}
{\Xi _Q} \simeq 0 \neq \matel
{\Omega _Q}{\bar Q \frac{i}{2} \sigma \cdot G Q}
{\Omega _Q} \; , 
\ee
since the light diquark system in $\Lambda _Q$ and 
$\Xi _Q$ carries spin zero, yet spin one for 
$\Omega _Q$. 

For the expectation value of the kinetic operator we have, 
as stressed above, a field theoretical inequality 
derived from the sum rules: 
\be 
\mu _{\pi}^2 > \mu _G^2
\ee 
There is significant evidence that it cannot exceed 
this lower bound by a lot. 

The expectation values of the dimension-six four-fermion 
operators, which provide the main motor driving 
differences in the decays widths of the various 
mesons of a given heavy flavour, cannot be 
deduced from first principles; their estimates 
suffer from still considerable uncertainties. 
Their size is usually calibrated by the so-called 
{\em factorization}  or {\em vacuum saturation ansatz}. 
There is a lively debate in the literature between 
the advocates and the agnostics \cite{PRO,CON}. 
There is some recent evidence \cite{SACH} 
from lattice simulations 
of QCD indicating that factorization apparently 
holds to better accuracy than anticipated by the 
agnostics. Two aspects that should not be in dispute 
(although it often goes unappreciated) are: 
\begin{itemize}
\item 
The dynamical content of the factorization ansatz very 
significantly
depends on the scale at which it is assumed; nonfactorizable 
contributions at one scale can become factorizable at a 
lower scale and vice versa! 
\item 
If factorization provides a valid approximation anywhere, 
it can 
be only at a low scale of around 1 GeV. 
\end{itemize}

\subsection{Lifetime Ratios}
Predictions on the lifetime ratios among beauty hadrons 
were inferred from the Heavy Quark Expansions a few years ago 
well before data of sufficient accuracy were available 
\cite{MIRAGE,BSTONE,PRO}: 
\be 
\frac{\tau (B^-)}{\tau (B_d)} \simeq 1 + 
0.05 \cdot \left( \frac{f_B}{200 \, {\rm MeV}}\right) ^2 
\label{BdBu}
\ee 
\be 
\frac{\bar \tau (B_s)}{\tau (B_d)} \simeq 1 \pm 
{\cal O}(0.01) 
\label{BdBs}
\ee 
where $\bar \tau (B_s)$ denotes the average lifetime of the 
two $B_s$ mass eigenstates. 
\be 
\frac{\tau (\Lambda _b)}{\tau (B_d)} \simeq 0.9 - 0.95
\label{BdBAR}
\ee 

More recently also the $B_c$ lifetime was predicted using 
the $B$ and $D$ lifetimes as input 
\cite{BCBIGI,BB}: 
\be 
\tau (B_c) \simeq 0.5 \; {\rm psec} 
\label{Bc} 
\ee 

Data from the LEP collaborations, CDF, SLD and CLEO yield 
as world averages 
\bea 
\frac{\tau (B^-)}{\tau (B_d)} &=&   1.07 \pm 0.03 \\ 
\frac{\bar \tau (B_s)}{\tau (B_d)} &=&   0.94 \pm 0.04 \\ 
\frac{\tau (\Lambda _b)}{\tau (B_d)} &=& 0.79 \pm 0.05 \\ 
\tau (B_c) &=& 0.46 \pm 0.17 \; {\rm psec} 
\label{BLIFESDATA} 
\eea

While the predictions on $\tau (B^-)/\tau (B_d)$, 
$\bar \tau (B_s)/\tau (B_d)$ and $\tau (B_c)$ 
are fully consistent with the data, the observed value 
of $\tau (\Lambda _b)$ is significantly lower than the 
predicted one. A few comments are in order to evaluate 
the situation: 
\begin{itemize}
\item 
The prediction of Eq.(\ref{BdBu}) has 
been criticized theoretically as not sufficiently 
conservative \cite{NEUSACH}; yet a recent 
lattice simulation of QCD yielded 
\be 
\frac{\tau (B^-)}{\tau (B_d)} = 
1.03 \pm 0.02 \pm 0.03  \; , 
\ee
which is fully consistent with Eq.(\ref{BdBu}).
\item 
Also Eq.(\ref{BdBAR}) has been criticised theoretically 
\cite{NEUSACH}. Other studies, however, find 
\cite{URIBOOST}: 
\be 
\frac{\tau (\Lambda _b)}{\tau (B_d)} = 1 - \Delta \; , 
\; \; \Delta \sim 0.03 - 0.12 \; , 
\ee 
which again is fully consistent with the original prediction 
of Eq.(\ref{BdBAR}), but hardly with the data. 
A recent careful reanalysis of four-quark operators 
\cite{PIRJOLURI} 
arrives at basically the same conclusion.  
No appealing fix for the prediction has been found 
within the framework of HQE. 
\item 
This discrepancy has prompted the radical ansatz that 
the weak widths of the various beauty hadrons scale 
with the fifth power of their hadronic mass rather 
than the $b$ quark mass \cite{ALTANO}: 
\be 
\Gamma (H_b) \propto G_F^2 M_{H_b}^5
\ee 
yielding 
\footnote{This feature was first noted in 
\cite{MIRAGE} well before there was any sign of a 
`short' $\Lambda _b$ lifetime.} 
\be 
\frac{\tau (\Lambda _b)}{\tau (B_d)} = 
\left(\frac{M_B}{M_{\Lambda _b}}\right)^5 \simeq 0.73 
\ee 
This ansatz is quite radical in  
that it cannot be reconciled with the OPE approach: 
for it manages to drive down the $\Lambda _b$ lifetime 
through a correction of order $1/m_b$ -- and actually 
one with a large coefficient: 
\be 
M_{\Lambda _b}^5 = (m_b + \bar \Lambda )^5 = 
m_b^5 \left( 1 + 5 \frac{\bar \Lambda }{m_b} + ... 
\right) 
\ee 
that is unequivocally anathema to the OPE! 
This is often referred to as a violation of 
local duality. On the phenomenological side one 
should note its prediction 
\be 
\frac{\bar \tau (B_s)}{\tau (B_d)}\simeq 
\left( \frac{M_{B_d}}{M_{B_s}} \right) ^5 
\simeq 0.92 \; , 
\ee 
i.e., smaller than the HQE prediction, Eq.(\ref{BdBs}). 
\item 
I prefer to wait and see what the CDF/D0 data in run II 
will yield before seriously contemplating such 
drastic measures. 
\end{itemize}
In passing I would like to add another comment: 
Predictions based on the 
{\em theoretical} framework of HQE implemented through the OPE 
are more unequivocal than those inferred from models with 
their ad-hoc assumptions as central elements. Then even 
a failure will teach us a meaningful, even if sad lesson.

\subsection{Semileptonic Transitions}

Semileptonic decays of beauty mesons represent a somewhat 
less complex dynamical scenario since factorization 
of the quark and the lepton bilinears is guaranteed to 
hold (although, in my judgement, this point is often 
exaggerated in the literature). Here I want to address 
one aspect of it only, although it is of central 
importance, namely the accurate extraction 
of the CKM parameters $|V(cb)|$ and $|V(ub)|$. 

\subsubsection{Total Semileptonic Widths}

HQE yields for the total semileptonic width 
$$  
\Gamma (B \ra l X_c) = 
\frac{G_F^2m_b^5 |V(cb)|^2}{192 \pi ^3} \times  
$$ 
\be 
\left[ z_0(m_c^2/m_b^2) \cdot  
\left( 1 - \frac{\mu _{\pi}^2 - \mu _G^2}{2m_b^2} \right) 
- 2 \left( 1 - \frac{m_c^2}{m_b^2} \right) ^4 
\frac{\mu _G^2}{m_b^2} - \frac{2\alpha _S}{3\pi} 
\cdot z_0^{(1)}(m_c^2/m_b^2) + ... \right] 
\label{GAMMASL} 
\ee
where the omitted terms are higher-order perturbative 
and/or power suppressed corrections; $z_0$ and 
$z_0^{(1)}$ are known phase space factors 
depending on $m_c^2/m_b^2$. 

The leading nonperturbative corrections of order 
$1/m_b^2$ are small reducing the width by about 
5 \% and the direct impact of higher-order 
power corrections is estimated to be rather 
negligible. They exert, however, a very considerable 
indirect influence. For Eq.(\ref{GAMMASL}) 
makes it obvious that the choice of the 
heavy mass $m_b$ is of crucial importance. 
As discussed before, usage of the pole mass 
would introduce considerable intrinsic uncertainties 
that are actually larger than the leading 
nonperturbative contributions. Instead one has to use 
the running mass evaluated at a low scale 
$\mu$ chosen around 1 GeV. 

Comparing this theoretical expresssion with the data 
leads to an extraction of $|V(cb)|$: 
\bea 
|V(cb)| = &0&.0419 
\sqrt{\frac{BR(B \ra l X_c)}{0.105}} 
\sqrt{\frac{1.55 \, {\rm psec}}{\tau (B)}} \times \\ 
&\times& \left( 
1 - 0.012 \cdot 
\frac{\mu _{\pi}^2 - 0.5 \, {\rm GeV^2}}
{0.1 \, {\rm GeV^2}} 
\right) 
\left( 
1 - 0.01 \cdot \frac{\delta m_b(\mu )}{50 \, {\rm MeV}}
\right) \times  
\\
&\times& \left( 
1 + 0.006 \cdot 
\frac{\alpha _S^{\rm \bar MS}(1\, {\rm GeV^2})  - 0.336}
{0.02} 
\right) 
\left( 
1 + 0.007 \cdot \frac{\bar \rho ^3}{0.1 \, {\rm GeV^3}}
\right) \; , 
\label{VCBINCL1}
\eea 
where $\bar \rho ^3$ reflects the contributions from the 
$1/m_b^3$ terms. These detailed list of possible uncertainties 
can be analyzed -- for details see \cite{HQEREV} -- 
yielding 
\bea 
|V(cb)| = &0&.0419 
\sqrt{\frac{BR(B \ra l X_c)}{0.105}} 
\sqrt{\frac{1.55 \, {\rm psec}}{\tau (B)}} \times \\ 
&\times& \left( 
1 - 0.012 \cdot 
\frac{\mu _{\pi}^2 - 0.5 \, {\rm GeV^2}}
{0.1 \, {\rm GeV^2}} 
\right) \times \\
&\times& \left( 
1\pm 0.015_{pert} \pm 0.01_{m_b} \pm 0.012_{nonpert} 
\right) 
\label{VCBINCL2} 
\eea 
I find it appropriate to combine various {\em theoretical} 
errors with all their correlations (and biases?) 
{\em linearly} rather than quadratically; thus I assign 
5\% as overall theoretical uncertainty to the extraction 
of $|V(cb)|$ from the total semileptonic width. 

Some authors in the past have claimed a considerably 
larger theoretical uncertainty on this extraction. 
They expressed the theoretical width in terms of the 
{\em pole} masses to which they somehow assigned a 
10\% uncertainty. Using the BLM resummation they 
found the coefficients of the $\alpha _S^2$ 
radiative corrections very large -- between 
$\simeq -10$ and $-20$ for $b\ra c$ and even 
$\simeq -30$ for $b \ra u$ -- reducing the overall width 
by rouhly 10\% or so. If that were the whole 
story, one would have to argue that the theoretical 
uncertainties are considerably larger than stated above; 
it would actually be legitimate to be concerned whether 
the results of such a procedure could be trusted at all! 

What was overlooked by those claims, however, is that those 
two uncertainties -- the one in $m_b$ and the one in 
the $\alpha _S$ expansion -- are highly correlated to 
each other and thus combine to create a much smaller 
overall error. This can be best seen by evaluating 
the masses at a scale around 1 GeV, as advocated before. 

The KM suppressed semileptonic width $B \ra l X_u$ 
can be expressed in terms of $|V(ub)|$ with almost as 
good an accuracy \cite{URIBU} 
\bea 
|V(ub)| = &0&.00465 
\sqrt{\frac{BR(B \ra l X_u)}{0.002}} 
\sqrt{\frac{1.55 \, {\rm psec}}{\tau (B)}} \times \\ 
&\times& \left( 
1\pm 0.025_{pert} \pm 0.03_{m_b} \pm 0.01_{nonpert} 
\right) 
\label{VUBINCL1} 
\eea  
\begin{itemize} 
\item 
since the energy release -- the inverse of which 
constitutes the expansion parameter -- 
is larger here than in $b\ra c$, 
\item 
and the dependance on $\mu _{\pi}^2$ is practically absent, 
\item 
whereas $m_b$ is less well-known than the mass splitting 
$m_b - m_c$ which largely controls $b \ra l X_u$.  
\end{itemize}
Pilot studies by ALEPH and DELPHI exhibit a width for 
$B \ra l X_u$ a bit below $2 \cdot 10^{-3}$ 
\cite{ACHILLE} suggesting 
\be 
|V(ub)| \sim 4 \cdot 10^{-3}
\ee 
somewhat larger, though consistent with what analyses of 
the exclusive modes 
$B \ra l \nu \pi$ and $B \ra l \nu \rho$ find where one 
has to invoke models for the hadronic form factors. 

The data on $\Gamma (B \ra l\nu  X_u)$ will improve 
significantly with input from CLEO, BELLE and 
BABAR. One particularly useful discriminator 
between $\Gamma (B \ra l\nu  X_u)$ and 
$\Gamma (B \ra l\nu  X_c)$ will be provided by 
measuring the spectra of the hadronic recoil masses 
for which models exist that implement 
all known constraints from QCD 
\cite{DIKEMAN1,DIKEMAN2,LIGETI}.

\subsubsection{$B \ra l \nu D^*$}

Introduction of the universal Isgur-Wise function was a 
crucial step in the evolution of heavy quark theory. 
It also provides great practical help in treating 
{\em exclusive} channels: 
it tells us that certain form factors --  
in particular the one for $B \ra l \nu D^*$ -- 
have to be normalized to unity in the infinite mass 
limit. 

Analyzing the data on $B \ra l \nu D^*$ 
and extrapolating them to the kinematical 
point of zero recoil yields  
\be 
|F_{D^*}(0) V(cb)| = 0.0339 \pm 0.0014 
\label{FDVCB}
\ee 
While $F_{D^*}(0) =1$ holds asymptotically, 
it will receive power suppressed and perturbative 
corrections for finite quark masses: 
\be 
F_{D^*}(0) = 1 + 
{\cal O}\left( \frac{\alpha _S}{\pi}\right) + 
{\cal O}\left( \frac{1}{m_c^2}\right) + 
{\cal O}\left( \frac{1}{m_cm_b}\right) + 
{\cal O}\left( \frac{1}{m_b^2}\right)
\ee 
The absence of $1/m_Q$ corrections here noted in 
passing in \cite{VS} was cast into the form 
of a theorem by Luke \cite{LUKE}. 

Essential information on this form factor can be 
inferred from the sum rules for axialvector 
currents: 
\be 
F_{D^*}(0) \simeq 0.91 - 
0.013 \cdot \frac{\mu _{\pi}^2 - 0.5 \, {\rm GeV^2}}
{0.1 \, {\rm GeV^2}} \pm 0.02_{excit} \pm 
0.01_{pert} \pm 0.025 _{1/m_Q^3}  
\ee 
where the number with the subscript $excit$ 
represents the estimate for how much 
the excitations beyond $D^*$ contribute to the 
sum rule. Putting everything together 
leads to  
\be  
F_{D^*}(0) \simeq 0.91 \pm 0.06 
\ee 
where I have again resisted the temptation to 
combine theoretical errors in quadrature! 

{}From the data stated in Eq.(\ref{FDVCB}) one then 
infers 
\be 
|V(cb)|_{excl} = 0.0377 \pm 0.0016|_{exp} \pm 0.0025 |_{theor}
\label{VCBEX}
\ee 
The two determinations in Eqs.(\ref{VCBINCL2}) and (\ref{VCBEX}) 
are systematically very different both in their experimental and 
theoretical aspects. Nevertheless they are quite consistent with 
each other with the experimental and theoretical uncertainties 
being very similar. A few years ago it would have seemed 
quite preposterous to claim such small theoretical uncertainties!

\subsubsection{Future Improvements}

I am quite confident that the uncertainties on 
$|V(cb)|$ can be reduced from the 
present 5\% level down to the 2\% level in the foreseeable future.  

$|V(ub)|$ (or $|V(ub)/V(cb)|$) is not known with an even remotely 
similar accuracy, and so far one has relied on models rather 
than QCD proper to extract it from data. Yet I expect  
that over the next ten years $|V(ub)|$  will be determined with a 
theoretical uncertainty below 10\% . It will be important to obtain 
it from systematically different semileptonic distributions and 
processes; Heavy Quark Theory provides us with the 
indispensable tools for combining the various analyses in a 
coherent fashion. 

This theoretical progress can embolden us to hope that in the end 
even $|V(td)|$ can be determined with good accuracy -- say 
$\sim 10 \div 15\%$ -- from 
$\Gamma (K^+ \ra \pi ^+ \nu \bar \nu )$, 
$\Delta m(B_s)$ vs. $\Delta m(B_d)$ or 
$\Gamma (B \ra \gamma \rho /\omega )$ vs. 
$\Gamma (B \ra \gamma K^* )$ etc. 

\section{The Cathedral Builders' Paradigm}
\subsection{The Paradigm}

The dynamical ingredients for numerous and multi-layered 
manifestations of CP and T violations do exist or are likely to exist. Accordingly one searches 
for them in many phenomena, namely in  
\begin{itemize}
\item 
the neutron electric dipole moment probed with ultracold 
neutrons at ILL in Grenoble, France; 
\item 
the electric dipole moment of electrons studied through the 
dipole moment of atoms at Seattle, Berkeley and Amherst in the US; 
\item 
the transverse polarization of muons in 
$K^- \ra \mu ^- \bar \nu \pi ^0$ at KEK in Japan; 
\item 
$\epsilon ^{\prime}/\epsilon _K$ as obtained from $K_L$ 
decays at FNAL and CERN and soon at DA$\Phi$NE in Italy; 
\item 
in decay distributions of hyperons at FNAL; 
\item 
likewise for $\tau$ leptons at CERN, the beauty factories and BES 
in Beijing; 
\item 
CP violation in the decays of charm hadrons produced 
at FNAL and the beauty factories; 
\item 
CP asymmetries in beauty decays at DESY, at the beauty 
factories at Cornell, SLAC and KEK, at the FNAL collider and 
ultimately at the LHC. 

\end{itemize} 
A quick glance at this list already makes it clear 
that frontline research on this topic 
is pursued at all high energy labs in the world -- and then some; 
techniques from several different branches of physics -- 
atomic, nuclear and high energy physics -- are harnessed in 
this endeavour together with a wide range of set-ups. 
Lastly, experiments are performed at the lowest temperatures 
that can be realized on earth -- ultracold neutrons -- and at the 
highest -- in collisions produced at the LHC. And all of that dedicated 
to one profound goal. 
At this point I can explain what I mean by the term 
"Cathedral Builders' Paradigm". 
The building of cathedrals required interregional collaborations, 
front line technology (for the period) from many different fields 
and commitment; it had to be based on solid foundations -- and 
it took time. The analogy to the ways and needs of high energy 
physics are obvious -- but it goes deeper than that. 
At first sight a cathedral looks 
like a very complicated and confusing structure with something 
here and something there. Yet further scrutiny reveals that 
a cathedral is more appropriately 
characterized as a complex rather than a complicated 
structure, one that is multi-faceted and multi-layered -- 
with a coherent theme! One cannot (at least for 
first rate cathedrals) remove any of its elements 
without diluting (or even destroying) its technical soundness and 
intellectual message. Neither can one in our efforts to come to grips 
with CP violation!  

\subsection{Summary}
\begin{itemize} 
\item 
We know that CP symmetry is not exact in nature since 
$K_L \ra \pi \pi $ proceeds and presumably because we  
exist, i.e. because the baryon number of the universe does 
{\em not} vanish. 
\item 
If the KM mechanism is a significant actor in $K_L \ra \pi \pi$ 
transitions then there must be large CP asymmetries in the decays 
of beauty hadrons. In $B^0$ decays they 
are naturally measured in units of 10 \%!  
\item 
Some of these asymmetries are predicted with high parametric 
reliability. 
\item 
New theoretical technologies will allow us to translate such 
parametric reliability into quantitative accuracy. 
\item 
Any significant difference between certain KM predictions for the 
asymmetries and the data reveals the intervention of New Physics. 
There will be no `plausible deniability'.  
\item 
We can expect 10 years hence the theoretical uncertainties in 
some of the 
predictions to be reduced below 10 \% . 
\item 
I find it likely that deviations from the KM predictions will show up 
on that level. 
\item 
Yet to exploit this discovery potential to the fullest one will have to 
harness the statistical muscle provided by beauty production 
at hadronic colliders. 
\end{itemize}

\subsection{Outlook}
I want to start with a statement about the past: 
{\em The comprehensive study of kaon and hyperon physics 
has been instrumental in guiding us to the Standard Model.}  
\begin{itemize}
\item 
The $\tau -\theta $ puzzle led to the realization that parity is not 
conserved in nature. 
\item 
The observation that the production rate exceeded the decay rate 
by many orders of magnitude -- this was the origin of the 
name `strange particles' -- was explained through postulating 
a new quantum number -- `strangeness' -- conserved by the strong, 
though not the weak forces. This was the beginning of the second 
quark family. 
\item 
The absence of flavour-changing neutral currents was incorporated 
through the introduction of the quantum number `charm', which 
completed the second quark family. 
\item 
CP violation finally led to postulating yet another, the third 
family. 
\end{itemize}
All of these elements which are now essential pillars of the Standard 
Model were New Physics at {\em that} time! 

I take this historical 
precedent as clue that a detailed, comprehensive and thus 
neccessarily long-term program on beauty physics will lead to a 
new paradigm, a {\em new} Standard Model! 

CP violation is a fundamental as well as mysterious phenomenon 
that we have not understood yet. This is not surprising: after all 
according to the KM mechanism CP violation enters through the 
quark mass matrices; it thus relates it to three central 
mysteries of the Standard Model: 
\begin{itemize}
\item 
How are fermion masses generated? 
\footnote{Or more generally: how are masses produced in 
general? For in alternative models CP violation enters through 
the mass matrices for gauge bosons and/or Higgs bosons.}  
\item 
Why is there a family structure?
\item 
Why are there three families rather than one?
\end{itemize} 
In my judgement it would be unrealistic to expect that these 
questions can be answered through pure thinking. I strongly 
believe we have to appeal to nature through experimental efforts to 
provide us with more pieces that are surely missing in the 
puzzle. CP studies are essential in obtaining the full dynamical 
information contained in the mass matrices or -- in the language 
of v. Eichendorff's poem quoted in the beginning, "to find the 
magic word" that will decode nature's message for us. 

Considerable progress has been made in theoretical engineering 
and developing a comprehensive CP phenomenology from 
which I conclude: 
\begin{itemize}
\item 
$B$ decays constitute an almost ideal, certainly optimal and unique 
lab. Personally I believe that even if no deviation 
from the KM predictions were uncovered, we would find 
that the KM parameters, in particular the angles of the 
KM triangle, carry special values that would give us 
clues about New Physics. Some very interesting 
theoretical work is being done about how GUT dynamics in 
particular of the SUSY (or Supergravity) variety operating 
at very high scales would shape the observable 
KM parameters. 
\item 
A comprehensive analysis of charm decays with special emphasis on  
$D^0 - \bar D^0$ oscillations and CP violation is a moral 
imperative! Likewise for $\tau$ leptons. 
\item 
A vigorous research program must be pursued for light 
fermion systems, namely in the decays of kaons and hyperons 
and in electric dipole moments. After all it is conceivable 
of course that no CP asymmetries are found 
in $B$ decays on a measurable level. 
Then we would know that the KM ansatz is {\em not} a 
significant actor in $K_L \ra \pi \pi$, that New Physics 
drives it -- 
but what kind of New Physics would it be? Furthermore even if 
large CP asymmetries were found in $B$ decays, it could 
happen that the signals of New Physics are obscured by 
the large `KM background'. This would not be the case 
if electric dipole moments were found or a transverse 
polarization of muons in $K_{\mu 3}$ decays. 
\item 
Close feedback between experiment and theory will be 
essential. 

\end{itemize}
 As the final summary: insights about Nature's 
Grand Design that can be obtained from a 
comprehensive and detailed program of CP studies 
\begin{itemize}
\item 
are of fundamental importance, 
\item 
cannot be obtained any other way and 
\item 
cannot become obsolete!

\end{itemize}

\bigskip 

{\bf Acknowledgments:} \hspace{.4em} 
I am deeply grateful to my collaborators A. Sanda, M. Shifman, 
N. Uraltsev and A. Vainshtein for their generosity in sharing 
their insights with me over so many years. I also would like 
to thank the organizers of this school -- 
in particular Profs. J.B. Choi and D.S. Hwang -- 
for their invitation 
to lecture in such a sophisticated and ancient 
metropolis, for making my stay so truly enjoyable and 
introducing me to Korea's fascinating history and culture. 
This work was supported by the National Science Foundation under
the grant numbers PHY 92-13313 and PHY 96-05080.

\eject


\begin{thebibliography}{99}
\bibitem{WINSTEIN} 
I have learnt this notation -- using $\Phi (\Delta S=1,2)$ rather 
than $\epsilon _K$ and $\epsilon ^{\prime}$ -- from B. Winstein.
\end{thebibliography}
\end{document}